\documentclass{aa}  
\usepackage{caption}
\usepackage{graphicx}
\usepackage{txfonts}
\usepackage{hyperref}
\usepackage{amsmath}
\usepackage{array}
\usepackage{natbib}
\usepackage{caption}
\usepackage{xcolor}
\usepackage{float}
\usepackage[utf8]{inputenc}
\usepackage{booktabs}
\usepackage{dblfloatfix}
\usepackage{placeins}
\usepackage{geometry}
\usepackage{graphicx}
\usepackage{caption}
\usepackage{titlesec}
\usepackage{subcaption}
\usepackage{placeins} 
\usepackage{orcidlink}

\newcommand{\nustar} {\textit{NuSTAR}\xspace}
\newcommand{\Ecyc}{\ensuremath{E_{\mathrm{cyc} }}}

\newcommand{\src}{\object{4U\,1538-52}}
\titlerunning{How the spin-phase variability of cyclotron lines shapes the pulsed fraction spectra} 
\authorrunning{D. K. Maniadakis et al.} 

\begin{document} 

\title{How the spin-phase variability of cyclotron lines shapes the pulsed fraction spectra: insights from \src}


\author{
    Dimitrios K. Maniadakis\orcidlink{0009-0008-1148-2320} \thanks{Email: \href{mailto:dim.maniad@gmail.com}{dim.maniad@gmail.com}} \inst{1,2}
    \and Ekaterina Sokolova-Lapa\orcidlink{0000-0001-7948-0470} \inst{3}
    \and Antonino D'Aì\orcidlink{0000-0002-5042-1036} \inst{1}
    \and Elena Ambrosi\orcidlink{0000-0002-9731-8300} \inst{1}
    \and Carlo Ferrigno\orcidlink{0000-0003-1429-1059} \inst{4,5}
    \and Giancarlo Cusumano\orcidlink{0000-0002-8151-1990} \inst{1}
    \and Alessio Anitra\orcidlink{0000-0002-2701-2998} \inst{2,6}
    \and Luciano Burderi\orcidlink{0000-0001-5458-891X} \inst{1,6}
    \and Melania Del Santo\orcidlink{0000-0002-1793-1050} \inst{1}
    \and Tiziana Di Salvo\orcidlink{0000-0002-3220-6375} \inst{2}
    \and Felix Fürst\orcidlink{0000-0003-0388-0560} \inst{7}
    \and Rosario Iaria\orcidlink{0000-0003-2882-0927} \inst{2}
    \and Valentina La Parola\orcidlink{0000-0002-8087-6488} \inst{1}
    \and Christian Malacaria\orcidlink{0000-0002-0380-0041} \inst{8}
    \and Peter Kretschmar\orcidlink{0000-0001-9840-2048} \inst{7}
    \and Fabio Pintore\orcidlink{0000-0002-3869-2925} \inst{1}
    \and Ciro Pinto\orcidlink{0000-0003-2532-7379} \inst{1}
    \and Guillermo Andres Rodriguez Castillo\orcidlink{0000-0003-3952-7291} \inst{1}
}

\institute{
INAF - IASF Palermo, via Ugo La Malfa 153, 90146 Palermo, Italy
\and
Dipartimento di Fisica e Chimica Emilio Segrè, Università degli Studi di Palermo, Via Archirafi 36, 90123 Palermo, Italy
\and
Dr. Karl Remeis-Sternwarte and Erlangen Centre for Astroparticle Physics, Friedrich-Alexander-Universität Erlangen-Nürnberg, Sternwartstr. 7, 96049 Bamberg, Germany
\and
Department of Astronomy, University of Geneva, Chemin d’Écogia 16, 1290 Versoix, Switzerland
\and
INAF, Osservatorio Astronomico di Brera, Via E. Bianchi 46, I-23807, Merate, Italy
\and
Dipartimento di Fisica, Università degli Studi di Cagliari, SP Monserrato-Sestu, KM 0.7, Monserrato (CA), 09042, Italy
\and
European Space Agency (ESA), European Space Astronomy Centre (ESAC), Camino Bajo del Castillo s/n, 28692 Villanueva de la Cañada, Madrid, Spain
\and
INAF - Osservatorio Astronomico di Roma, Via Frascati 33, 00076 Monte Porzio Catone (RM), Italy
}

\date{ }
  \abstract
   {}
    {We aim to study the energy-dependent pulse profile of the X-ray accreting pulsar \src\ and its phase-dependent spectral variability, with particular emphasis on the behavior around the cyclotron resonant scattering feature at  $E_\mathrm{cyc} \sim 21 \mathrm{keV}$.}
    {We analyze all available \nustar\ observations of \src.
    We decompose energy-resolved pulse profiles into Fourier harmonics to study their energy dependence. Specifically, we compute pulsed fraction spectra, cross-correlation, and lag spectra, identifying discontinuities and linking them to features in the phase-averaged spectra.
    We perform both phase-averaged and phase-resolved spectral analyses to probe spectral variability and its relation to pulse profile changes.
    Finally, we interpret our findings via physical modeling of energy- and angle-dependent pulse profile emission, performing radiative transfer in a homogeneous slab-like atmosphere under conditions relevant to \src. The emission is projected onto the observer’s sky plane to derive expected observables.}
    {In contrast to the \textit{dips} in pulsed fraction spectra observed in other sources (e.g., Her X-1), we find a broad \textit{bump} near the cyclotron resonance energy in \src. This increase is driven primarily by phase-dependent spectral variability, especially by strong variations in cyclotron line depth across different phase intervals.
    We interpret the observed contrast between \textit{dips} and \textit{bumps} in various sources as arising from phase-dependent variations of cyclotron line depth relative to the phase-modulated flux.

    We model the X-ray emission from an accreting neutron star and find that our simulations indicate high values of both the observer’s inclination and the magnetic obliquity, along with a ${\sim}10$–$15^\circ$ asymmetry between the locations of the magnetic poles. Assuming this geometry, we satisfactorily reproduce the observed pulse profiles and introduce general trends in the observables resulting from the system's geometry.
    }
   {}
   \maketitle

\section{Introduction}
\label{intro}

High-mass X-ray binary (HMXB) pulsars are systems that host a strongly magnetized neutron star (NS) that accretes material from a companion star. As the material falls towards the NS, it couples with the magnetic field lines and is channeled into the polar caps. At low accretion rates (\(\dot{M} \lesssim 10^{17} \text{ g s}^{-1}\)), the accreted material is stopped near the NS surface, forming a hotspot rather than an extended column. In this regime, the emission is primarily thermal and originates from the heated polar cap.  

At high accretion rates (\(\dot{M} \gtrsim 10^{17} \text{ g s}^{-1}\)), radiation pressure becomes significant, leading to the formation of a radiation-dominated shock above the NS surface. This shock decelerates the inflowing material, creating a column structure where matter settles more gradually onto the NS surface. The column height increases with the accretion rate, influencing the observed X-ray emission pattern \citep{Basko1976}. The accretion power is converted into radiation via thermal and non-thermal processes such as synchrotron, bremsstrahlung, and the inverse-Comptonization of all of the aforementioned photon fields. Since the rotational axis is usually not aligned with the magnetic one, the X-ray emission will appear pulsated to a distant observer.

X-ray spectral analysis is one of the standard tools to infer the properties of X-ray pulsars (XRPs). Besides the phase-averaged spectral analysis (averaging over an observational exposure time much longer than the spin period), one can also study phase-dependent spectral variations in the so-called phase-resolved (or spin-resolved) spectral analysis. To translate photon arrival times into phases (i.e, rotational phases of the NS, defined in the range from 0 to 1, where 0 corresponds to a reference phase such as the minimum of the pulse profile), one usually adopts a timing solution correcting for the orbital parameters of the system (if known) and, when significantly detected, the spin derivative terms.  

The shape of the pulse profile (PP) is the result of the combination of many fundamental properties of the NS (e.g., emission anisotropy and light bending) in a non-linear manner. The majority of the pulsed photons originate from the accretion flow near the magnetic poles of the pulsar. X-ray photons emitted from the surface are redistributed, after interacting with the relativistic particles of the accreted material, via the aforementioned non-thermal radiative processes. This interaction occurs in the presence of strong magnetic and gravitational fields of the XRP. The resulting PPs are thus energy-dependent and, in addition, they are luminosity dependent \citep{Wang1981}. 

For low luminosities, the photons escape primarily along the magnetic axis (pencil beam). However it is theorized that beyond a certain luminosity threshold, a radiative shock is formed, causing photons to escape from the sides of the accretion column (fan beam) rather than the upward direction, dramatically changing the PPs \citep[][and references therein]{Becker2022}.  

XRPs often exhibit cyclotron-resonant scattering features (CRSF) in their spectra, appearing as absorption-like features in the $10$--$100\,\mathrm{keV}$ range \citep[see][for a review]{Staubert2019}. These arise near the NS magnetic poles, where electrons in the intense magnetic field are confined to quantized Landau levels. The energy spacing of these levels is given by
\begin{equation}
\label{equation_cycl_formula}
E_{\text{cyc}} = \frac{n}{(1 + z)} \frac{\hbar e B}{m_e c} \approx \frac{n}{(1 + z)} \, 11.6 \, \text{[keV]} \times B_{12}, 
\end{equation}
where \textit{z} is the gravitational redshift, \textit{n} the Landau quantum state and \textit{$B_{12}$} the magnetic field strength in units of $10^{12}$ Gauss. When photons at these energies interact with electrons, they undergo resonant scattering, leading to a reduction in observed flux at $E_{\mathrm{cyc}}$. 
CRSFs (and harmonics) provide the only direct method to estimate the surface magnetic field of a NS, as the line-forming region is believed to be very close to the stellar surface. Observations reveal a complex phenomenology: (i) the profile of the fundamental line is non-Gaussian, (ii) CRSFs vary with pulse phase (e.g., Cen X-3), (iii) their energy shifts with luminosity (both positively and negatively) \citep{Staubert2019}, and (iv) they correlate with the spectral continuum. These behaviors suggest intricate physical processes, with ongoing theoretical efforts seeking to explain their non-linearity and dependence on accretion dynamics \citep{Poutanen2013, Loudas2024}.

\src\ is a HMXB pulsar consisting of a NS in orbit with the B0Iab supergiant companion QV Nor. The source is known for its persistent X-ray emission, with notable flaring activity attributed to the close proximity and low eccentricity ($e$\,=\,0.18) of the binary system \citep{Hemphill2014}.  
The orbital characteristics of \src\ are well-defined, with an orbital period of approximately 3.728 days \citep{Hemphill2019} and evidence of eclipses lasting about 0.2 days \citep{Rodes-Roca2009}. The orbital inclination is well constrained at 67$^\circ$(1) \citep{Falanga2015}.  The source shows also super-orbital modulation at a period of 14.9130\,$\pm$\,0.0026 days, consistent with four times the orbital period \citep{Corbet2021}. 

The pulsar has a spin period of approximately 530 seconds \citep{Davison1977}. The NS displayed many episodes of torque reversals \citep{Malacaria2020}, with the latest reversal from spin-down to spin-up at the end of 2019, although the general increase in frequency appears to be modulated by many intermittent, months-long spin-down episodes\footnote{\url{https://gammaray.msfc.nasa.gov/gbm/science/pulsars/lightcurves/4u1538.html}}, indicative of the complex interactions between the NS and the wind of the companion \citep{Karino2020, Sharma2023, Hu2024}. Timing studies have shown that the PPs remain invariant before and after torque reversals, suggesting a stable oblique rotator configuration \citep{Hu2024}.


Early studies of 4U 1538-522 X-ray spectra revealed the presence of cyclotron resonance scattering features (CRSFs)
at around 20 $\mathrm{keV}$ \citep{Clark1990}. A possible first harmonic was reported by \cite{Robba2001} at 51 $\mathrm{keV}$ and by \cite{Rodes-Roca2009} at 47 $\mathrm{keV}$. \citet{Hemphill2016} reported a secular change of the CRSF energy, approximately 1.5 $\mathrm{keV}$ over a span of 8.5 years. \citet{Varun2019} reported on the pulse-phase dependence of the spectral parameters using a 50 ks observation with \texttt{ASTROSAT}. They noted a maximum relative shift of the cyclotron line energy of 13\% (\Ecyc \space in the 20--23 $\mathrm{keV}$ range), a variation of the CRSF depth correlated with the pulse shape and the spectral continuum. 

In addition, spectral analysis has identified various emission lines, particularly in the iron K$\alpha$ band. The presence of absorption features, such as one at 2.1 $\mathrm{keV}$, has also been investigated, providing insights into the material surrounding the NS and its magnetic field \citep{Rodes-Roca2014}.

The study of energy dependence of the PPs together with the low energy position of the CRSF, allowed to set constraints on the geometry of the source. \citet{Clark1990} was the first to perform phase-resolved spectroscopy, revealing the variable nature of the CRSF and modeling the energy-dependent PPs using the framework proposed by \citet{Meszaros1985a}. Later, \citet{Cominsky1991} obtained similar results to those of \citet{Clark1990}, fitting the PPs with a model similar to that of \citet{Parmar1989} and finding evidence of a magnetic dipolar asymmetry. Further refinements to the accretion geometry were made by \citet{Bulik1992} and \citet{Bulik1995}, who developed detailed models that fully accounted for both relativistic effects and the combined intensities of radiation and magnetic fields. Their findings suggested an asymmetric configuration of the magnetic polar caps, which were neither equal in size nor strictly antipodal. \\
\par

\begin{figure}
    \centering
    \hspace*{-1cm} 
    \includegraphics[width=0.9\columnwidth]{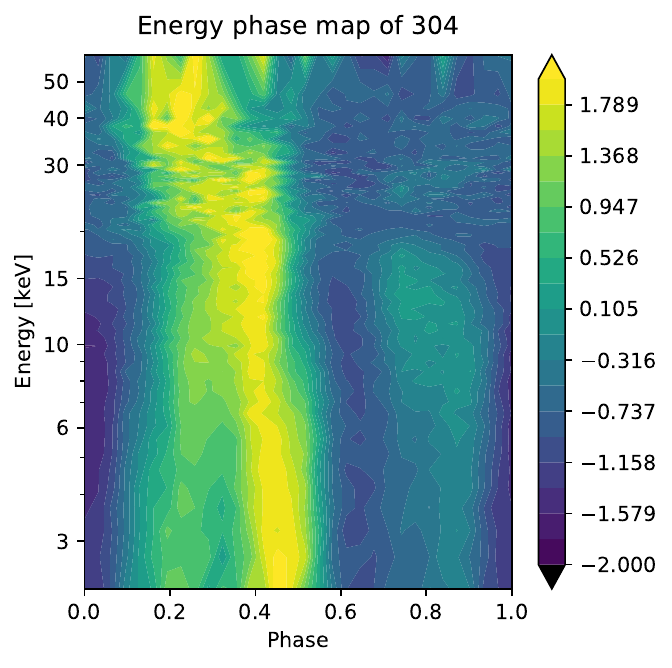}
    \caption{Energy-phase map of the observation 304, generated with 32 phase bins and a minimum S/N of 15: normalized pulse profiles at equally spaced logarithmic energy intervals. Each pulse in the bin was normalized by subtracting its mean and dividing by the standard deviation.}

    \label{mappa304}
\end{figure}

\citet{Ferrigno2023} developed a pipeline tailored for PPs analysis of \nustar\ data, that first showcased the utility of modelling, even phenomenologically, the pulsed fraction spectrum (PFS) to infer PP changes correlated with the CRSFs. Local changes in the PFS were observed at energies corresponding to features in the X-ray spectrum. Specifically, for all X-ray pulsars analyzed in that work (Her\,X-1, 4U 1626-67, Cen\,X-3, and Cep\,X-4), a \textit{dip} in the PFS was detected at the energy of the cyclotron absorption feature. In contrast, the PFS of the X-ray pulsar V 0332\,+\,53 \citep{D'Ai2025} exhibits two distinct and narrow \textit{bumps} that were interpreted as indicative of a more complex CRSF profile with extended wings. 

In this work, we aim to further investigate the connection between the PP properties and the spectral features of the source \src, using all the available \nustar data. The four observations have comparable luminosities, as reported in Table \ref{table_observations}, allowing for a coherent analysis of the pulse phase–dependent behavior across epochs. We focus on achieving the best possible modeling of the PFS and find that it differs significantly from previous cases. Notably, we do not observe either a \textit{dip} or narrow emission wing-like features at the cyclotron line energy. Instead, our modeling reveals a broad \textit{bump} at energies around the spectral energy of the CRSF, highlighting a connection between this feature and the spin-dependent spectral variability of the source. Finally, we explore possible explanations for the diverse phenomenology observed in the PFS of different sources, particularly around CRSF energies.

\section{Observations and Data Analysis}

We used data from the \nustar\ satellite \citep{Harrison2013} which collect X-ray events through two independent and similar detectors, the Focal Plane Module A (FPMA) and B (FPMB). \textit{NuSTAR} has observed  \src\ four times to date (see the observations log in Table \ref{table_observations}). The first observation in August 2016 (30201028002) shows the system during a full eclipse passage. The second (30401025002) and third observations (30602024002), taken in May 2019 and in February 2021, respectively, show the source in the persistent state. In the final observation (30602024004), which occurred just a few days after the third, an eclipse was observed again during the second half of the observation. 
Hereafter, we will refer to the single observations using the label shortcuts 302, 304, 306-2, and 306-4.

\begin{table*}
\caption{Observations log}
\renewcommand{\arraystretch}{1.3}
\centering
    \begin{tabular}{lcccccr@{}lc}
        \hline
        ObsID & Exposure & Total counts & Flux\tablefootmark{a} & L\tablefootmark{b} & $\nu$ 
        & $T_\text{ref}$\tablefootmark{c}\\
              & ks & 10$^{4}$ & $10^{-10}$ erg/cm$^2$/s & $10^{36}$ erg/s & mHz & seconds  \\
        \hline
        302 & 18.24 & 21.4 & 5.2  & 2.7 &  1.899 & 208639296.615 \\
        
        304  & 36.16 & 81.7 & 8.8   & 4.6 &  1.897 & 294531749.622\\
        
        306-2 & 20.89 & 34.3 & 6.7   & 3.5 & 1.902 &351157346.471 \\
         
        306-4  & 14.80  &  22.2 & 4.7 & 2.4 &  1.901 & 351650375.669 \\
        \hline
    \end{tabular}
    \tablefoot{
    \tablefoottext{a}{Absorbed flux range: 4–50 $\mathrm{keV}$.}
    \tablefoottext{b}{Isotropic bolometric (absorbed) luminosities assuming a distance 6.6 kpc \citep{Bailer-Jones2018}.}
    \tablefoottext{c}{Seconds since January 1st, 2010, 00:00:00 UTC.}
    }
    \label{table_observations}
\end{table*}

We processed the data  using the \textit{NuSTAR} Data Analysis Software (\texttt{nustardas}, v. 1.9.7, as in HEASoft v. 6.31), and the most recent \textit{NuSTAR} calibration files (CALDB v20240104). For higher-level scientific products, we used our dedicated toolkit \textsc{nustarpipeline}\footnote{\url{https://gitlab.astro.unige.ch/ferrigno/nustar-pipeline}}. After re-processing the data to clean level 2 data stage, we defined a circular source region centered on the most accurate X-ray coordinates of the source, with a radius of 2 arc-minutes, and a background region of the same size in an area free of any contaminating sources for both the FPMA and FPMB detector images. 

We excluded, for every single observation, time intervals where the source exhibited clear signs of \textit{dips}, flares, noticeable spectral changes, including the eclipse events in the observations 302 and 306-4. To do this, we defined the hardness ratio between the source light curves within the 3–7 $\mathrm{keV}$ and 7–30 $\mathrm{keV}$ energy bands, using 0.1-second time bins and then examined the time evolution. They were subsequently re-binned to maintain a minimum signal-to-noise ratio (S/N) of 10. If no major variations were detected, we focused on the flux changes in the re-binned light curves, constructing a histogram of the overall 3–30 $\mathrm{keV}$ rates. Assuming a log-normal distribution for the histogram, we fitted a Gaussian in logarithmic space and filtered out intervals that deviated by more than 5 $\sigma$ from it (see Appendix \ref{gtis_app}).

\subsection{Timing and pulse profile analysis}
\label{timinganalysis}

For the timing and the PPs analysis, we used our pipeline, tailored for \textit{NuSTAR} data, described in detail in \citet{Ferrigno2023}. 
Hereby we summarize the main steps of the analysis for every ObsID.
First, the arrival times of each event are converted to the Solar System barycenter frame. We perform a Lomb-Scargle (LS) search in order to identify coherent signals at the light curve of the energy range of 3 -- 7 $\mathrm{keV}$, usually sufficient to find pulsations in case of a bright X-ray pulsar, and define the preliminary spin frequency ($P_\mathrm{spin}$). After verifying this value for consistency with the values reported in the literature, we refine the spin frequency by performing an epoch folding search within a small frequency interval around $\pm5\%$ of the preliminary $P_\mathrm{spin}$. We report the refined frequencies in Table \ref{table_observations}. We also correct the arrival times for the binary motion of the NS by using the orbital parameters \footnotemark available from the \textit{Fermi Gamma-ray Burst Monitor} (GBM) online database \citep{Malacaria2020}. 

To study the PPs properties as a function of energy, we generated multiple energy–phase matrices with different numbers of phase bins for each observation, folding the data using the best period spin value. We computed separate source and background matrices for the two detectors, FPMA and FPMB, and then sum the resulting matrices. The energy bin spacing is set to match the intrinsic energy resolution of the FPM. For each energy bin, we calculate the S/N of the pulse while accounting for background events and sum adjacent bins to reach a minimum S/N of 15 across the \nustar\ energy band. For a quick-look visualization of these matrices, we produced
color-intensity heat maps of the energy-resolved profiles (Fig. \ref{mappa304})

\footnotetext{
    \small We adopted the following orbital parameters of the system:
    orbital period: $3.7283749$ days; period derivative: $0 \, \text{days/day}$; $T_{\pi/2}$: $2451442.489$ (JED); $a \sin(i)$: $56.600 \, \text{light-seconds}$; longitude of periastron passage: $64.00^\circ$; eccentricity: $0.174$.}

\begin{figure}
     \centering
     \hspace*{-0.3cm} 
     \includegraphics[width=1.0\columnwidth]{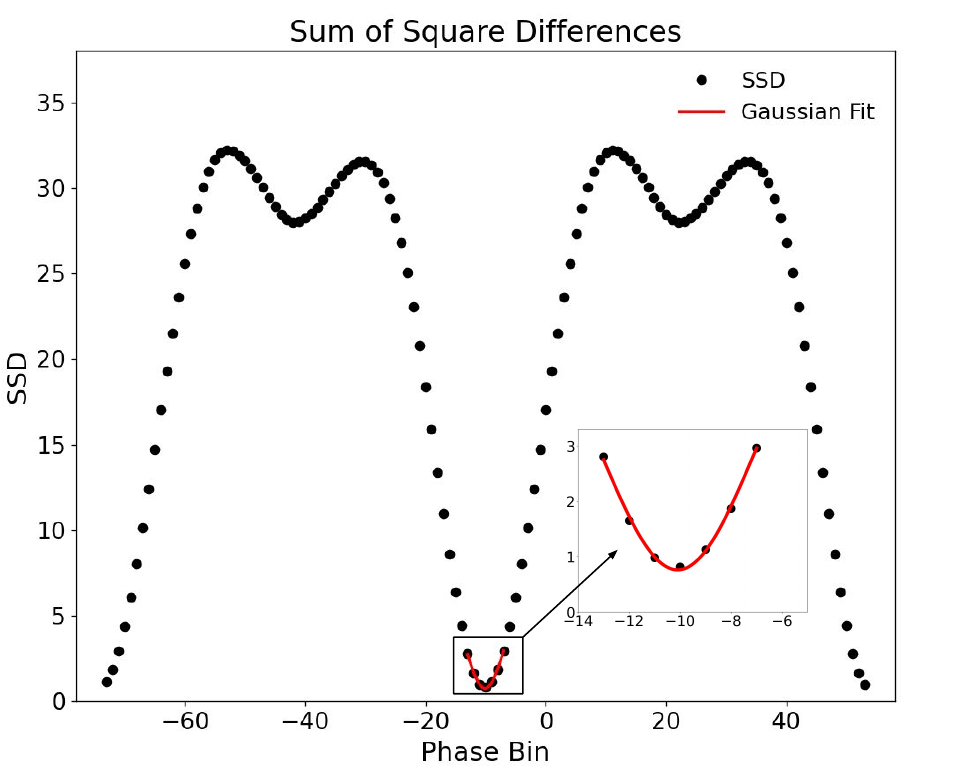}
     \caption{Alignment of the energy-phase matrices. The plot shows the sum of squared differences (SSD) values calculated for each phase shift between the PP of a given observation and the reference PP (ObsID 304). The minimum SSD value corresponds to the best alignment, and the points around this minimum are fitted with a Gaussian to refine the phase shift determination and account for statistical fluctuations. The centroid of the Gaussian provides the precise phase shift used for alignment.}

     \label{fig:SSDs}
\end{figure}

\begin{figure}
    \centering
    \includegraphics[width=0.9\columnwidth]{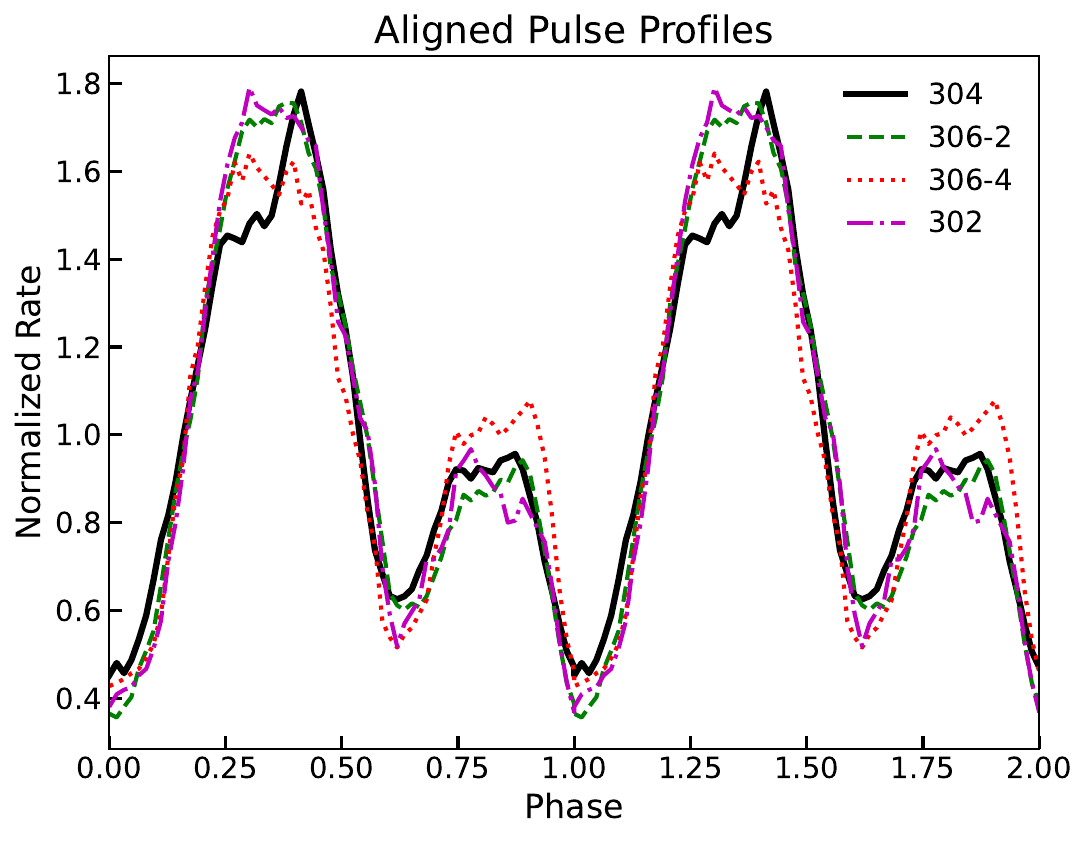}
    \caption{Aligned PPs in the 3--70 $\mathrm{keV}$ band (errors not
    shown for clarity's sake). Two periods are shown for better clarity.}
    \label{fig:aligned_PPs}
\end{figure}

We do not attempt to find a unique timing solution to phase-connect the pulses of the different observations, and, as no statistical evidence is found, no spin derivative term is taken into account for the single observation. However, the PPs of the different observations are very similar, regardless of the energy range from which they are extracted. We assume therefore that the source phase-dependent properties remain also similar, when the profiles are correctly aligned.

In order to align the PPs of the four different ObsIDs, we take as reference PP the ObsID 304, which is the one with the highest statistics. We extracted the energy-averaged 3--70 $\mathrm{keV}$ PP using 64 phase bins. For this PP, the zero phase corresponds to the center of the bin with the minimum normalized rate value. To align the phases of the other PPs, we calculated the sum of the squared differences (SSD) between the PP to be aligned and the reference one. For each of the 64 phase bins, we compute the squared difference between the respective values of the two PPs and sum these differences over all bins. We then shift the profile by one phase bin and repeat the process, resulting in 64 SSD values for each observation. We visually identify the phase bin corresponding to the minimum value and refine the phase measurement value further: we model the points around the minimum with a constant plus a Gaussian line to reduce statistical fluctuations. The best-fitting centroid value of the Gaussian provides the phase shift, and by using this value, we determine the new time of reference for the aligned profiles (see Fig. \ref{fig:SSDs}). The PPs of the different observations aligned according to this method, are shown in Fig.\ref{fig:aligned_PPs}.

We pass from a qualitative description of pulse-energy dependence to a more quantitative analysis using the pulsed fraction spectra (PFS), following the methods described in \citet{Ferrigno2023}. In brief, to calculate the pulsed fraction value of a single energy bin, we apply a fast Fourier transform (FFT) on each energy-resolved pulsed profile, defined by the re-binned energy-phase matrix. The PF value is derived using a harmonic decomposition, given by the equation
\begin{equation}\label{eq:fft}
\mathrm{PF}_{\mathrm{FFT}} = \frac{\sqrt{2\sum_{k=1}^{N_\mathrm{harm}} |A_k|^2}}{|A_0|},
\end{equation}
where $|A_0|$ is the average value of the PP (the zero term of the FFT transform), and $|A_k|$ represents the amplitude of the $k$-th harmonic. We extend the number of harmonics necessary to describe the pulse with a statistical acceptance level of at least 10\%, which typically corresponds to 3 to 5 harmonics. Even though other definitions exist for calculating PF values, such as the min-max method, which relies on the relative difference between the maximum and minimum flux of the pulse profile, \citet{Ferrigno2023} demonstrated that the FFT method is the most robust and least biased with respect to the number of phase bins used.

The uncertainty for each PF point is assigned with a bootstrap method. We generate 1000 faked profiles, assuming Poisson statistics for the counts registered in each phase bin, we calculate the PF value for the generic faked profile, and from this Monte-Carlo sample, we derive the standard deviation and assume it as the uncertainty at 1-$\sigma$ confidence level.

We then model the broad energy dependence of the PFS using a phenomenological polynomial function. To keep the polynomial order as low as possible, we split the full dataset into a low-energy and a high-energy band.  The splitting energy is treated as a free parameter within the 10–15 $\mathrm{keV}$ range and is chosen as the energy corresponding to a zero in the first derivative of an interpolating function \citep{Ferrigno2023}. The polynomial order is the minimum one to ensure a $p$-value greater than 0.05. We used Gaussian profiles to model local features in both the low- and high-energy bands. 

\begin{figure*}[htbp]
    \centering
    \includegraphics[width=0.9\textwidth]{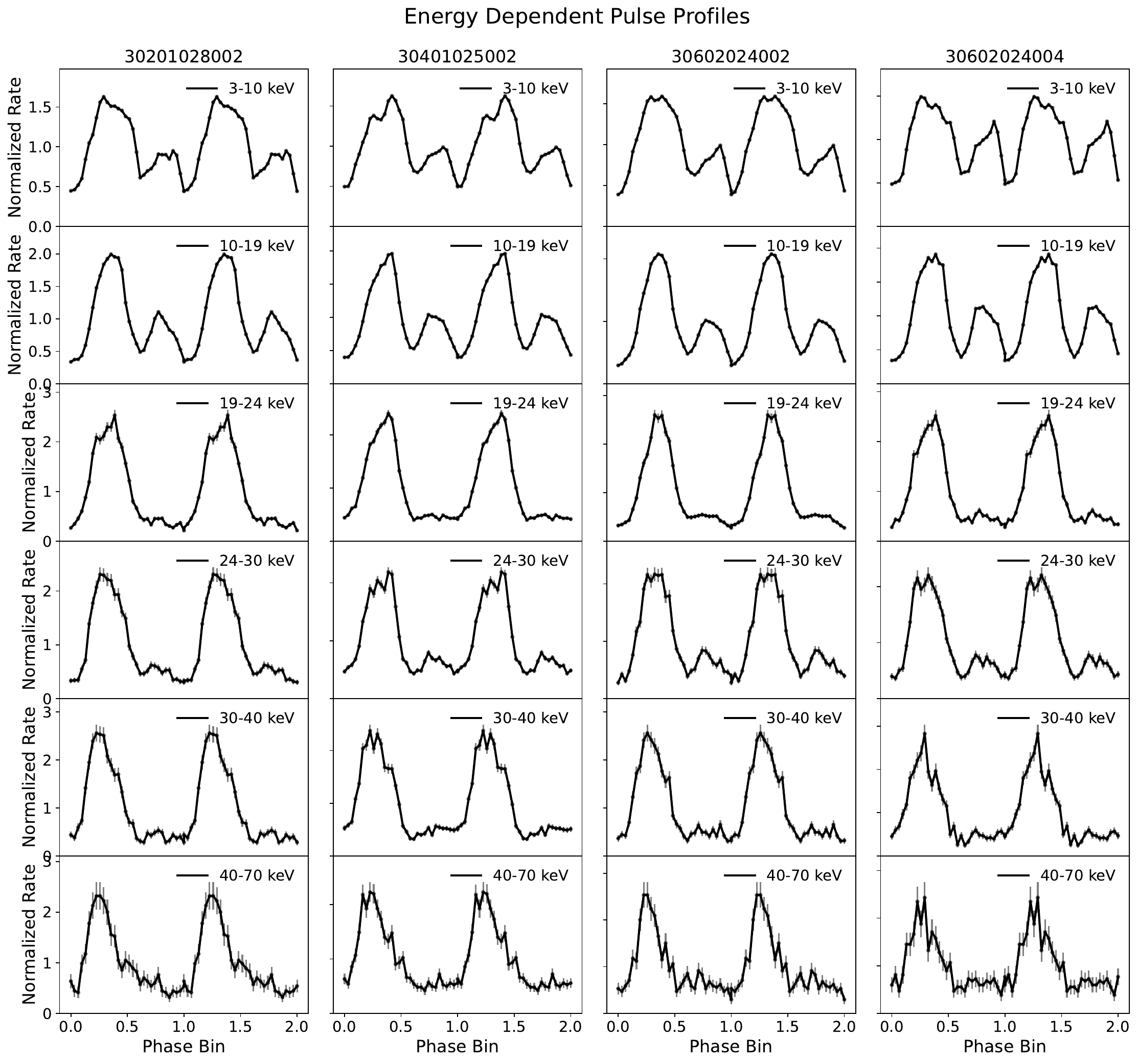}
    \caption{Energy-selected PPs for all examined observations.}
    \label{all_pps}
\end{figure*}

We use the \texttt{lmfit} package (v1.3.2) for the PFS fitting, with parameter uncertainties estimated via MCMC sampling using the \texttt{emcee} package (v3.1.4). The algorithm from \citet{Goodman2010} is employed with 50 walkers, 600 burn-in steps, and 6000 total steps. Best-fit parameters were derived from the 50th percentile, with uncertainties from the 16th and 84th percentiles.
Optimal parameters, such as an S/N of 15 and 32 phase bins, were chosen for this source to balance spectral resolution and S/N. 
Lastly, we compute the correlation and lag spectra for each observation using the methods described in \citet{Ferrigno2023}. We use as PPs of reference the 3--70 $\mathrm{keV}$ and the 35--70 $\mathrm{keV}$, but excluding in this template PP computation the energy bin of the PP we want to cross-correlate.

\subsection{Spectral analysis}
\label{spectralsub}

We generate the ancillary response and redistribution matrix files using \texttt{numkarf} and \texttt{numkrmf}, respectively. The source spectra are processed using the \texttt{ftgrouppha} tool\footnote{\url{https://heasarc.gsfc.nasa.gov/lheasoft/help/ftgrouppha.html}}, following the optimization algorithm described by \citet{Kaastra2016} and maintaining a minimum S/N of five for ObsID 304 and a minimum S/N of three for the ObsID 306-2. We only analyze the spectra of ObsIDs 304 and 306-2, as these observations provide the longest time windows when the source is in a persistent state (no eclipses), thus giving sufficient statistics for also conducting a constraining phase-resolved analysis.

Spectral analysis results from these \nustar observations have been previously reported in the literature \citep{Hemphill2019, Sharma2023, Hu2024, Tamang2024}. Therefore, rather than attempting to determine the best-fit model for the entire spectrum, our primary focus is on determining the CRSF parameters while ensuring they remain uncorrelated with the continuum parameters. To achieve this, we performe a Bayesian fit on the re-binned spectra in the 4–50 $\mathrm{keV}$ energy range, which was determined by the quality of the data.

We first perform a fit using a standard minimization algorithm, then explore the neighboring parameter space using a Monte Carlo Markov Chain (MCMC) method. Specifically, we employ the Goodman \& Weare algorithm \citep{Goodman2010} with 200 walkers, a burn-in phase of 20000 steps, and a total length of 200000 steps. We use the W statistic, which is analogous to the Cash statistic (C-stat) when a background dataset is provided, assuming Poisson distributions for both source and background spectra \citep[see Sect. 7 in][]{Arnaud2011}. To ensure the robustness of our results, we visually inspected the fit statistic distribution and parameter corner plots to verify chain convergence and confirm that the cyclotron line parameters remain uncorrelated with those of the continuum. All spectral fitting is performed using \texttt{xspec}.

To account for systematic differences between the FPMA and FPMB instruments of \nustar, we included a multiplicative constant in the model. This constant was fixed at 1 for the FPMA data, while it was left as a free parameter for the FPMB data to allow for calibration adjustments between the two detectors. In our fits, the constant for the FPMB remained close to 1, indicating minimal cross-calibration discrepancies between the two instruments and ensuring a consistent flux scale across both modules.

\begin{figure*}[ht]
    \centering
    \includegraphics[width=0.9\textwidth]{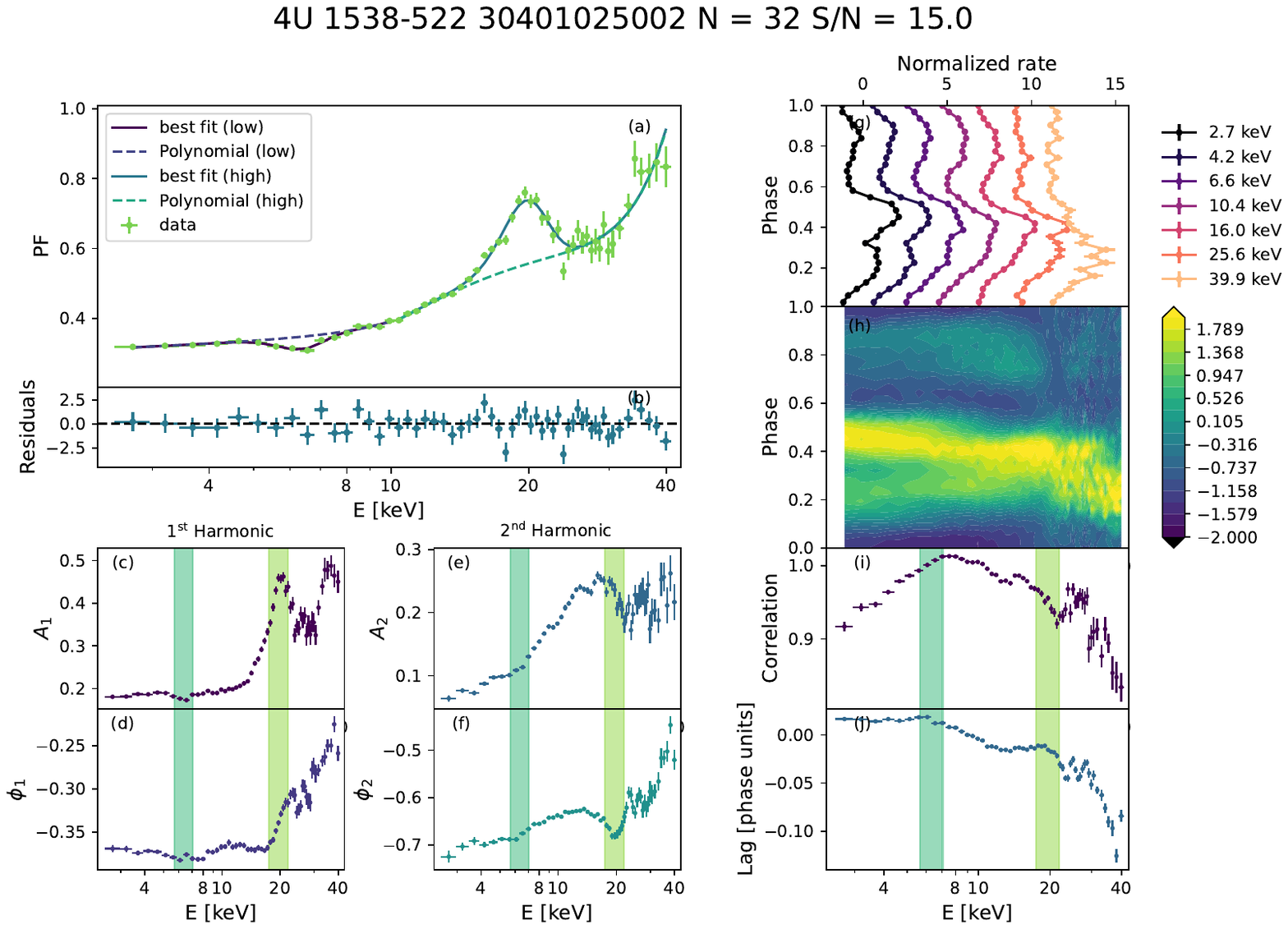}
    \caption{Summary plot for ObsID 304 using 32 phase bins and a S/N 15. Panel a: pulsed fraction (green points) along with its best-fit model (solid lines); polynomial fits are also included for comparison. Panel b: residuals from the fit. Panels c–f: phases and normalized amplitudes of the first ($A_1$, $\phi_1$) and second harmonic ($A_2$, $\phi_2$), respectively. The vertical shaded regions denote the energy ranges and widths of the Gaussian functions fitted to the pulsed fraction. Panel g: a selection of normalized PPs at equally spaced logarithmic energy intervals, horizontally offset for clarity. Each pulse in the bin was normalized by subtracting its mean and dividing by the standard deviation. Panel h: color-map display of the normalized PPs as a function of energy, with 20 evenly spaced contour lines drawn for reference. Panel i: cross-correlation between the PP at each energy band and the overall average profile. Panel j: corresponding phase lags. The colored vertical regions are consistent with those in panels d–f.}
    \label{SUMMARYPLOT304}
\end{figure*}

\section{Results of timing and pulse profile analysis} \label{resultstiming}

The PPs of all four observations display similar behavior, as illustrated in Fig. \ref{fig:aligned_PPs}. The 3.0--70.0 $\mathrm{keV}$ energy-averaged PP for all ObsIDs reveals an asymmetric double-peaked shape. An exception is observed in the first peak of ObsID 304, which appears more structured, while the peak's width remains similar across the other three observations. To analyze the energy-dependence of the PPs, we generate energy-resolved PPs in the 3--10, 10--19, 19--24, 24--30, 30--40, and 40--70 $\mathrm{keV}$ energy ranges (see Fig. \ref{all_pps}).

The amplitude of the secondary peak is strongly energy-dependent: it is well detected below 19 $\mathrm{keV}$, nearly undetectable in the 19--24 $\mathrm{keV}$ range, reappears in the 24--30 $\mathrm{keV}$ range, and then vanishes at higher energies, consistently with results reported by \cite{Clark1990}. Notably, the 19--24 $\mathrm{keV}$ energy range corresponds to the CRSF energy, whose centroid energy is approximately 21 $\mathrm{keV}$. This trend is consistent across all four observations, even though in the 302 observation the reappearance of the second peak is barely detectable. No significant phase shift of the secondary peak is observed. In contrast, the amplitude of the first peak shows less energy dependence, although its position shifts to earlier phases as the energy increases. These broad-band, energy-resolved PPs clearly demonstrate the strong dependence of the pulse profiles on the spectral shape.

\subsection{Pulsed fraction spectra fit results}

We fit phenomenologically the PFS following the methodology described in Section \ref{timinganalysis}, dividing the fitted range into a soft and a hard band split at around 11 $\mathrm{keV}$. In all four observations, the PFS exhibit a similar trend. PF values begin at approximately 0.3 in the low-energy band (around 3 $\mathrm{keV}$) and gradually increase up to 0.8 at 19.5 - 20 $\mathrm{keV}$. Subsequently, the PF values decrease, only to rise again after 30 $\mathrm{keV}$, reaching again values close to or higher than 0.8 in the highest energy band (above 40 $\mathrm{keV}$). Notably, there is a clear \textit{dip} in the PF values at 6.4 $\mathrm{keV}$, the energy corresponding to the K$\alpha$ iron line \citep{Rodes-Roca2014}. This \textit{dip} is clearly detected in the observations with the most robust statistics, 304 and 306-2, and is modeled using a negative Gaussian profile, while in the other two observations, 302 an 306-4, there are no significant residuals in this energy range.

To model the PFS, we explored two different phenomenological approaches: fitting a polynomial combined with a positive broad Gaussian centered at approximately 20 $\mathrm{keV}$, and a negative broad Gaussian centered around 25 $\mathrm{keV}$. In all four PFS examined, the model incorporating the positive Gaussian consistently provided a better fit than the one with the negative Gaussian (see Fig. \ref{SUMMARYPLOT304}, Fig. \ref{SUMMARYPLOT306}, and Appendix \ref{timing_plots_302_3064}). The corresponding fit results are summarized in Table \ref{PFSfits}. Across all fits, the centroid energy of the positive Gaussian lies between 19.5 and 20.5 $\mathrm{keV}$, approximately 1 $\mathrm{keV}$ below the cyclotron line energy in the corresponding spectra \citep{Hu2024}. The Gaussian width is found to be between 2.2 and 2.9 $\mathrm{keV}$, which is comparable to the observed width of the CRSF, of around 3 $\mathrm{keV}$, thus suggesting a close link between this feature in the PFS  and the CRSF. Given this, we aim to investigate the physical origin of this \textit{bump} in the PFS and its potential connection to the cyclotron line.

\subsection{Harmonic decomposition: energy-dependence of the amplitudes}

\begin{figure}[htbp]
    \centering
    \hspace{-0.7cm}
    \includegraphics[width=0.90\columnwidth]{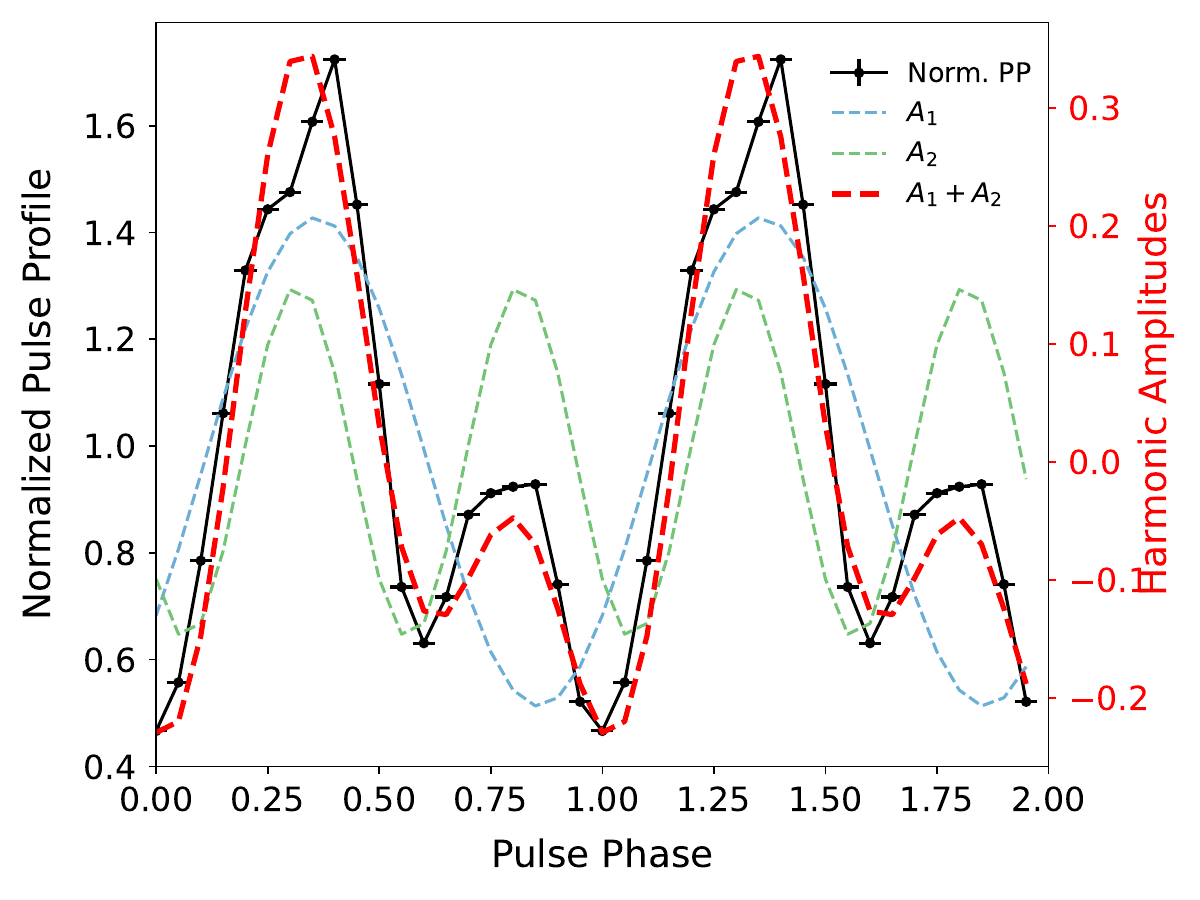}
    \caption{ The total 3--70 keV pulse profile of ObsID 304 and the first two Fourier harmonics from its FFT decomposition.}
    \label{harmonic_PP}
\end{figure}

The contribution of the first two harmonics of the Fourier decomposition dominate the 3--70 keV energy averaged pulse profile in all the four observations, as shown in Fig. \ref{harmonic_PP} with the truncated decomposition of ObsID 304.
The energy-dependent behavior of the first two harmonics of the FFT is similar across all four observations. The amplitude of the first harmonic ($A_1$), computed as the Fourier amplitude normalized to the mean count rate of the PP, follows the general trend of the PFS, showing increased values around 20 $\mathrm{keV}$, followed by a decrease and then an increase at energies above 30 $\mathrm{keV}$. In addition, a decrease is observed at the iron line energy in the two observations with the highest statistics, 304-2 and 306.
The second harmonic amplitude ($A_2$) also shows a increase, peaking at energies just below 20 $\mathrm{keV}$, which is lower than the broad feature in the PFS. However, this rise is more gradual and less pronounced compared to the first harmonic. At energies corresponding to the broad feature in the PFS, the second harmonic shows a slight \textit{dip} for a few $\mathrm{keV}$ before briefly rising again (see Fig. \ref{SUMMARYPLOT304}, Fig. \ref{SUMMARYPLOT306}, and Appendix \ref{timing_plots_302_3064}).

\subsection{Cross-correlation and lag spectra}

We computed the energy-resolved cross-correlation (CC) spectra for all four observations to evaluate the energy-dependent coherence of the PPs with respect to reference PPs. In the summary plots (see Fig. \ref{SUMMARYPLOT304}, Fig. \ref{SUMMARYPLOT306}, and Appendix \ref{timing_plots_302_3064}), we show the CC and lag spectra using the 3-70 $\mathrm{keV}$ averaged PP as template. 

The behavior across all observations exhibits similar patterns. Specifically, the cross-correlation shows a steady increase up to approximately 10 $\mathrm{keV}$, followed by a decrease until 20 $\mathrm{keV}$, which aligns with the broad feature observed in the PFS. Beyond 20 $\mathrm{keV}$, the values rise again until around 30 $\mathrm{keV}$. After this point, in observations 306-2 and 306-4, the data points become too sparse to discern any clear trend. In the other two observations, the cross-correlation values decrease. 

In the lag spectra at the cyclotron line energy, we observe that global trend of the lag values drastically changes, from a relatively flat trend to a steep decline of the lag values towards earlier phases. 

\begin{figure}[ht]
    \centering
    \hspace{-1cm}\includegraphics[width=0.8\columnwidth]{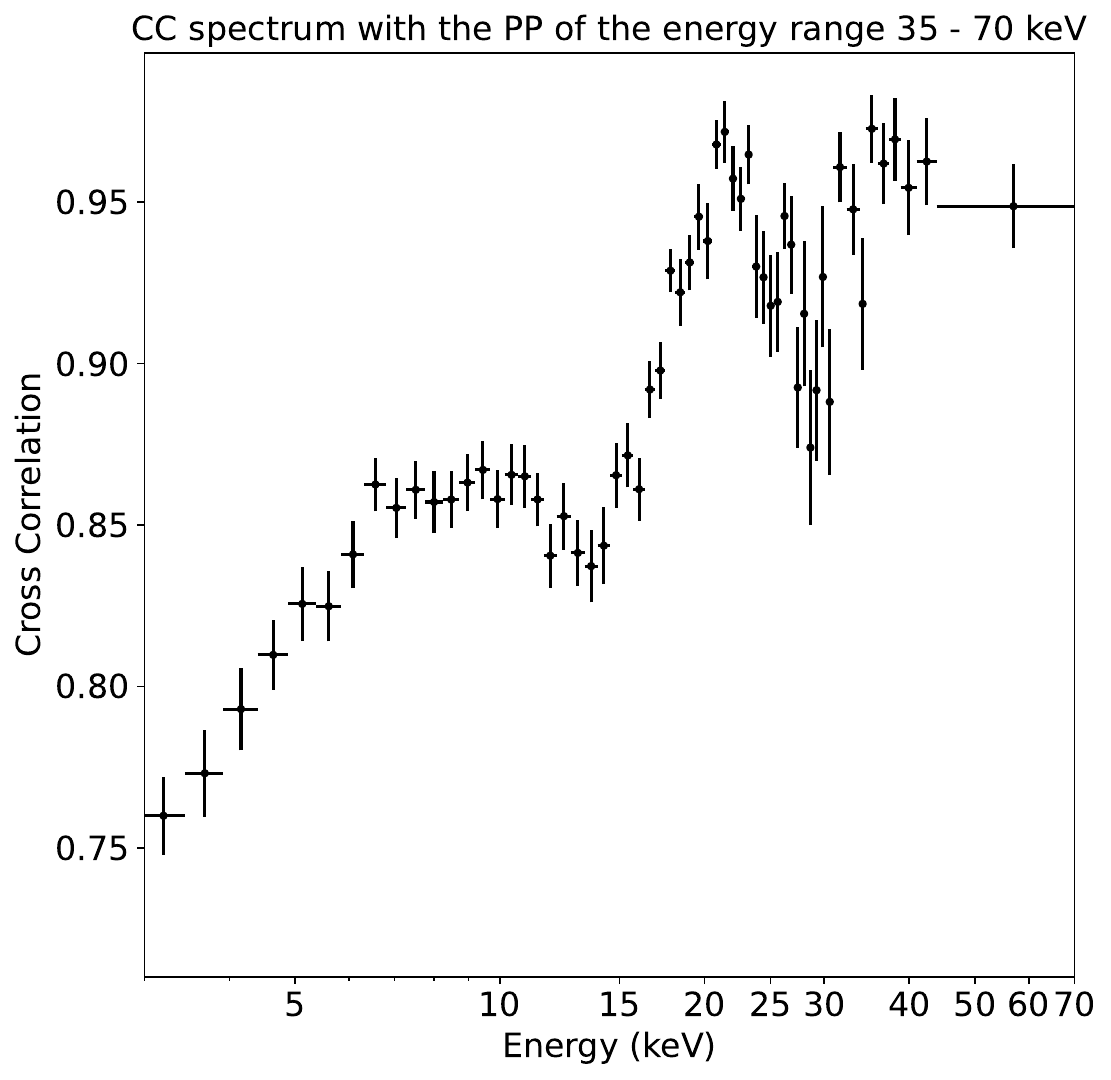}
    \caption{Cross-correlation spectrum extracted during ObsID 304 using the PP from the 35--70 $\mathrm{keV}$ energy range as template. }
    \label{304correlation}
\end{figure}

We also found the CC spectrum to be particularly informative when using the PP of energies above 35 $\mathrm{keV}$ as a template, as shown in Fig. \ref{304correlation}. In this case, the CC spectrum more clearly highlights the strong correlation between the PP extracted around the CRSF energy range, where the \textit{bump} on the PFS is observed, and the highest energy band.

\begin{table}[h!]
    \centering
    \resizebox{1.02\columnwidth}{!}{

    \renewcommand{\arraystretch}{1.4} 
    \begin{tabular}{lcccc}

    \toprule
    Parameter (Unit) & 302 & 304 & 306-2 & 306-4 \\
    \midrule
    
    $E_{\rm bump}$ (keV) & $20.03 \pm 0.10 $  & $19.71 \pm 0.06$  & $19.69 \pm 0.08$ &  $20.53 \pm 0.15$  \\
    $\sigma_{\rm bump}$ (keV) & $ 2.89_{-0.06}^{+0.05}$  &  $2.24 \pm 0.08$ &  $2.81_{-0.09}^{+0.07}$ & $2.48\pm 0.16$\\
    $A_{\rm bump}$ & $1.03 \pm 0.05$  &  $ 1.01\pm 0.14$ & $1.62 \pm 0.10$ & $1.10 \pm 0.13$\\
    
    \midrule
    
    $E_{\rm Fe}$ (keV) & -  &  $6.38  \pm 0.05$ &  $6.47  \pm 0.06$ &  - \\
    $\sigma_{\rm Fe}$ (keV) & - &  $0.79_{-0.07}^{+0.09}$ &  $0.89_{-0.10}^{+0.14}$ &  - \\
    $A_{\rm Fe}$ &  - & $-0.075_{-0.014}^{+0.011}$ & $-0.099_{-0.014}^{+0.011}$ &  - \\
    
    \midrule

    $E_{\rm split}$ (keV) & 10.01 & 10.02 & 11.23 & 10.4 \\
    $n_{\mathrm{pol,low}}$ & 2 & 3 & 1 & 3 \\
    $n_{\mathrm{pol,high}}$ & 2 & 3 & 2 & 3 \\

    \midrule

    $\chi^2_{\rm red}$ (low) / dof &   1.48/13 & 1.20 /9 & 0.95/13 & 0.75/13  \\
    $\chi^2_{\rm red}$ (high) / dof & 1.26/29  &  1.51/35 & 1.01/24 & 1.35/22 \\
    \bottomrule
    \end{tabular}
    }    
    \caption{Best-fit parameters for the modeling of the PFS across all four observations.}
    \label{PFSfits}
\end{table}


\section{Results of spectral analysis} \label{spectral_section}

To model the broadband continuum emission we used the \texttt{newhcut} component, first introduced by \citet{Burderi2000}. This model represents a power-law with a high-energy cut-off, where a smoothing function in the form of a polynomial, is applied near the cut-off energy. The width of the polynomial, which determines the energy range over which the transition from the power-law to the cutoff occurs, is typically fixed at 5 $\mathrm{keV}$. Preliminary fit results indicated that this component alone was inadequate to fit the broadband 4--50 $\mathrm{keV}$ spectrum, therefore we added to the model a soft blackbody component (\texttt{bbody}), which is often required in modeling the spectra of bright accreting X-ray pulsars as also done in the literature for this source in particular \citet{Sharma2023}. Since our analysis is limited to energies above 4 $\mathrm{keV}$, we fixed the spectral parameters that are less constrained due to the lack of the softer spectral coverage, i.e., the black-body temperature at 1 $\mathrm{keV}$  and the value of the equivalent hydrogen absorption column (modeled using the \texttt{tbabs} component in \texttt{xspec}), $N_\textrm{H}$\,=\,1 $\times 10^{22}$ \, \text{cm}$^{-2}$, both values taken from the fits in \citet{Sharma2023}.

We found that the CRSF shape is statistically better modeled using a Lorentzian profile (\texttt{cyclabs} in \texttt{xspec}) rather than a Gaussian profile (\texttt{gabs} in \texttt{xspec}). The latter was more susceptible to spectral degeneracy between parameters governing the line shape and the continuum in some of the phase-resolved spectra (see also details in Section \ref{PRS}). The iron line complex was modeled using a single Gaussian emission line (\texttt{gauss} in \texttt{xspec}).

\begin{figure}[htbp]
    \centering
    \includegraphics[width=0.9\columnwidth]{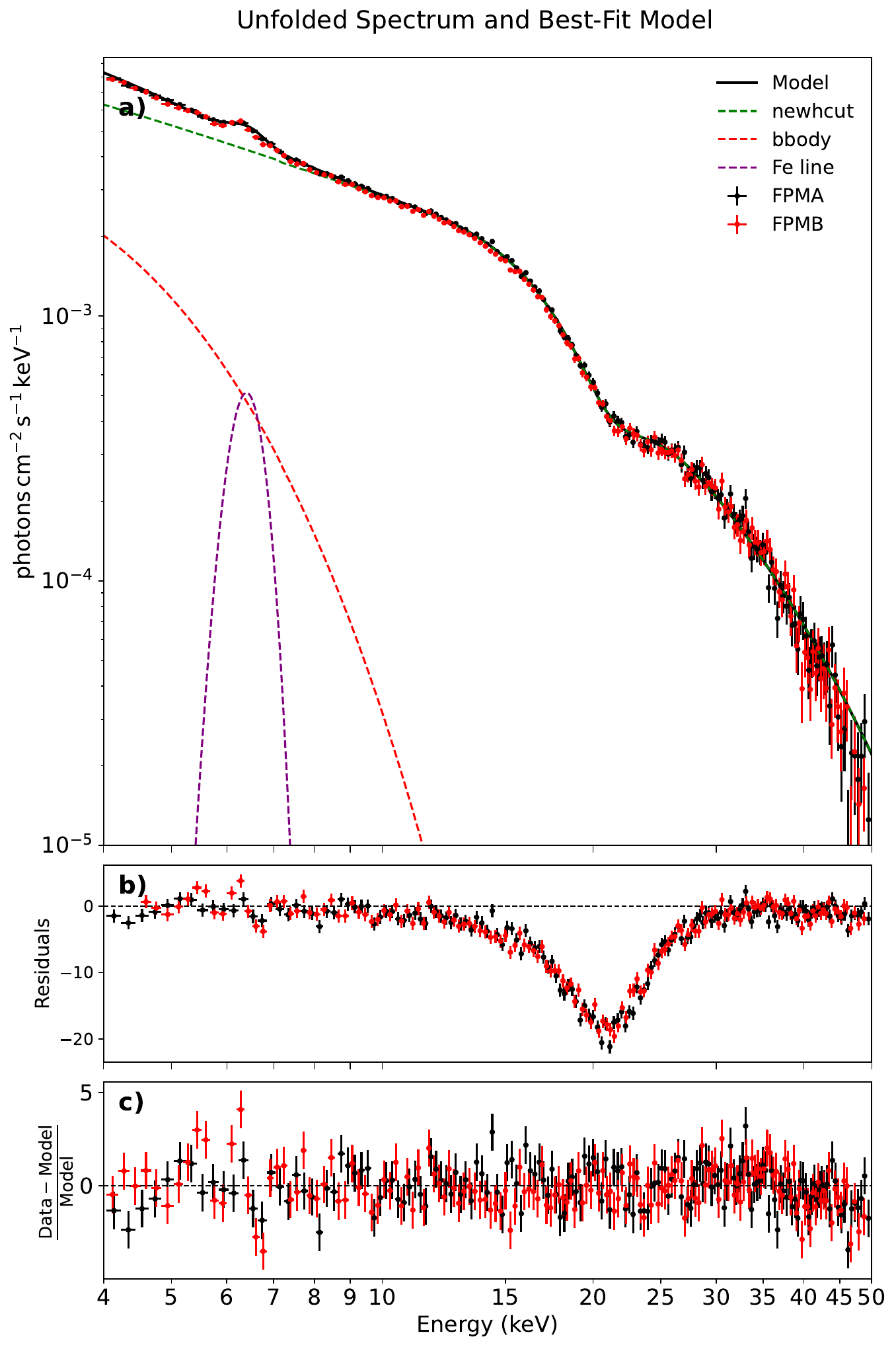}
    \caption{a) data and best-fitting model for the phase-averaged spectrum of ObsID 304.
    b) residuals of the best-fitting model when the depth 
    of the cyclotron line is set to zero. c) residuals of the best-fitting model.}
    \label{PA_304}
\end{figure}

\begin{table}[h!]
    \centering
        \caption{Best-fit results for the phase-averaged spectra}
    \label{tab:phase_averaged_comparison}
    \renewcommand{\arraystretch}{1.4} 
    \begin{tabular}{lrr}
    \toprule
    Parameter (Unit) & 304 & 306-2 \\
    \toprule 
    $N_\mathrm{H}$ $(10^{22}$ cm$^{-2})$ & (1) & (1) \\
    $E_\mathrm{cut}$ (keV) & $14.11_{-0.14}^{+0.19}$ & $15.4_{-0.2}^{+1.4}$ \\
    $E_\mathrm{fold}$ (keV) & $10.64 \pm 0.10$ & $10.27_{-0.75}^{+0.14}$ \\
    $\Gamma$  & $0.902 \pm 0.018$ & $0.885_{-0.052}^{+0.018}$ \\
    $E_{\mathrm{width}}$ & (5) & (5)\\
    $N_{\mathrm{newhcut}}$  & $0.0230 \pm 0.0010$ & $0.0168_{-0.0017}^{+0.0007}$ \\
    $\mathrm{kT}$ (keV) & (1) & (1) \\

    $N_\mathrm{bbody}$  & $(8.8 \pm 0.5) \times 10^{-4}$ & $(6.5_{-0.4}^{+0.8}) \times 10^{-4}$ \\
    \bottomrule
    $E_\mathrm{cyc}$ (keV) & $20.84_{-0.09}^{+0.07}$ & $20.51_{-0.33}^{+0.10}$ \\
    $\sigma_\mathrm{cyc}$ (keV) & $3.42_{-0.12}^{+0.17}$ & $3.86_{-0.14}^{+0.50}$ \\
     $D_\mathrm{cyc}$  & $0.580 \pm 0.012$ & $0.68_{-0.02}^{+0.13}$ \\
    \bottomrule
    $E_{\mathrm{Fe}}$ (keV) & (6.4) & (6.4) \\ 
    $\sigma_\mathrm{Gauss}$ (keV) & $0.35_{-0.03}^{+0.04}$ & $0.20_{-0.04}^{+0.07}$ \\
    $N_\mathrm{Gauss}$ & $(4.6 \pm 0.4) \times 10^{-4}$ & $(3.0_{-0.3}^{+0.4}) \times 10^{-4}$ \\
    \bottomrule
    $C_{\mathrm{FPMB}}$  & $0.979 \pm 0.002$ & $1.011 \pm 0.004$ \\
    Flux\tablefootmark{a} & 8.8  &  6.7 \\ 
    \bottomrule
    W-stat / d.o.f.  & $1.172 / 312$ & $1.036 / 297$ \\
    \bottomrule
    \label{PAS_table}
    \end{tabular}
    \tablefoot{
    \tablefoottext{a}{Absorbed flux range: 4–50 $\mathrm{keV}$ in units of $10^{-10}$ erg cm$^{-2}$ s$^{-1}$.}}
\end{table}

As reported in Sect. \ref{intro} a CRSF  harmonic has been reported for this source by several authors \citep{Robba2001, Hemphill2019}, though we did not find strong evidence of it in the two examined phase-averaged spectra. Although the addition of an absorption feature at the spectral model, at around 50 $\mathrm{keV}$ does improve the fit, its energy position is poorly constrained while its width takes large values greater than 10 $\mathrm{keV}$. This suggests that the feature likely models residuals in the continuum rather than representing an absorption feature at these energies. The fits are shown in Fig. \ref{PA_304} for ObsID 304 and in \ref{306PA} for ObsID 306-2. The results of these phase-averaged fits are reported in table \ref{PAS_table}. 

\subsection{Phase-Resolved Spectral Analysis} \label{PRS}

\begin{figure}[htbp]
    \centering
    \includegraphics[width=0.9\columnwidth]{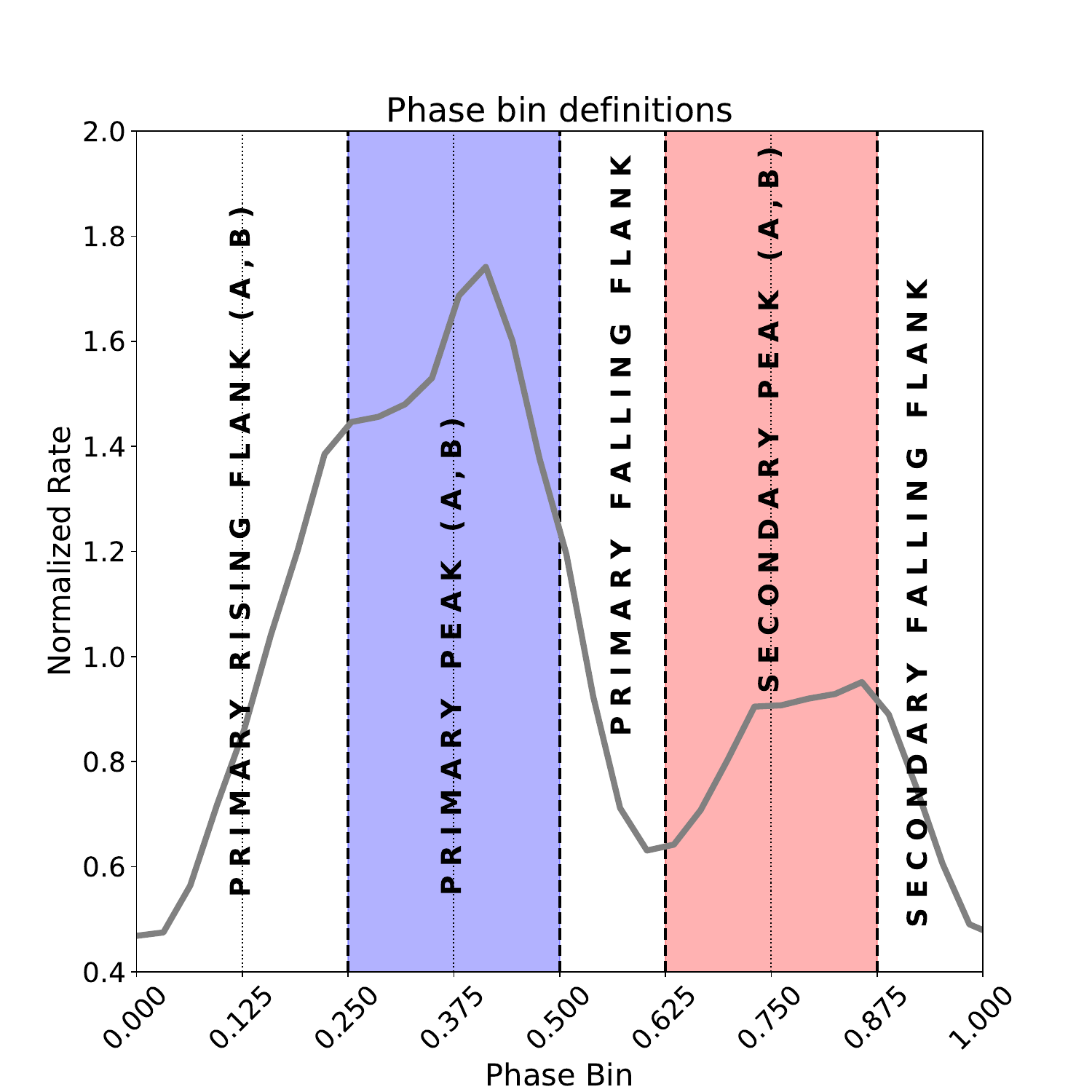}
    \caption{Definition and naming of the phase bins used in the analysis.}

    \label{phasebinsnames}
\end{figure}

To investigate the phase-dependent spectral variations of the cyclotron line, we performed an independent phase-resolved spectral analysis for the two observations, 304 and 306-2. We found that a satisfactory compromise between spectral resolution and statistics is obtained by using eight evenly spaced phase bins. This also ensured a good coverage for energies up to 40 $\mathrm{keV}$ for each phase-selected spectrum. The phase bins are defined as shown in Fig. \ref{phasebinsnames} and will be referenced accordingly from now on.




\begin{figure}[htbp]
    \centering
    \includegraphics[width=0.9\columnwidth]{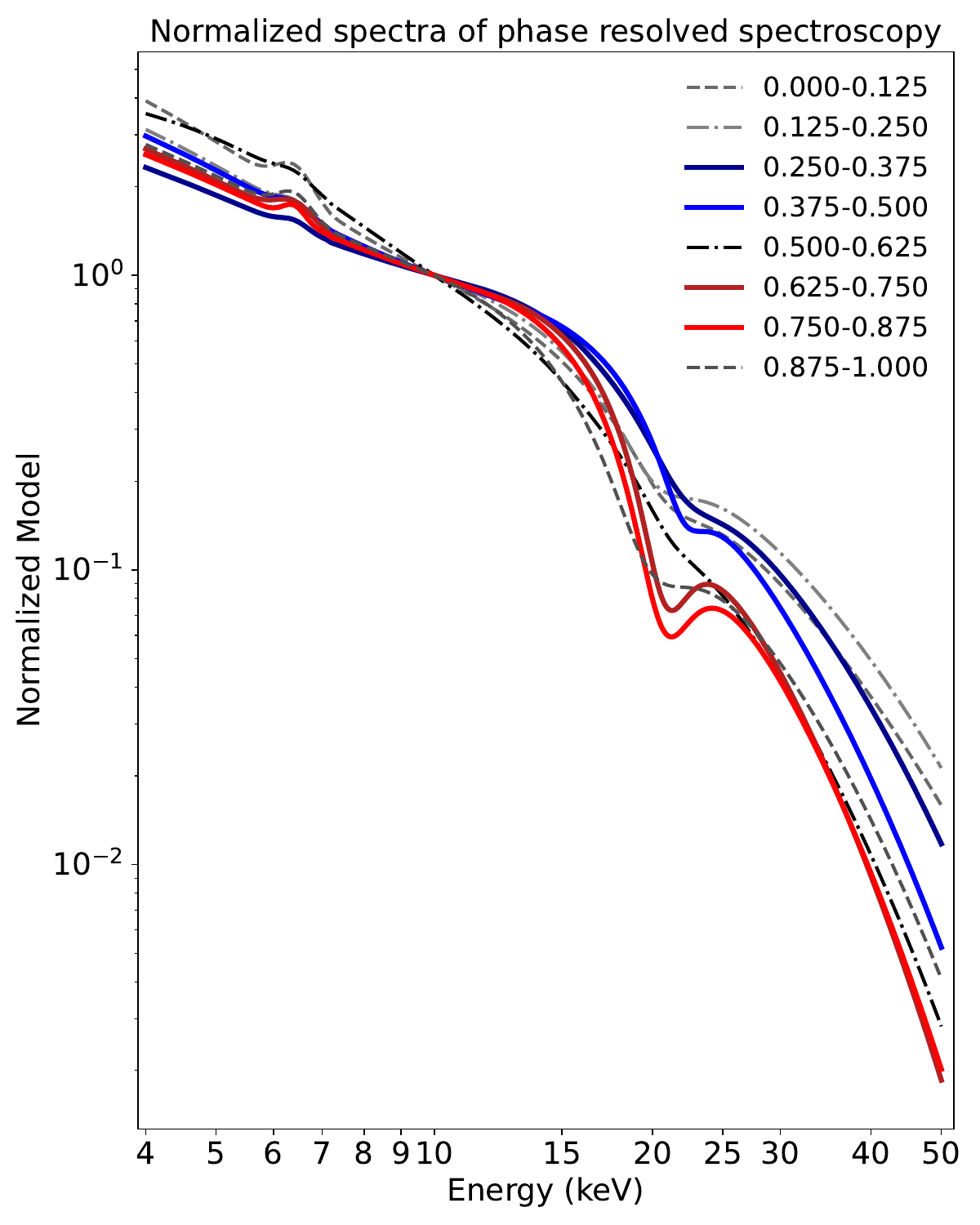}
        \caption{Phase-dependent spectral variability for ObsID 304: each curve represents one of the best-fitting model from the eight phase-selected spectra. Models are normalized to unity at the reference energy of 10 $\mathrm{keV}$. The blue and red curves correspond to spectra extracted in the phase intervals of the primary and secondary peak, respectively. The black spectrum shows a very shallow cyclotron line in the primary falling flank. The gray spectra correspond to the primary rising flank and secondary falling flank.
        }
    \label{PRmodels_304}
\end{figure}

We used the same spectral model of the phase-averaged analysis to fit the phase-resolved spectra. The variability of the best-fitting parameters is shown in Fig.\ref{fig:PRparameters}. The plot of all best-fit models from the phase-resolved analysis (normalized for clarity to unity at 10 $\mathrm{keV}$) can be seen at Fig. \ref{PRmodels_304}, which illustrates the variability of both the continuum and the cyclotron line.

The continuum parameters exhibit significant phase variability, with similar trends in both observations. The photon index is higher in the primary rising and secondary falling flanks, whereas the folding energy decreases before starting to rise again in the secondary falling flank. The normalization of the blackbody component closely follows the shape of the energy-averaged PP.

The only phase bin with significantly different parameter values in the two observations is the primary falling flank. For instance, in observation 304, the flux of the blackbody component is the highest among all phase bins, while in observation 306-2, it is the lowest and constrained only by an upper limit. Conversely, the normalization of the cut-off power law shows the opposite trend. Additionally, satisfactory fits are generally obtained when the temperature of the blackbody component is fixed at 1 $\mathrm{keV}$, while for just this phase bin we had to set it to 1.5 $\mathrm{keV}$ in order to achieve a good fit. These findings suggest stronger degeneracy and  inter-correlation of the parameters for this specific phase bin. Regarding the CRSF in both observations, it is only marginally detected in this particular phase bin, whereas it is detected in all the other bin intervals.

Although the Lorentzian profile provides a better fit for the CRSF shape compared to the Gaussian absorption profile, it remains poorly constrained in certain phase-resolved spectra. To restrict the parameters space degeneracy, we chose to freeze the width of the cyclotron line to the phase-averaged value when the fit tended to converge to nonphysical very broad values, such as in the primary rising flank, in the secondary falling flank, and just in the case of ObsID 306-2, in the primary peak A. 

In the primary falling flank, we found that the CRSF is only marginally detected. In ObsID 304, although the fit could constrain a value for the line width, it had a large uncertainty. In contrast, in ObsID 306-2, where the line depth is even lower, the line width had to be frozen to the value determined for the same phase bin in the ObsID 304 spectrum to stabilize the fit.


The general variability of the CRSF between the primary and secondary peaks is arguably the most intriguing observational fact, particularly the modulation of the line's depth, as clearly shown in Fig. \ref{PRmodels_304}, where the best-fitting models of the phase bins of the primary and secondary peak are colored in blue and red respectively, to better highlight the different shapes.

\begin{figure}[htbp]
    \centering
    \includegraphics[width=0.8\columnwidth]{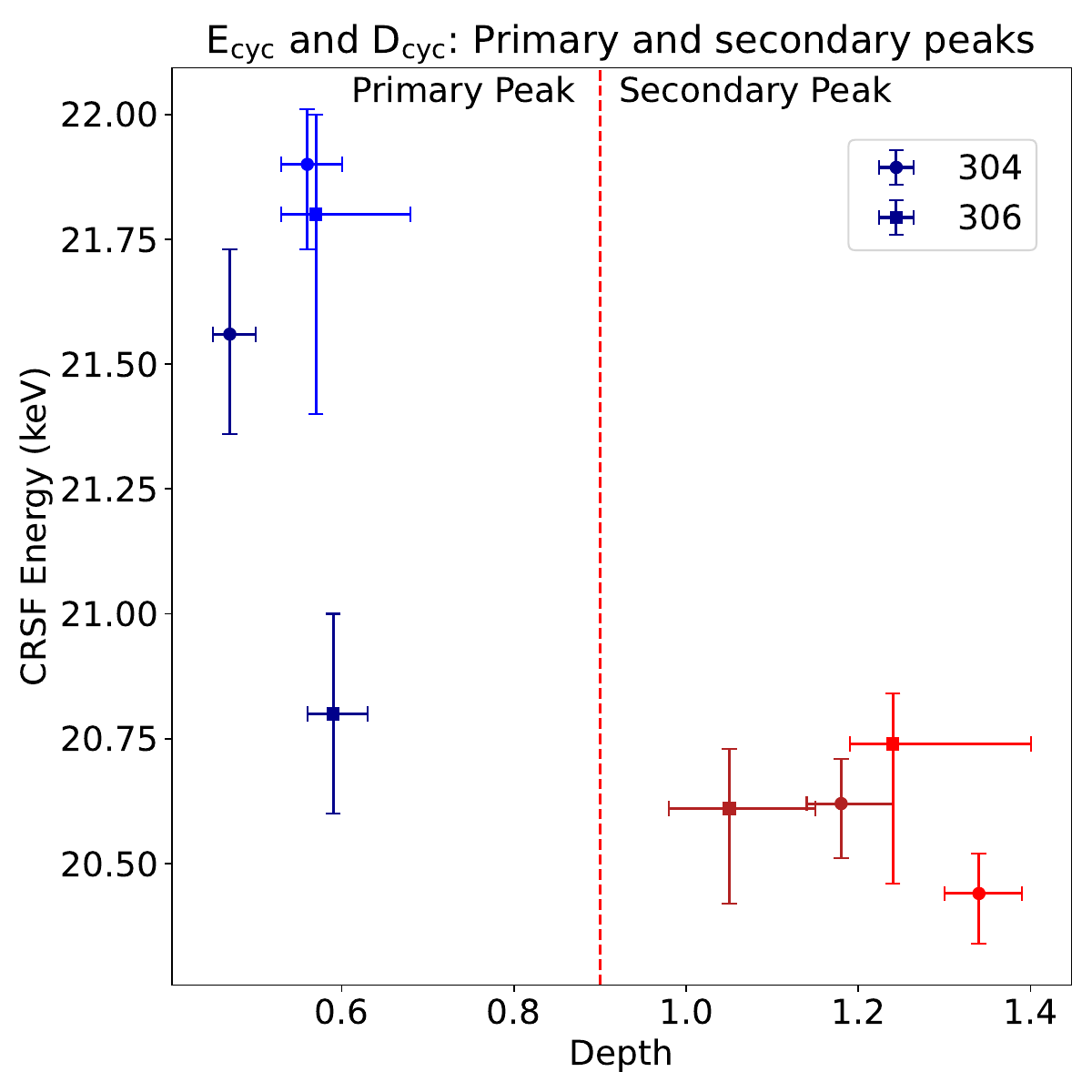}
    \caption{CRSF energy versus line depth values for the phase bins 
    corresponding to the primary and secondary peaks (0.250–0.500 and 0.625–0.875).}
    \label{ene_vs_tau}
\end{figure}

The CRSF energy, $E_\mathrm{cyc}$, exhibits significant phase-dependent shifts. The highest values, reaching up to ${\approx}$22 $\mathrm{keV}$, are observed during the primary peak phases in both observations, while the lowest values, ranging between ${\approx}$19.1 and ${\approx}$19.5 $\mathrm{keV}$, occur in the secondary falling flank, approximately 3 $\mathrm{keV}$ lower than in the primary peak. The line depth is significantly greater in the secondary peak than in the primary peak. The difference in energy and depth values between the two peaks, $D_\mathrm{cyc}$, is illustrated in Fig. \ref{ene_vs_tau}, where it can be seen that the values corresponding to the two pulse peaks tend to cluster in opposite regions of the diagram.
 

\section{Discussion}

The relation between the CRSF and the energy resolved PPs of \src\ has been extensively studied in the past. \citet{Clark1990} was the first to not only detect the CRSF in the X-ray spectrum but also to highlight its phase-dependent variability using data from \textit{Ginga}, along with an early attempt to physically model the PP shape changes with energy. A more detailed analysis was done by \citet{Bulik1995}, where they showcased the impact of general relativistic effects on the phase-dependent spectral features of 4U 1538-52. By employing model calculations of inhomogeneous magnetized neutron star atmospheres and fitting \textit{Ginga} data, they determined geometric and magnetic parameters of the system. Their results suggested that the polar caps are unequal and non-antipodal, indicating either a misaligned or distorted magnetic field structure. Additionally, the study found that relativistic effects significantly influence the observed flux and pulse profiles, with evidence for a bent or off-center magnetic axis. A similar conclusion was reached by \citet{Burderi2000} for Cen X-3, reinforcing the idea that deviations from a simple dipolar geometry may be a common feature in accreting pulsars.

In this work, we analyze the PPs of \src\ using data from \nustar, which provides improved sensitivity and broadband coverage in the hard X-ray band. With its high spectral resolution, \nustar allows for a more detailed investigation of the PP morphology and the spectral phase-dependent behavior of the CRSF.  We examined the persistent emission of the source (i.e. outside of the eclipse) across the four \nustar observations. No significant variation in the accretion regime was noted (the average source flux varied by a factor of less than two among the observations and the spectral shape remained similar) and the analysis of pulse-phase dependent behavior across epochs produced consistent results. We focused on how the CRSF characterizes and modifies the energy-resolved PPs, by extracting the analytical PPs elements from the energy-phase matrix from the available \nustar observations. In particular we emphasize here the importance of modeling the PFS to gain insight in the spin-dependence of the CRSF parameters. We found consistent results for all the examined observations, except for eclipse intervals where the low S/N prevents a detailed investigation.  By generating energy-resolved PPs across different energy bands, we found that the secondary peak of the PP nearly disappears at the CRSF energy and then reappears at higher energies, consistent with the findings of \citet{Clark1990}. In the PFS, we observed a significant increase of the PF values just before the CRSF energy, followed by downtrend after the CRSF energy. We modeled the PFS phenomenologically using a polynomial, a negative Gaussian to the iron line energy and a positive, moderately broad, Gaussian in the energy range of the CRSF. The best-fitting parameters of the Gaussian resulted in a width ranging from 2.2 to 2.9 $\mathrm{keV}$ across the four analyzed ObsIDs. Such values are comparable to the width of the CRSF found spectrally. The centroid energy, around 20 $\mathrm{keV}$, is 1 $\mathrm{keV}$ lower than the spectral value (but we note that the spectral centroid energy systematically depends on the functional shape used to fit the CRSF, namely Lorentzian profiles tend to downshift measured energies with respect to Gaussian profiles). Furthermore, when using the energy-resolved PP above 35 $\mathrm{keV}$ as a template for constructing the CC spectrum, we observed a maximum at the CRSF energy, consistent with the disappearance of the secondary peak at these energies. Additionally, phase-resolved spectroscopy revealed that the cyclotron line depth differs significantly between the primary and secondary peak: in the phase interval of the secondary peak, it is much greater than in the phase interval of the primary peak. The centroid energy also differs, ranging from around 20.7 to nearly 22 $\mathrm{keV}$ at the primary peak and from around 20.4 to 20.7 $\mathrm{keV}$ at the secondary peak. The cut-off and e-fold energies of the continuum model also vary significantly between the two peaks, as the spectrum of the secondary peak drops off more rapidly above 35 $\mathrm{keV}$ compared to that of the primary peak, consistent with the PP analysis.

\begin{figure}[htbp]
    \centering
    \includegraphics[width=0.9\columnwidth]{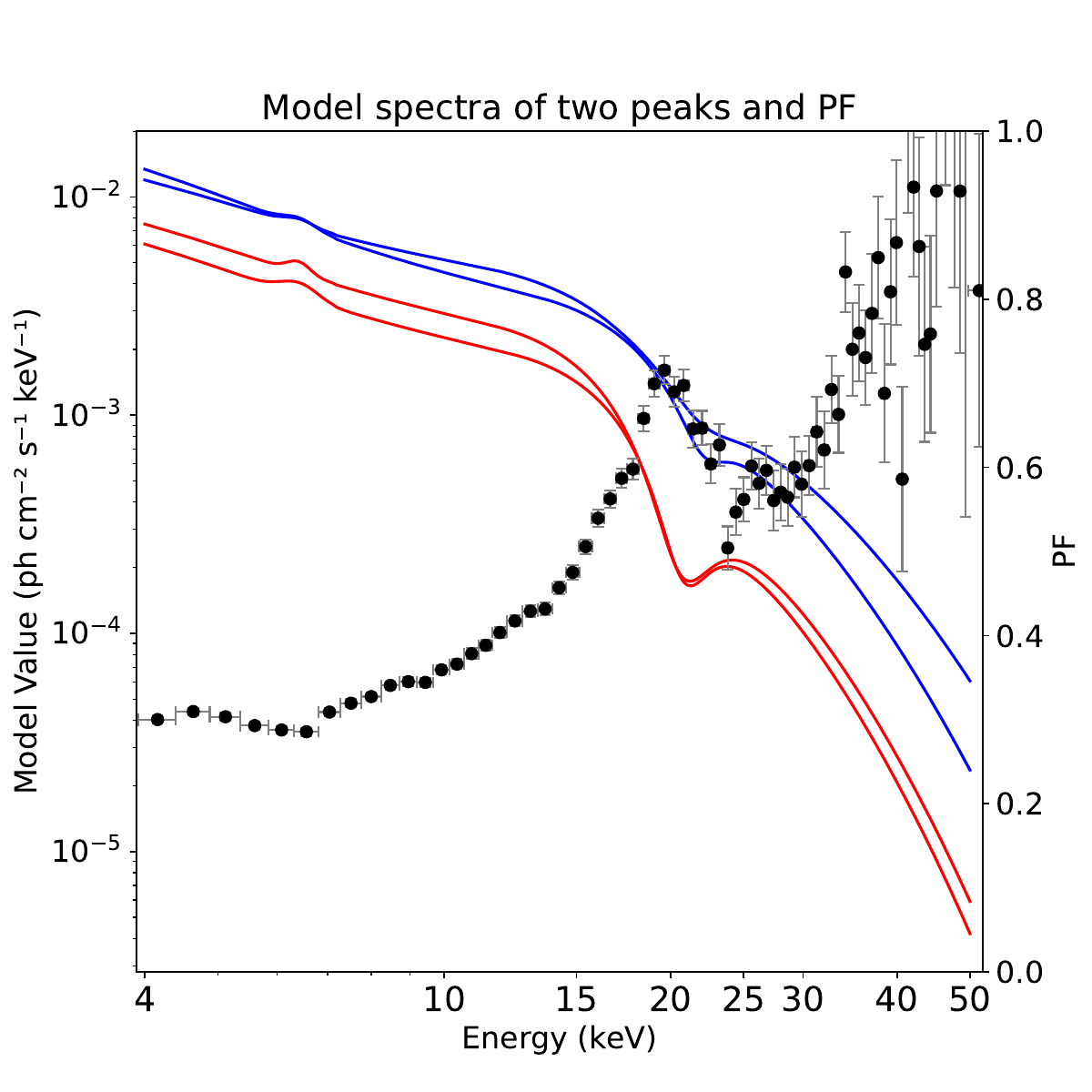}
         \caption{PFS of observation 304, calculated using 8 phase bins and an S/N of 5 for consistency with the phase-resolved spectral analysis, overplotted with the primary (blue) and secondary (red) peak spectra. }
    \label{PRmodelsPF_304}
\end{figure}

\subsection{Explaining the dips and bumps at the CRSF energy in the pulsed fraction spectra}
\label{bumpdisc}


\begin{figure*}[htbp]
    \centering
    \includegraphics[width=0.8\textwidth]{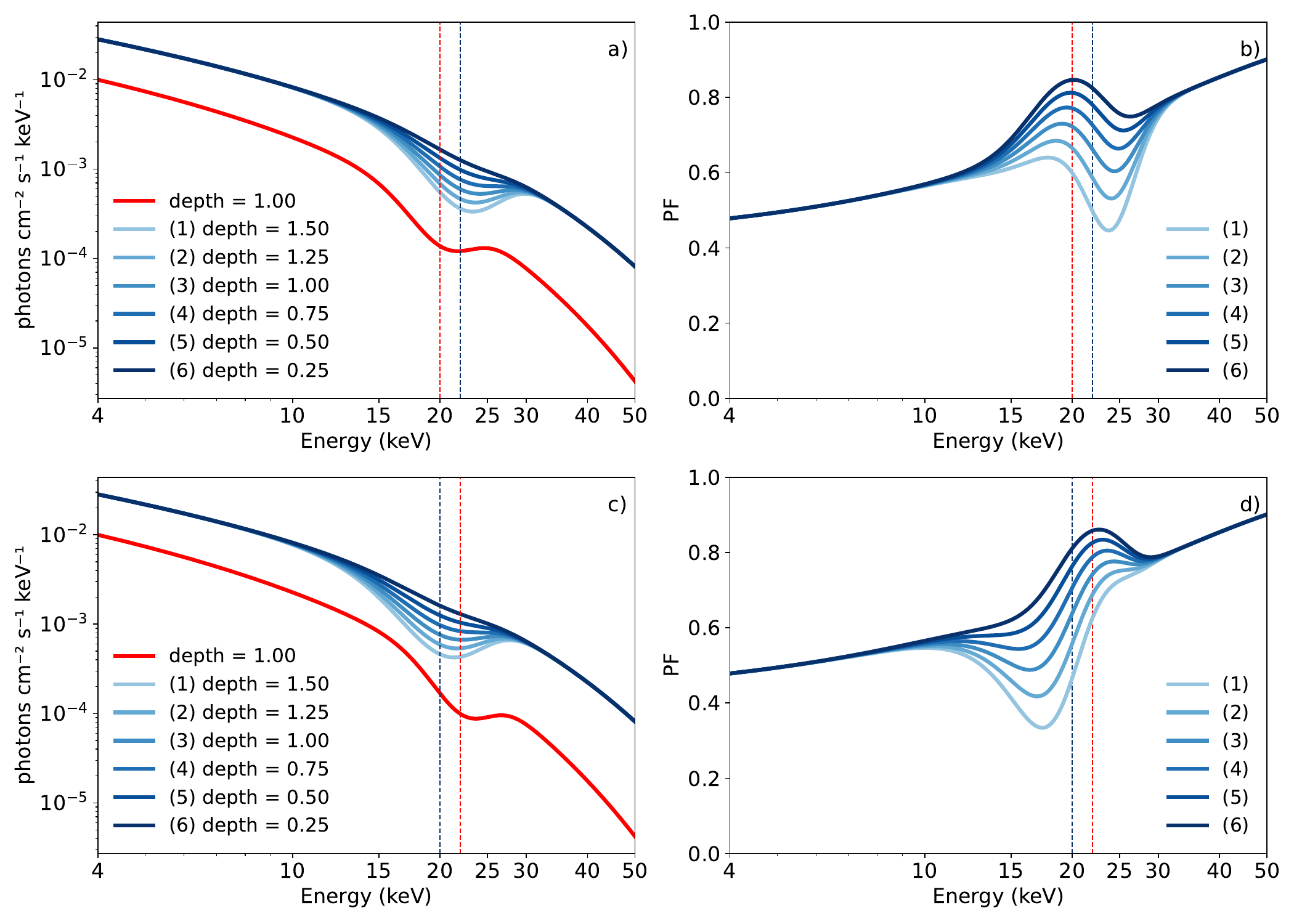}
        \caption{
        a) Simulated spectra for the primary and secondary peaks, modeled by a power-law continuum with an exponential energy cut-off. The cut-off energy is set at 12 $\mathrm{keV}$ for the primary peak spectrum (blue) and 8 $\mathrm{keV}$ for the secondary peak spectrum (red). The absorption line in the secondary peak spectrum is centered at 20 $\mathrm{keV}$ with a width of 3 $\mathrm{keV}$, while the absorption line in the primary peak spectrum is at 22 $\mathrm{keV}$ with a width of 4 $\mathrm{keV}$. b) Simulated PFS of the spectra shown in (a) using the CV proxy. The red and blue vertical lines highlight the energies of the absorption feature in the secondary and primary peaks, at 20 and 22 $\mathrm{keV}$, respectively.
        c) Simulated spectra for the primary and secondary peaks, where the energies of the absorption features have been exchanged between the two. d) Simulated PFS of the spectra shown in (c) using the CV proxy.
    }
    \label{simuPF}
\end{figure*}

The combined analysis of the PFS, PPs, and phase-resolved spectroscopy provides clear hints that the \textit{bump} in the PFS of \src\ arises from the strong spin-phase variability of the CRSF. This is illustrated in Fig. \ref{PRmodelsPF_304}, where we present the best-fitting models of the spectra of the primary (0.250–0.500) and  secondary (0.625–0.875) peaks, over-plotted with the PFS. The PFS is calculated using eight phase bins and a S/N of five to maintain consistency with the phase-resolved spectroscopy. 

A key question that arises is why the PFS at the cyclotron energy deviates from the trends reported in the literature for other sources. Unlike the \textit{dips} noted by \citet{Ferrigno2023} or the two narrow \textit{bumps} observed adjacent to the cyclotron line in V0332\,+\,53, as described by \citet{D'Ai2025}—interpreted as cyclotron line wings—\src\ does not follow either of these patterns. For this discussion, we aim to recreate the conditions that lead to a clear \textit{bump}, or a \textit{dip}, through a simple phase-dependent modeling of the energy spectra.

We, first, attempt to reproduce the PFS of \src, with a simple calculation. We assume to have two spectra at different flux levels, described by a simple power-law and an energy cut-off, and we insert an absorption feature on these two spectra, as shown in Fig. \ref{simuPF} (left). To roughly mimic the observed behavior of \src, we insert an absorption feature at 20 $\mathrm{keV}$ (width: 3 $\mathrm{keV}$) into the lower-flux (secondary peak) spectrum, and another absorption feature at 22 $\mathrm{keV}$ (width: 4 $\mathrm{keV}$) into the higher-flux (primary peak) spectrum. Additionally, the primary peak spectrum is assigned a higher energy cut-off (12 $\mathrm{keV}$) compared to the secondary peak spectrum (8 $\mathrm{keV}$). We simulate multiple cases by varying the depth of the absorption feature in the primary peak.

To quantify the PFS for each case, we adopt the Coefficient of Variation (CV) as a proxy for the value of the pulsed fraction value of each energy bin, defined as $\text{CV} = \sigma/\mu$, where $\sigma$ and $\mu$ are the standard deviation and the mean value of the points in an energy bin. The CV is thus roughly equivalent to the root-mean-square (RMS) calculation method, which, as demonstrated by \citet{Ferrigno2023}, is analogous to the FFT approach. For the purposes of this exercise, we do not take into account model uncertainties. The results of this simple exercise are illustrated in Fig. \ref{simuPF}.

In Fig. \ref{simuPF} (a) and (b), we observe that when the absorption line in the primary peak is very shallow, a distinct \textit{bump} appears around the line energy, very similar to what we see on the PFS of \src. However, as the depth of the line increases across the different cases (in Fig. \ref{simuPF}, from darker to lighter blue colors), the \textit{bump} becomes less pronounced, its peak shifts to lower energies, and a \textit{dip} emerges at higher energies. As the absorption line in the spectrum of the primary peak becomes deeper than that of the spectrum in the secondary peak, the \textit{dip} becomes more prominent and the \textit{bump} becomes progressively suppressed. On the other hand, when the energy of the cyclotron line in the spectrum of the primary peak is lower than that of the secondary peak spectrum, the \textit{dip} and \textit{bump} would appear in opposite locations, with the \textit{dip} occurring at lower energies and the \textit{bump} at higher energies, as shown in Fig. \ref{simuPF} (c) and (d). Therefore, modeling the PFS phenomenologically with only negative or positive Gaussian features may work only as a good first-order estimation, as the observed profile can get more complicated. 

Additionally, we observe that the higher cut-off energy in the primary peak spectrum relative to the secondary peak spectrum leads to an upward trend in the PFS. This is because the spectrum of the secondary peak falls off more rapidly at high energies, thus causing the secondary peak to disappear in the pulse profile, which increases the spectral variability between the two spectra at higher energies, and therefore the PF values.

In the case of \src, the fit described in Section~\ref{resultstiming} performs reasonably well, with reduced $\chi^2$ values ranging from 1 to 1.5 across all four ObsIDs. We attempted alternative fitting strategies for the PFS, including a linear or second-degree polynomial continuum combined with a positive Gaussian near ${\approx}$\,20 $\mathrm{keV}$ and an additional negative Gaussian at ${\approx}$\,25 $\mathrm{keV}$. However, these models resulted in poorer fit statistics, indicating that the additional negative Gaussian did not lead to a better description of the feature.

It is important to note that the above exercise—based on only two spectra (primary and secondary peaks)—is a simplification. The total spin-dependent spectral variability of an X-ray pulsar cannot be reduced to just two spectra, and the resulting PFS is more complex than what is captured in our simplified two-component case. Nevertheless, this exercise is still effective in illustrating how phase-dependent features such as \textit{bumps} and \textit{dips} emerge in the PFS, and in explaining why the energy at which they appear may not exactly coincide with the spectral energy of the cyclotron line. This framework provides a useful reference, especially for relatively straightforward cases such as \src.

Another simple case is Her X-1, one of the sources analyzed by \citet{Ferrigno2023}. In this source, a pronounced \textit{dip} is observed in the PFS near the CRSF energy. Phase-resolved spectroscopy by \citet{Furst2013} revealed that the line's depth varies strongly with pulse phase. In the off-pulse phases, the CRSF is shallow and centered at ${\sim}$\,35 $\mathrm{keV}$, while at the pulse peak, its depth increases by a factor of about three, and its centroid shifts to ${\sim}$\,38 $\mathrm{keV}$. This behavior is opposite to what is observed in \src, where the CRSF is deeper at low flux and shallower at high flux, i.e., the correlation between spin-dependent flux and line depth is reversed.

Based on this simple exercise, we would thus expect the PFS of Her X-1 to show a more prominent \textit{dip} rather than a \textit{bump}, and for the dip to appear at higher energies than the spectral value of the CRSF. This is indeed consistent with the findings of \citet{Ferrigno2023}, who report a PFS dip at $40.44 \pm 0.15$,$\mathrm{keV}$, compared to the CRSF centroid of $37.4 \pm 0.2$,$\mathrm{keV}$ measured by \citet{Furst2013} using the same dataset.

Consequently, the PFS effectively serves as an indicator of underlying spectral variability and provides additional constraints on the CRSF, which might otherwise be missed due to spectral degeneracy between the continuum and CRSF parameters. A systematic study of the different behaviors observed in the PFS and the phase-resolved CRSF spectral variability goes beyond the scope of the present work, though we have already identified and demonstrated the key drivers of the apparent diversity in pulsed fraction trends around the cyclotron line energies. Additionally, we have shown that the upward trend of the PFS can be attributed to the lower high-cut energy in the spectrum of the secondary peak compared to that of the primary—or, equivalently, to the disappearance of the secondary peak at higher energies.

\subsection{Physical modeling simulations}

To understand the possible origin of the reported phase-energy behavior of the flux of the source, we perform physical modeling of emission from magnetic poles of a neutron star and qualitative comparison with observations. Here, we focus on general trends in energy-dependent behavior of PPs, variability of the fundamental cyclotron resonance, and the PFS around these energies. Specifically, we aim to address the formation of the two separate peaks in the low-energy (or energy-integrate) PPs, the reduction of the secondary peak near the observed CRSF centroid energy, the general appearance of the fundamental CRSF between the main and the secondary peak, and the resulting bump in the PFS.

We follow the same setup as presented in \citet{Meszaros1985}, performing polarized radiative transfer calculations in a homogeneous, strongly magnetized slab-like atmosphere. The atmosphere is self-emitting, with magnetic Compton scattering and bremsstrahlung, including the first cyclotron resonance and the first-order relativistic corrections. For these simulations, we use the \texttt{FINRAD} code \citep{Sokolova-Lapa2021, Sokolova-Lapa2023}. Within this model, the set of parameters determining the emission is the following: the cyclotron energy, corresponding to the magnetic field, $E_\mathrm{cyc}$, the electron plasma temperature, $kT_\mathrm{e}$, the electron number density, $n_\mathrm{e}$, and the total Thomson optical depth, $\tau_\mathrm{T}$. We obtain similar emission patterns to those described in \citet{Meszaros1985a}: a slab of the depth $\tau_\mathrm{T}\sim20$ is transparent close to the magnetic field direction ($\theta \lesssim 10^\circ$) for photons of soft energies, $2$--$6\,\mathrm{keV}$, but has much higher opacity in this direction for photons of higher energies. At the same time the polarization-averaged differential flux $\mathrm{d}F=I(\theta, E)\cos{\theta}$ falls down towards $90^\circ$ more rapidly in the vicinity of the cyclotron energy than in the low-energy continuum.

\begin{figure}
    \centering
    \hspace*{-1cm} 
    \includegraphics[width=0.9\columnwidth]{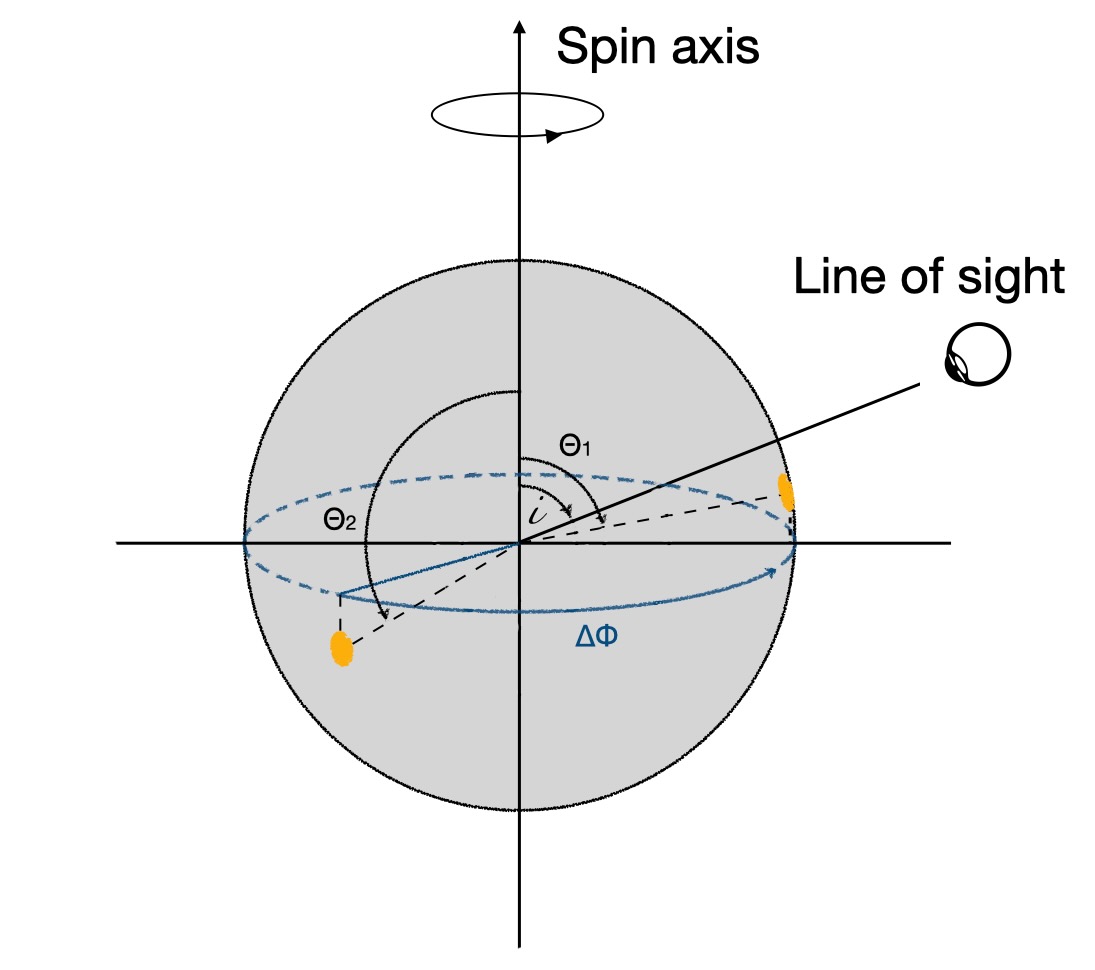}
    \caption{Geometry adopted for \src, showing the neutron star, magnetic poles at $\Theta_1 = 80^\circ$ and $\Theta_2 = 105^\circ$, azimuthal separation $\Delta\Phi = 168^\circ$, and inclination to the observer’s line of sight $i = 67^\circ$.}

    \label{geometry}
\end{figure}

The predicted $\mathrm{d}F$ allows us to set the emission from each surface element of two spot-like regions on the surface of the neutron star, associated with its magnetic poles. To obtain the projection of the emission onto the plane of a remote observer, we use the ray-tracing code by \citet[][Falkner et al., submitted]{Falkner2018}, which calculates photon trajectories from a rotating neutron star in the Schwarzschild metric. The observed flux depends on the locations of the poles with respect to the rotation axis, $\Theta_1$ and $\Theta_2$, the azimuthal separation of the poles, $\Delta \Phi$, and the inclination of the rotation axis to the observer's line of sight, $i$. These parameters are referred to in the following as ``geometric'' ones. Here, we consider only the standard values for the neutron star mass, $M_\mathrm{NS}=1.4 M_\odot$ and radius, $R_\mathrm{NS}=12\,\mathrm{km}$, resulting in the surface gravitational redshift of $1+z\approx1.24$.

The preliminary study showed that for a fixed $E_\mathrm{cyc}$, the main driver of the observed phase and energy variability is geometric parameters. We fix $E_\mathrm{cyc}=25\,\mathrm{keV}$, which results in observed redshifted cyclotron energy of ${\approx}20\,\mathrm{keV}$. In addition, we set the other parameters for the internal emission: $kT_\mathrm{e}=6\,\mathrm{keV}$, $n_\mathrm{e}=10^{23}\,\mathrm{cm}^{-3}$, and $\tau_\mathrm{T}=20$. Within the framework of the adopted model, the energy-integrated PPs, corresponding to the observed two clearly separated peaks of different heights (see, e.g., Fig.~\ref{fig:aligned_PPs}), occur only in a localized region of the parameter space, requiring $i\gtrsim 45^\circ$ and $45^\circ \lesssim \Theta_1 \lesssim 90^\circ$ simultaneously. At the same time, only geometries with $|i-\Theta_1|\lesssim25^\circ$ and $\Theta_1\lesssim85^\circ$ result in a significant reduction of the secondary peak in PPs obtained near the fundamental CRSF energy. Consistent with previous studies \citep{Bulik1992, Bulik1995}, we conclude that the positions of the two peaks in the PPs of \src\ require a slight asymmetry in the pole locations of ${\approx}10$--$15^\circ$. Here, we qualitatively reproduce some of the observables and illustrate them with a showcase below. A more detailed exploration of the parameter space will be presented in future work focused on the physical model.

\begin{figure*}
\centering
\includegraphics[height=28\baselineskip]{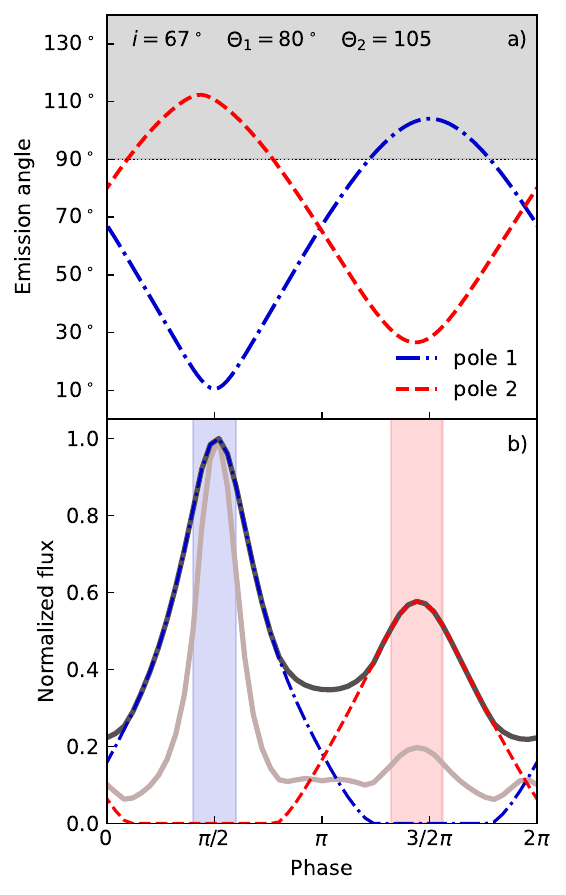}
\includegraphics[height=28\baselineskip]{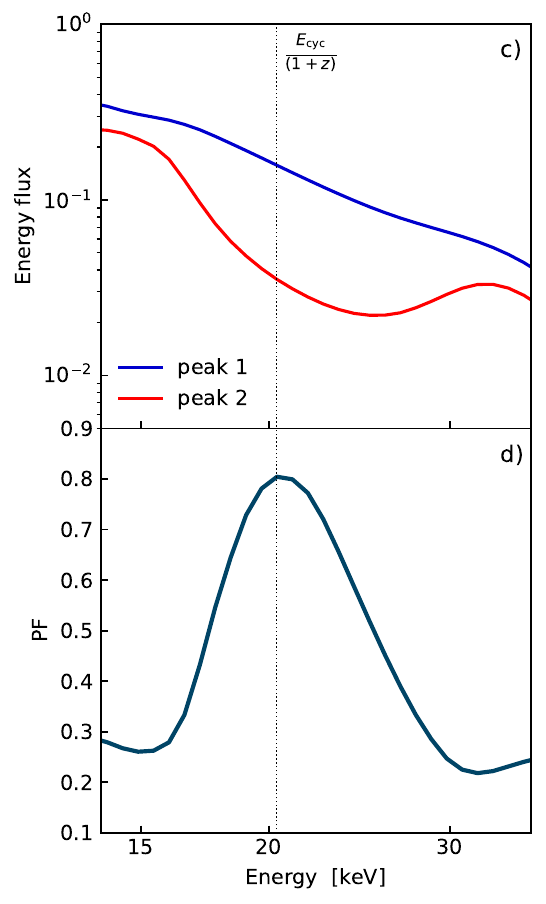}
\caption{Phase-dependent emission angle and various observables modeled for a selected geometry: $i=67^\circ$, $\Theta_1=80^\circ$, $\Theta_2=105^\circ$, $\Delta \Phi = 168^\circ$. The mass and radius of the neutron star are $M_\mathrm{NS}=1.4M_\odot$ and $R_\mathrm{NS}=12\,\mathrm{km}$, the size of spot-like poles is $r_0=300\,\mathrm{m}$. The emission from each emitting surface element is calculated for: $E_\mathrm{cyc}=25\,\mathrm{keV}$, $kT_\mathrm{e}=6\,\mathrm{keV}$, $n_\mathrm{e}=10^{23}\,\mathrm{cm}^{-3}$, $\tau_\mathrm{T}=20$. a) Emission angle, visible for a remote observer at each rotation phase. The angle is calculated for a center of a pole on the neutron star surface in the Schwarzschild metric (blue dashed-dotted line corresponds to pole~1, red dashed line -- to pole~2). The shadowed region shows when a pole is obscured by the neutron star. b) Energy-integrated ($5$--$80\,\mathrm{keV}$) pulsed profile (black thick line) for the same geometry, with individual pole contributions. Light-gray line shows the PP in the energy range $19$--$24\,\mathrm{keV}$, corresponding to the core of the fundamental CRSF in the observer's rest frame. c) Modeled energy flux near the fundamental line. The blue and red lines are obtained for the central regions of the primary (``peak 1'') and secondary (``peak 2'') peaks in the PPs; the respective ranges of phases are marked by shadow regions in panel~b. The black dotted line marks the redshifted value for the cyclotron energy. d) PFS in the same energy range as in panel~c.}
\label{fig:modemis}
\end{figure*}

\begin{figure*}[htbp]
    \centering
    \begin{subfigure}{0.37\textwidth}
        \centering
        \includegraphics[width=\linewidth]{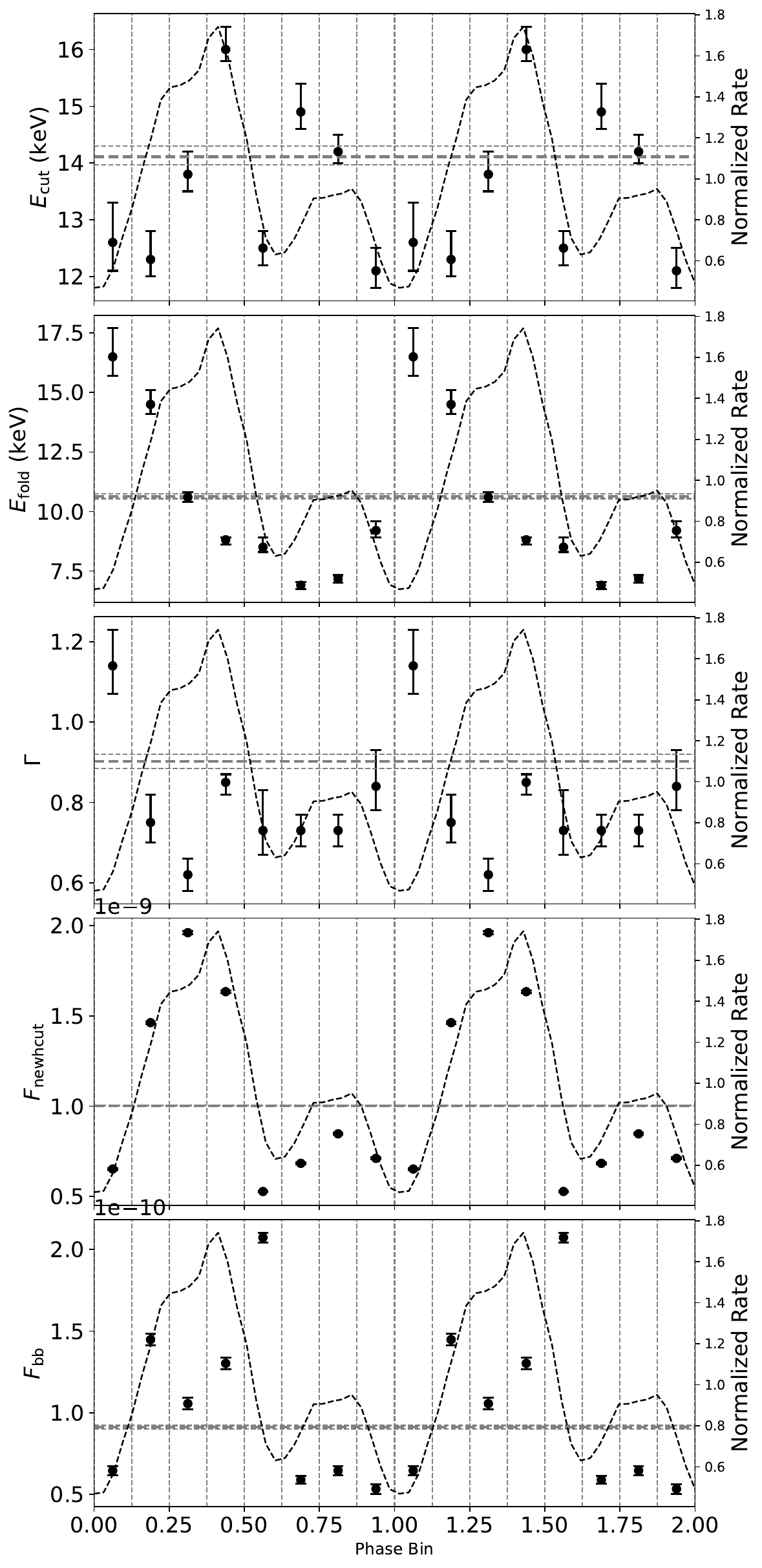}
    \end{subfigure}
    \hspace{0.05\textwidth}  
    \begin{subfigure}{0.37\textwidth}
        \centering
        \includegraphics[width=\linewidth]{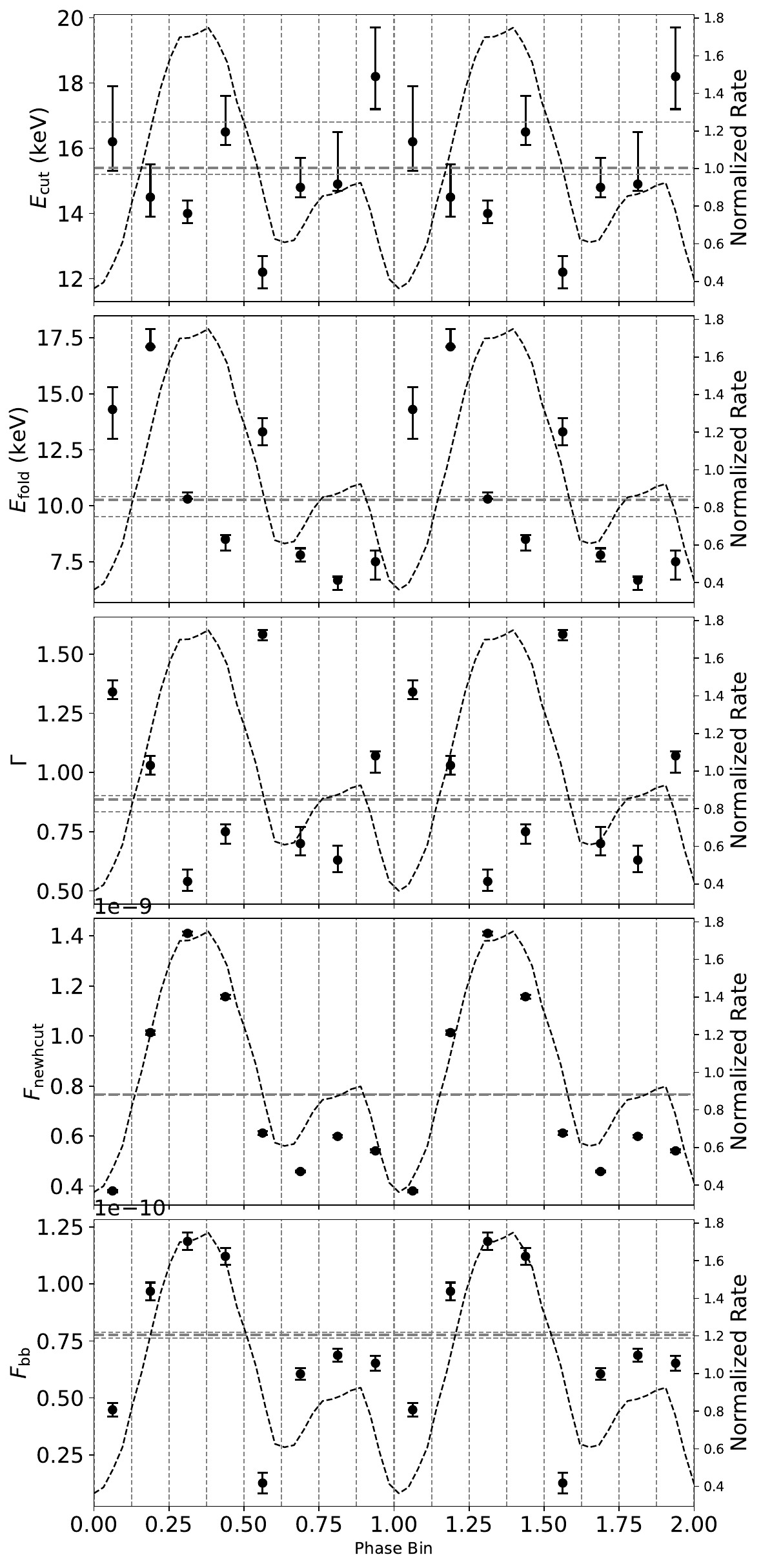}
    \end{subfigure}
    
    \vskip\baselineskip  
    
    \begin{subfigure}{0.37\textwidth}
        \centering
        \includegraphics[width=\linewidth]{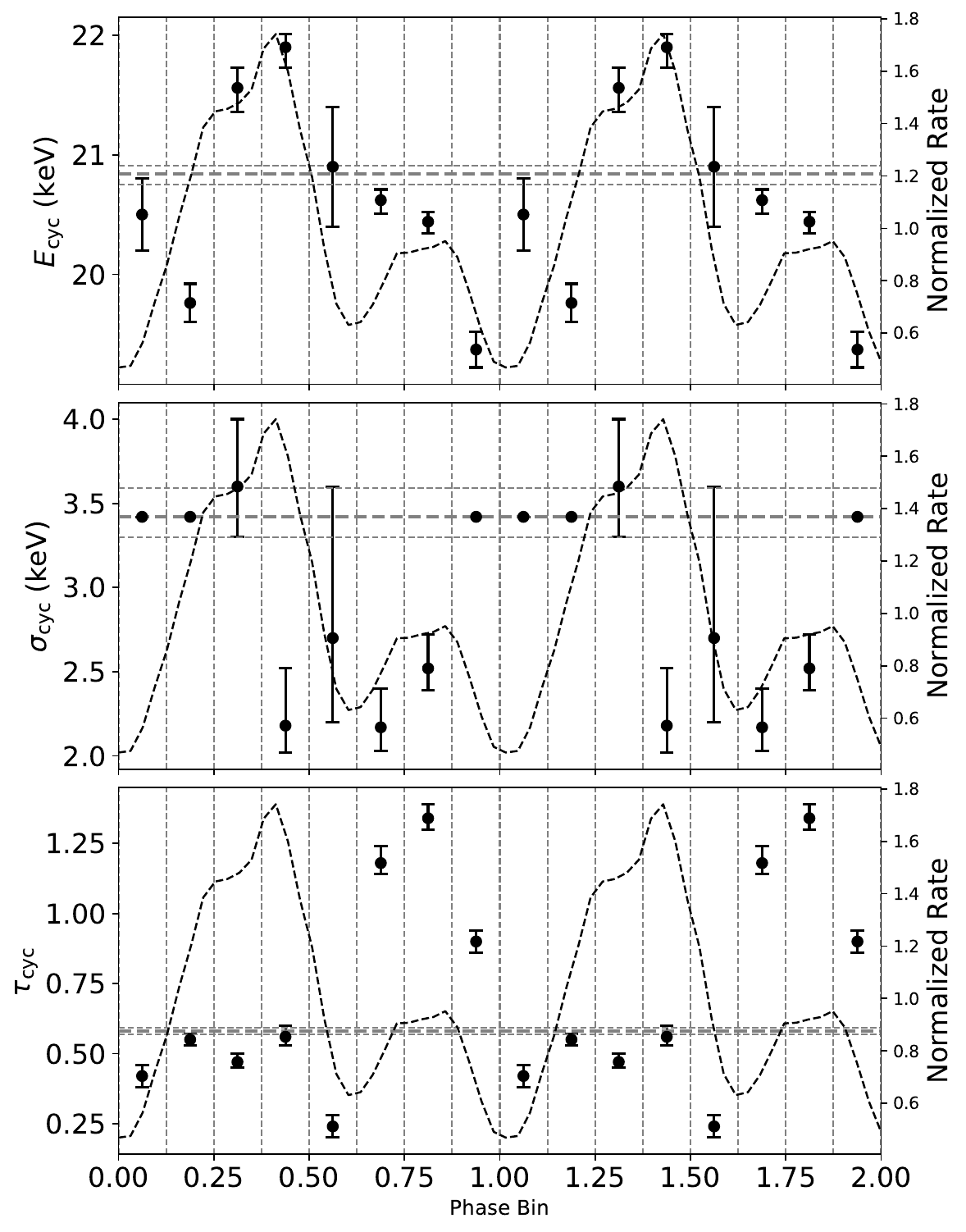}
    \end{subfigure}
    \hspace{0.05\textwidth}  
    \begin{subfigure}{0.37\textwidth}
        \centering
        \includegraphics[width=\linewidth]{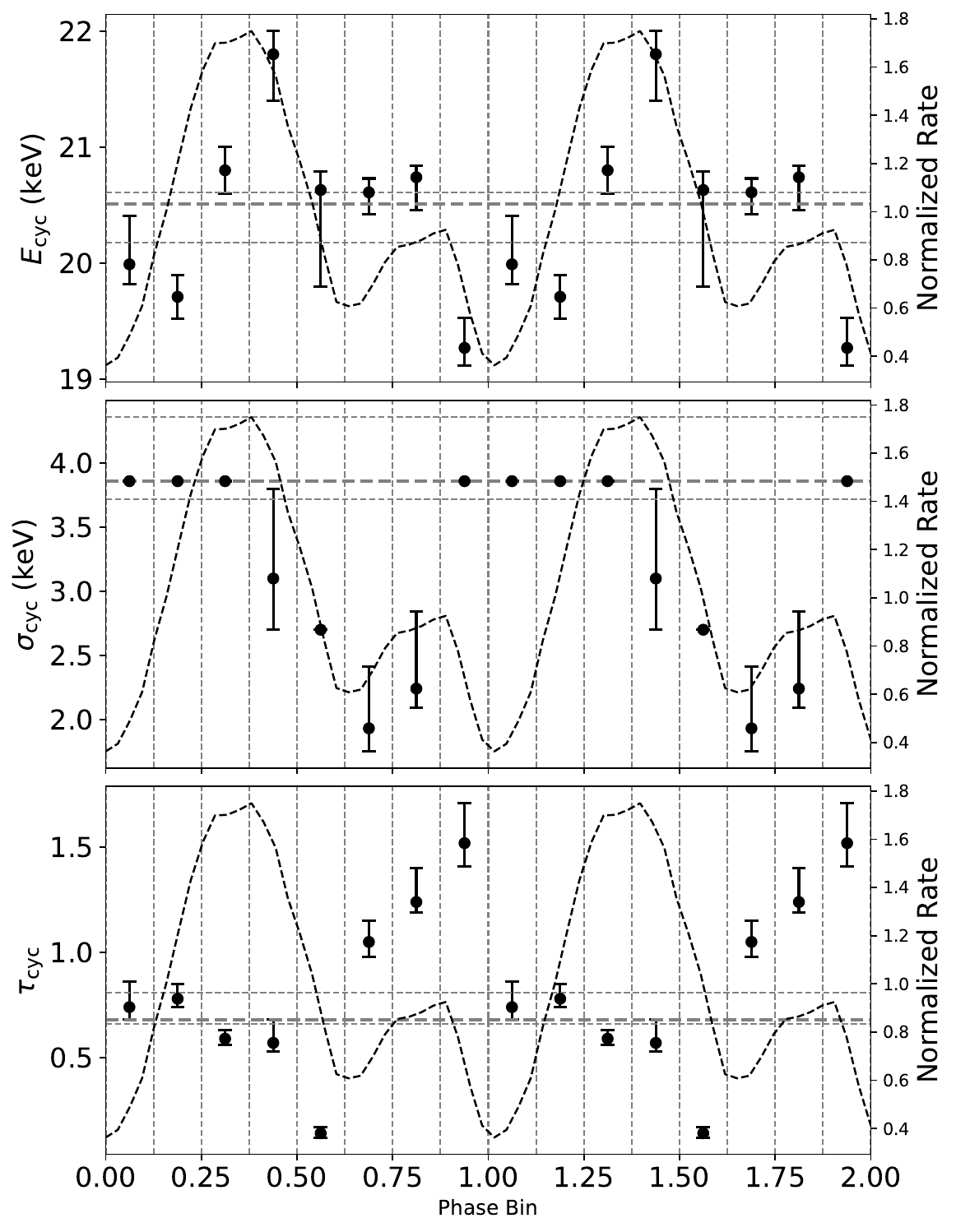}
    \end{subfigure}
  
    \caption{Phase-resolved parameters}
    \label{fig:PRparameters}
\end{figure*}

Our simulations indicate that, given a small size of the poles on the neutron star surface (with a radius of about a few hundred meters), the geometric parameters and the internal emission patter, $I(\theta)$, determine the observed spectral flux variations. The angle, under which the observer sees each of the poles at a certain rotation phase, is fully determined by the geometry and the compactness of the neutron star. For a selected geometry, which, according to our simulations, is appropriate for \src, $i=67^\circ$ (fixed to the inclination of the system), $\Theta_1=80^\circ$, $\Theta_2=105^\circ$, $\Delta \Phi = 168^\circ$, we show the phase dependency of the emission angle in Fig.~\ref{fig:modemis}(a) and the adopted geometry in Fig.~\ref{geometry}. This variation is a crucial component for understanding the observed cyclotron line behavior. The scattering cross sections exhibit strong anisotropy dominated by angle-dependent Doppler broadening near the cyclotron resonance \citep[see, e.g.,][]{Schwarm2017}. As a result, the closer the emission angle is to the magnetic field direction, along which the thermal motion of electrons is not restricted, the shallower the line profile observed \citep{Meszaros1985a}.

Figure~\ref{fig:modemis}(b) shows PPs for the chosen geometry. As indicated by the dashed-dotted blue and dashed red lines, each of two peaks in the profile is fully dominated by the contribution from one of the poles. From Fig.~\ref{fig:modemis}(a), it is clear that near the first peak, we observe photons emitted from pole~1 at an angle of ${\approx}10^\circ$ to the magnetic field, while near the secondary peak, photons are emitted from pole~2 at ${\approx}30^\circ$. This difference in the emission angle results in a significantly more shallow profile of the fundamental CRSF observed at phases near the main peak (Fig.~\ref{fig:modemis}, c). As was shown in Sect.~\ref{bumpdisc}, this fact leads to a pronounced bump in the PFS, which we show in Fig.~\ref{fig:modemis}(d). The PF peaks at slightly lower energies than the geometric center of the observed lines, indicating the largest difference between the line fluxes. Due to the complex redistribution occurring during the formation of broad lines near the continuum cut-off, one can expect some variation of the centroid energies of the lines seen under different angles. Specifically, a simple fit with a Gaussian absorption in the two spectra of Fig.\ref{fig:modemis}(c) gives a centroid energy of $24.5 \pm 0.2$ $\mathrm{keV}$ for the spectrum of the primary peak, and $22.7 \pm 0.4$ $\mathrm{keV}$ for that of the secondary peak. The bump appears at lower energies, consistent with the expectations discussed in subsection\ref{bumpdisc}. 


A similar physical model for the emission was previously adopted to describe spectra \citep{Bulik1992} and pulse profiles \citep{Bulik1995} of \src. \citet{Bulik1992} included the fully relativistic Compton scattering cross sections and obtained observed spectral fluxes for different rotation phases. They considered a more complex shape of emission regions -- spherically curved polar caps, but omitted the general relativistic effects on photon trajectories. The latter effect was included in \citet{Bulik1995} who fitted pulse profiles of the source at different energies. The emission was calculated from an inhomogeneous atmosphere, but with the same non-relativistic cross sections as used in the current work. We would like to note that in the qualitative study performed here, the light-bending effect plays a more crucial role in shaping the general behavior of observables than the relativistic effects in the scattering cross sections. The latter is not expected to dramatically alter the emission pattern from magnetized plasma but does noticeably affect the width of the cyclotron resonances, generally making them narrower and deeper \citep{Alexander1989, Bulik1992}.
Light bending plays a crucial role in the observability of emitting regions on (and near) the neutron surface, causing radiation beaming even from an isotropically emitting spot and significantly increasing the contribution from the back side of the star \citep{pechenick1983hot}. In general, it leads to a decrease of the PF, except in extreme
cases of high compactness or extended accretion columns, where photons bent through large angles create a bright ring around the neutron star, resulting in a spike in the light curve (\citep{pechenick1983hot, Falkner2016, Markozov2024}. For the geometry adopted here (see Fig. \ref{geometry}), our simulations show that although the overall shape of the PPs and PFS is preserved in the flat space-time projection, accounting for light bending changes the pulse
fraction across the whole spectrum (on average by ${\sim}40\%$) and makes the minima in pulse profiles significantly shallower.

Although earlier works incorporated more complex physics — such as relativistic cross sections in \citet{Bulik1992} and self-consistent atmospheric structure calculations in \citet{Bulik1995} — the otherwise similar models to the one presented were generally unable to describe the data of \src\ in a statistically satisfactory manner without introducing additional components of ambiguous physical interpretation. For example, even with relativistic cross sections, the shape of the cyclotron line in \citet{Bulik1992} deviated significantly from the observed one, prompting the authors to adopt a mixed, arbitrary weighted contributions from emitting surface elements with different magnetic fields, covering a wide range of cyclotron energies, $E_\mathrm{cyc}=16$--$26\,\mathrm{keV}$ to obtain the total flux. \citet{Bulik1995} also encountered the difficulty of simultaneously describing a deep shape of the cyclotron line and the phase modulation of the flux. They introduced an additional ``nonmagnetic'' component summed with the magnetic one. The latter determined the location of the cyclotron line, while the nonmagnetic component played role of a continuum contribution, reducing the line depth. Taking into account these facts, together with the high quality of the data presented in the current work and similar state of the physical modeling, we do not aim to perform phase-dependent spectral fitting, which would inevitably require introduction of additional components. Instead, we argue that the general trends observed in the PPs, spectral flux, and PFS of the source can be qualitatively described and well understood within the framework of a physical model based on thermal resonant Comptonization in the slab-like emission regions and its relativistic projection onto the observer's plane.

\subsection{Hints of a cyclotron harmonic in a single phase bin}

Our phase-resolved spectral analysis of ObsID 304 revealed residuals around 40 $\mathrm{keV}$ in the phase bin of the primary peak A (0.375-0.500), shown in Fig.~\ref{fig:grid304} of the Appendix \ref{spectral_plots}. Modeling these residuals with an additional absorption feature using a Lorentzian profile, whose width was fixed to that of the first harmonic, as it could not be otherwise constrained, improves the reduced chi-squared from 243.9/234 to 217/232. The best-fit parameters indicate a line energy at $40.0^{+1.3}{-1.1}$,$\mathrm{keV}$ with a depth of $0.74^{+0.28}{-0.26}$, corresponding to a detection significance of approximately $2.6\sigma$, which is suggestive but not yet statistically compelling.

Given that the energy of the absorption feature is almost exactly twice the fundamental cyclotron line energy, one possible origin could be the harmonic of the cyclotron line. This specific phase bin appears to be unique among the four \nustar observations: as discussed in Section \ref{resultstiming}, the first peak in observation 304 exhibits a more structured profile, showing a higher count rate at a phase of approximately 0.4. In contrast, the other three observations do not display this increase in count rate at the same phase.

We argue that the higher statistics at this phase might facilitate the detection of this potential harmonic line, whereas for the same phase bin of the other observations, the non-detection is mainly due to lack of statistics. A harmonic CRSF has been reported for this source before \citep{Robba2001, Rodes-Roca2009}, though at significantly higher energies, suggesting that the feature detected here may have a different origin or correspond to a different accretion regime.

Finally, it should be kept in mind that the formation of the second harmonic may not mirror that of the fundamental line. As shown by \citet{Schwarm2017b} (Fig. 5), the first and second harmonics reach their maximum depths at different angles relative to the magnetic field, which can be attributed to differences in redistribution and the ratio of $E_{\mathrm{cyc}}^n$ to the plasma temperature. It is therefore possible that the second harmonic emerges more clearly at certain viewing angles or magnetic configurations, which could be met only in specific pulse phases—such as in this particular bin of ObsID 304.

\section{Conclusions}

We have analyzed all available \nustar\ data of \src\ to investigate the energy-resolved pulse profile variations and the phase-dependent spectral variability, with a particular focus on the fundamental CRSF. Through decomposing the pulse profiles into Fourier harmonics and computing the pulsed fraction spectra, cross-correlation, and lag spectra, we identified key spectral features associated with the CRSF. Our timing and pulse profile analysis revealed that the secondary peak of the PP disappears at the CRSF energies, which correspond to where a broad \textit{bump} is observed in the PFS. Phase-resolved spectroscopy further shows that the CRSF is significantly deeper during the phase bins of the secondary peak, consistent with the disappearance of the secondary peak at these energies. 

We demonstrated that broad \textit{bumps} and \textit{dips} observed in the PFS around the CRSF energy are a natural consequence of strong phase-dependent variations in the CRSF parameters. In the case of \src, the observed \textit{bump} arises because the CRSF is much deeper in the low-flux phase bins, while the high-flux bins show a much shallower feature. Conversely, in sources like Her X-1, where \textit{dips} are seen in the PFS, the CRSF is shallower during the off-pulse (low-flux) phases and deeper during the pulse peak. Through a simple exercise, we also showed that the energy of these features in the PFS does not necessarily coincide with the spectral energy of the CRSF when the line energy itself varies with spin phase. Specifically, if the CRSF energy increases with flux, as seen in both Her X-1 and \src, then the location of the \textit{dip} or \textit{bump} in the PFS will shift accordingly: if the line is deeper at high flux, a \textit{dip} will appear at higher energies; if it is shallower, a \textit{bump} will appear at lower energies. This naturally explains, for instance, why the \textit{dip} in the PFS of Her X-1 is observed at a higher energy than the CRSF centroid, while the \textit{bump} in the PFS of \src\ appears at a lower energy. Furthermore, variations in the cut-off energy as a function of the spin-dependent flux can account for the overall upward or downward trends observed in the PFS. These findings underscore the importance of the PFS as a diagnostic tool for CRSFs, providing an independent and direct way to probe their spectral spin-dependent behavior. The PFS can give immediate information on the relationship between the CRSF strength as a function of the spin-dependent flux, enabling a more direct interpretation of the underlying accretion geometry and magnetic field structure. 

To provide a qualitative description to the observed energy- and phase-dependent phenomena, we also performed modeling of emission from strongly magnetized plasma and studied the effect of geometry -- the magnetic pole locations and the observer's line of sight -- on pulse profiles, spectra, and energy dependency of the pulsed fraction. Our model indicates the high observer's inclination and magnetic angle, with a difference between the two angles $|i-\Theta_1|\lesssim25^\circ$ based on a significant reduction of the secondary peak in the pulse profile. Within the frame of the model, the behavior of the CRSF at the two peaks associated with the two poles, is explained by observing photons emitted under different angles to the magnetic field at these phases. This results in a shallow line with a slightly higher centroid energy at the primary peak and a deeper line with a lower energy at the secondary peak. In turn, this leads to the discussed variation of the PFS at energies near the fundamental CRSF.
The behavior of the CRSF described here favors line formation under thermal Comptonization. In contrast, the angular dependence of line parameters formed under the influence of bulk Comptonization is more complex and is associated with significantly larger shifts in centroid line energies \citep[see, e.g.,][their Chapter 5]{Schwarm2017b}. We found a similar asymmetry in the location of the two poles as reported before, ${\sim}10$--$15^\circ$. Both, previous studies \citep{Bulik1992, Bulik1995} and the offset from a pure dipole, indicate that the conditions at the two poles, that is, the characteristic magnetic field strength and the plasma temperature, might slightly differ. We would like to note that to describe the qualitative behavior observables, the two identical poles assumed in this work were sufficient. However, for the future studies aiming at the more careful and statistically acceptable description of the data, this and other effects will likely need to be considered.

\begin{acknowledgements} 

This research was supported by the International Space Science Institute (ISSI) in Bern, through the ISSI Working Group project \href{https://collab.issibern.ch/neutron-stars/}{Disentangling Pulse Profiles of (Accreting) Neutron Stars} and by the European Space Agency (ESA) through the \href{https://www.cosmos.esa.int/web/esdc/visitor-programme}{Archival Research Visitor Program}.\\

AD, EA, GC, VLP acknowledge funding from the Italian Space Agency, contract ASI/INAF n.I/004/11/4. 

CP, MDS, AD, FP, GR acknowledge support from SEAWIND grant funded by the European Union - Next Generation EU, Mission 4 Component 1 CUP C53D23001330006.

FP, AD, MDS, GR and CP aknowledge the INAF GO/GTO grant number 1.05.23.05.12 for the project OBIWAN (Observing high B-fIeld Whispers from Accreting Neutron stars).

EA aknowledges support from the INAF MINI-GRANTS number 1.05.23.04.04 for the project PPANDA (Pulse Profiles of Accreting Neutron Stars deeply analyzed). 

AA acknowledges financial support from ASI-INAF Accordo Attuativo HERMES Pathfinder operazioni n. 2022-25-HH.0

ESL acknowledges support from Deutsche Forschungsgemeinschaft grant WI 1860/11-2.

We made use of Heasoft and NASA archives for the \nustar data.
We developed our own timing code for the epoch folding, orbital correction,
building of time-phase and energy-phase matrices.
This code is based partly on available Python packages such as:
\texttt{astropy} \citep{AstropyCollaboration2013},
\texttt{lmfit} \citep{lmfit},
\texttt{matplotlib} \citep{Hunter2007},
\texttt{emcee} \citep{Foreman-Mackey2013},
\texttt{stingray} \citep{Huppenkothen2019},
\texttt{corner} \citep{Foreman-Mackey2016},
\texttt{scipy} \citep{Virtanen2020}.
An online service that reproduces our current results is available on the Renku-lab platform of the Swiss Science Data Centre at \href{https://renkulab.io/projects/carlo.ferrigno/ppanda-light/sessions/new?autostart=1}{this link}.\\

\end{acknowledgements}



\appendix
\titlespacing*{\section}{0pt}{*0}{*0} 

\onecolumn

\section{Summary plots of the pulse profile properties of 302, 306-2 and 306-4}
\label{timing_plots_302_3064}

\FloatBarrier

\begin{figure*}[ht]
    \centering
    \includegraphics[width=0.75\textwidth]{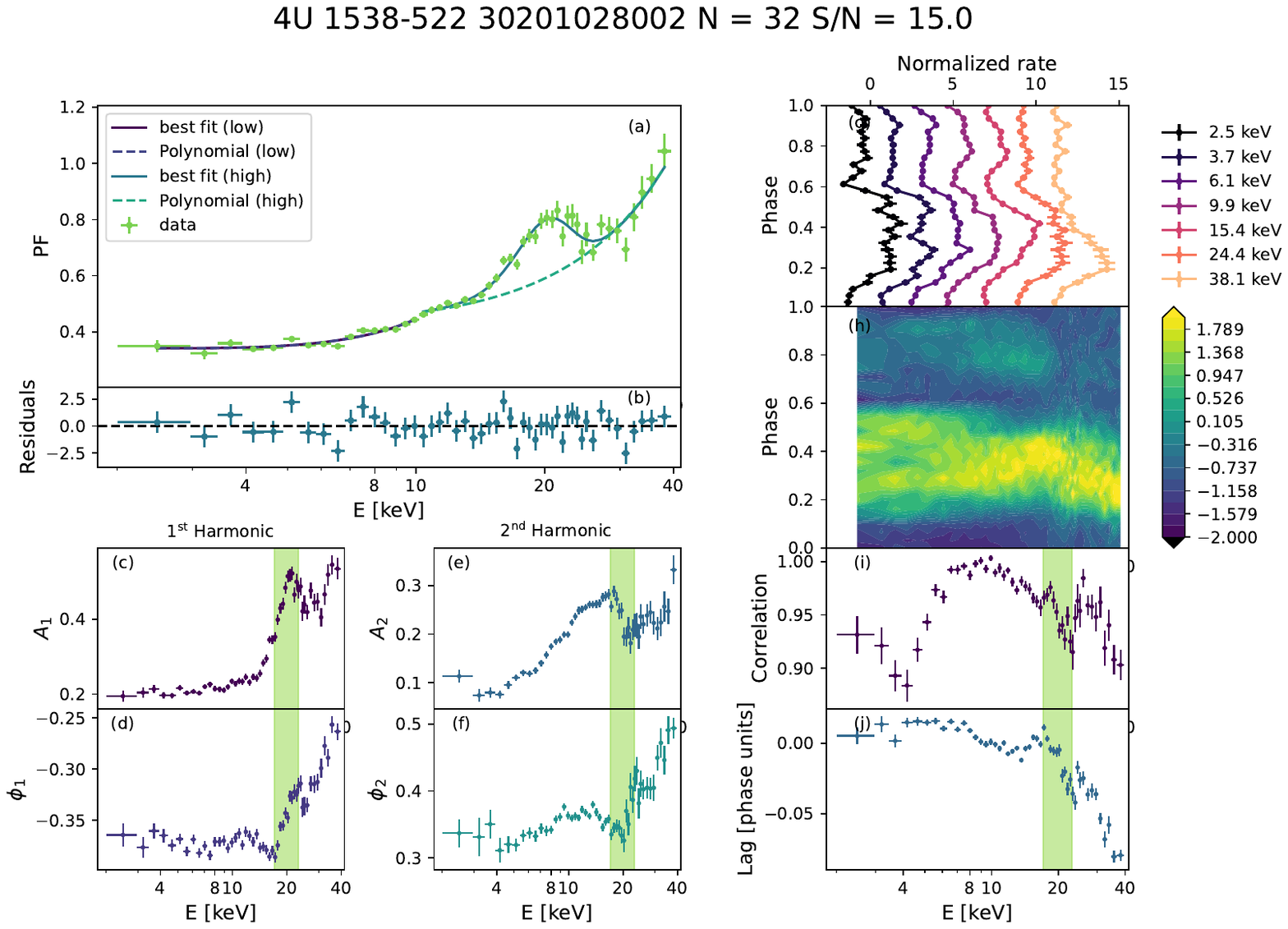}
    \caption{PP properties for 4U 1538-522 in ObsID 302  (see the caption of Fig.\ref{SUMMARYPLOT304}),
    for 32 phase bins and a S/N of 15.}
    \label{SUMMARYPLOT302}
\end{figure*}

\begin{figure*}[ht]
    \centering
    \includegraphics[width=0.75\textwidth]{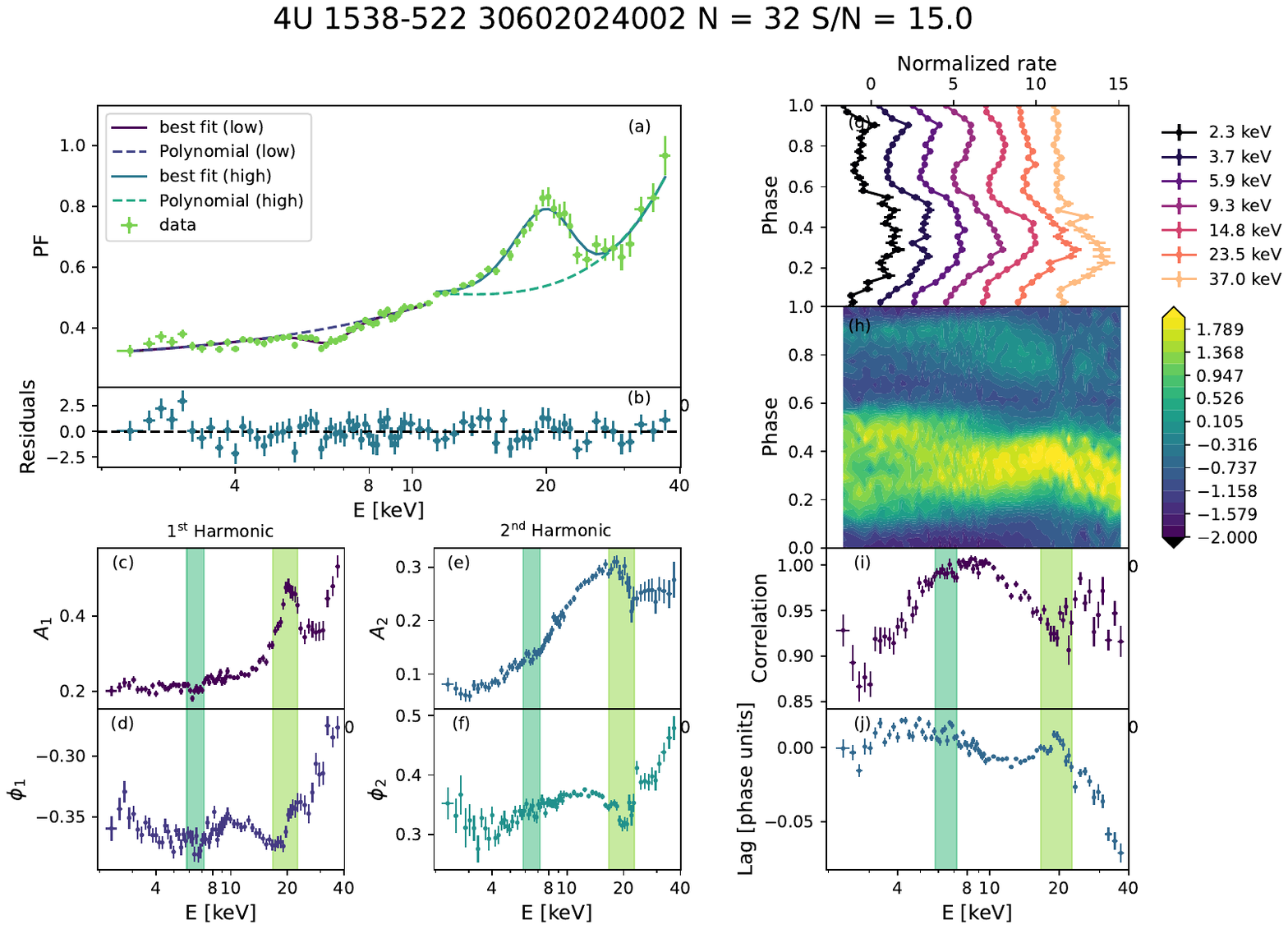}
    \caption{PP properties for 4U 1538-522 in ObsID 306 (see caption of Fig.\ref{SUMMARYPLOT304}), 
    for 32 phase bins and a S/N of 15.}
    \label{SUMMARYPLOT306}
\end{figure*}

\begin{figure*}[ht]
    \centering
    \includegraphics[width=0.75\textwidth]{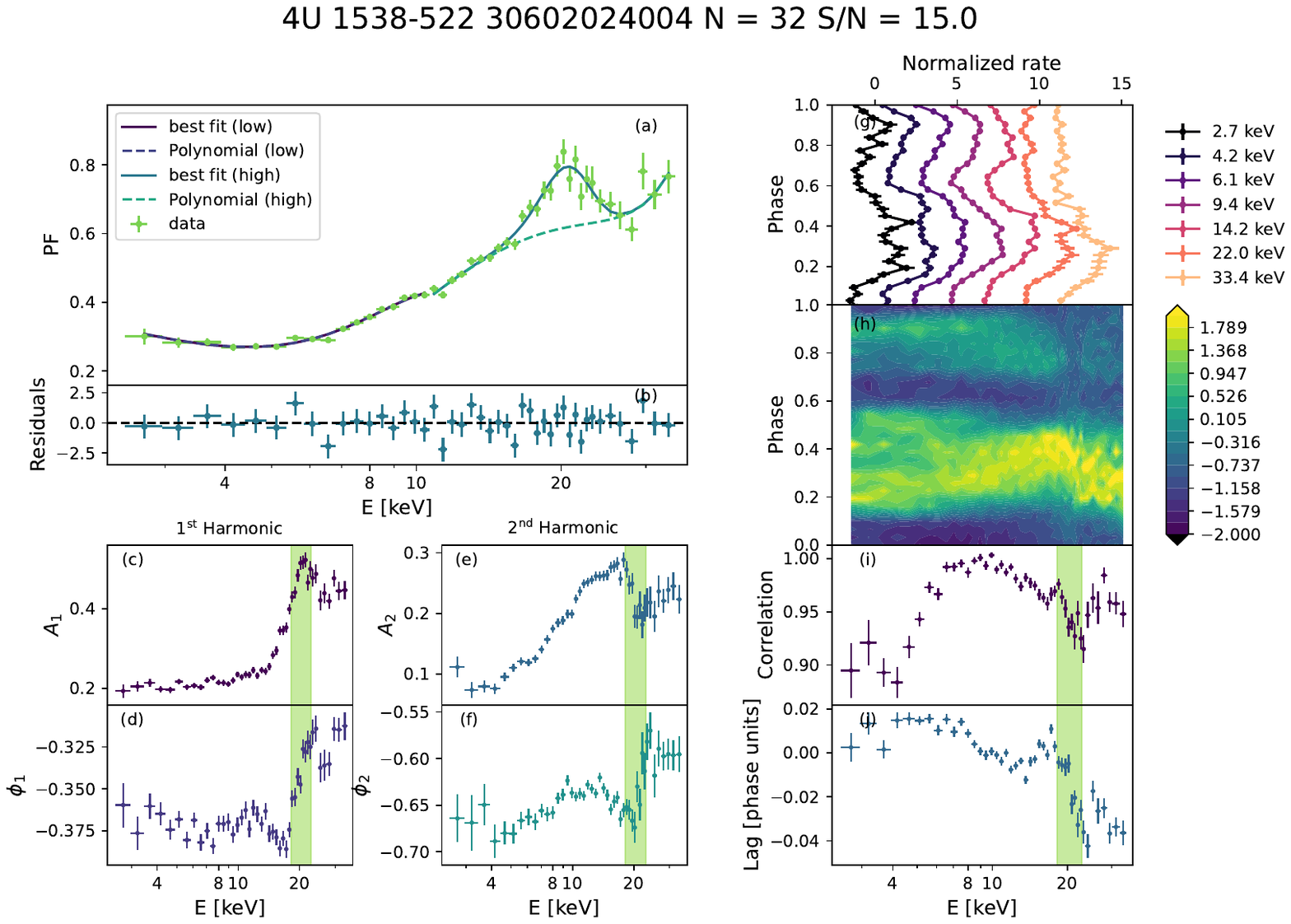}
    \caption{PP properties for 4U 1538-522 in ObsID 306-4  (see caption of Fig.\ref{SUMMARYPLOT304}),
    for 32 phase bins and a S/N of 15.}
    \label{SUMMARYPLOT306-4}
\end{figure*}

\FloatBarrier

\section{Plots of spectral analysis}
\label{spectral_plots}

\FloatBarrier

\begin{figure}[H] 
    \centering
    \includegraphics[width=0.85\textwidth]{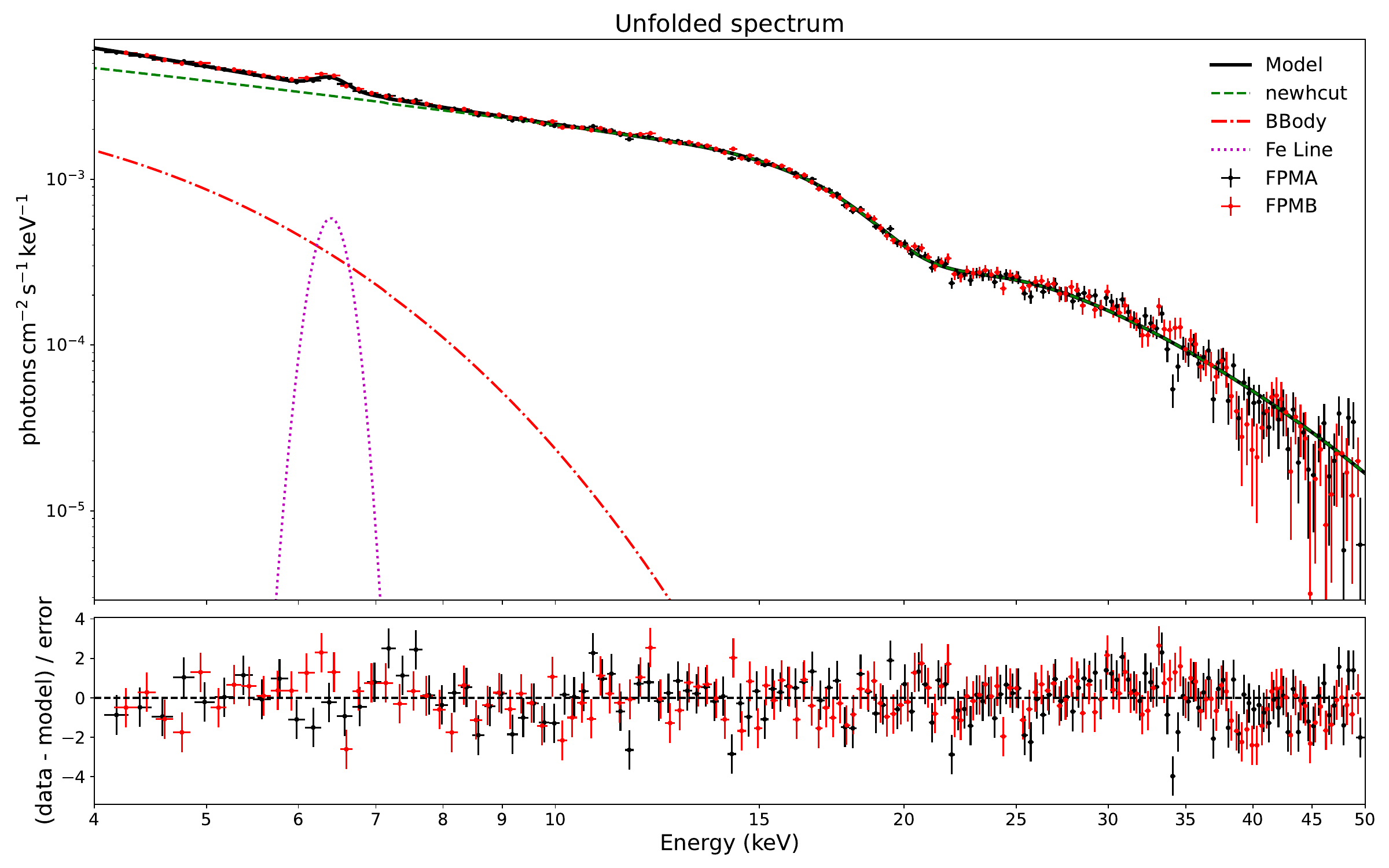}
    \caption{Observation 306-2 PA}
    \label{306PA}
\end{figure}


\subsection*{Phase-resolved spectral analysis of 304}

\begin{figure}[H]
    \centering
        \begin{subfigure}{0.45\textwidth}
            \centering
            \includegraphics[width=\linewidth]{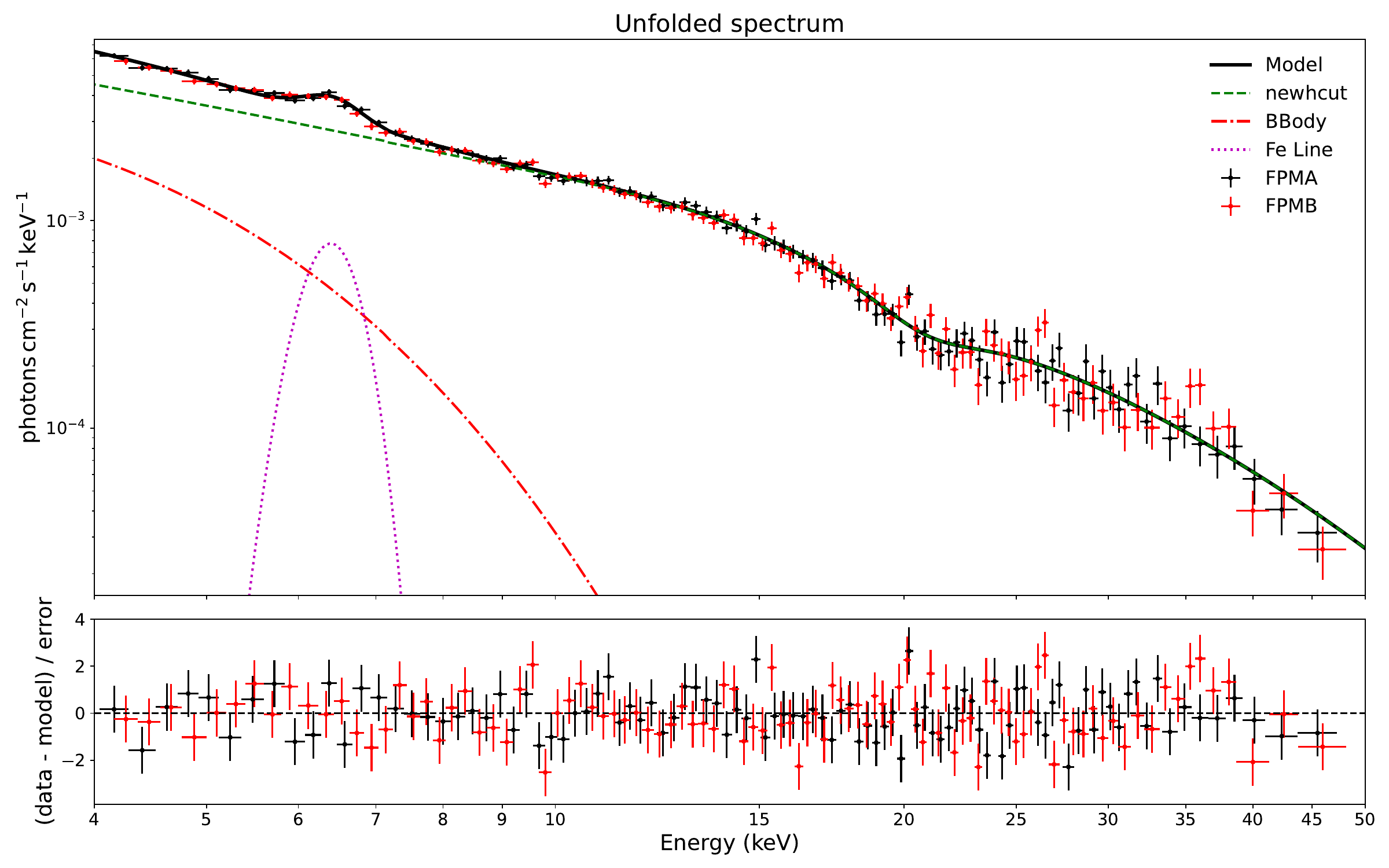}
        \end{subfigure}
        \hfill
        \begin{subfigure}{0.45\textwidth}
            \centering
            \includegraphics[width=\linewidth]{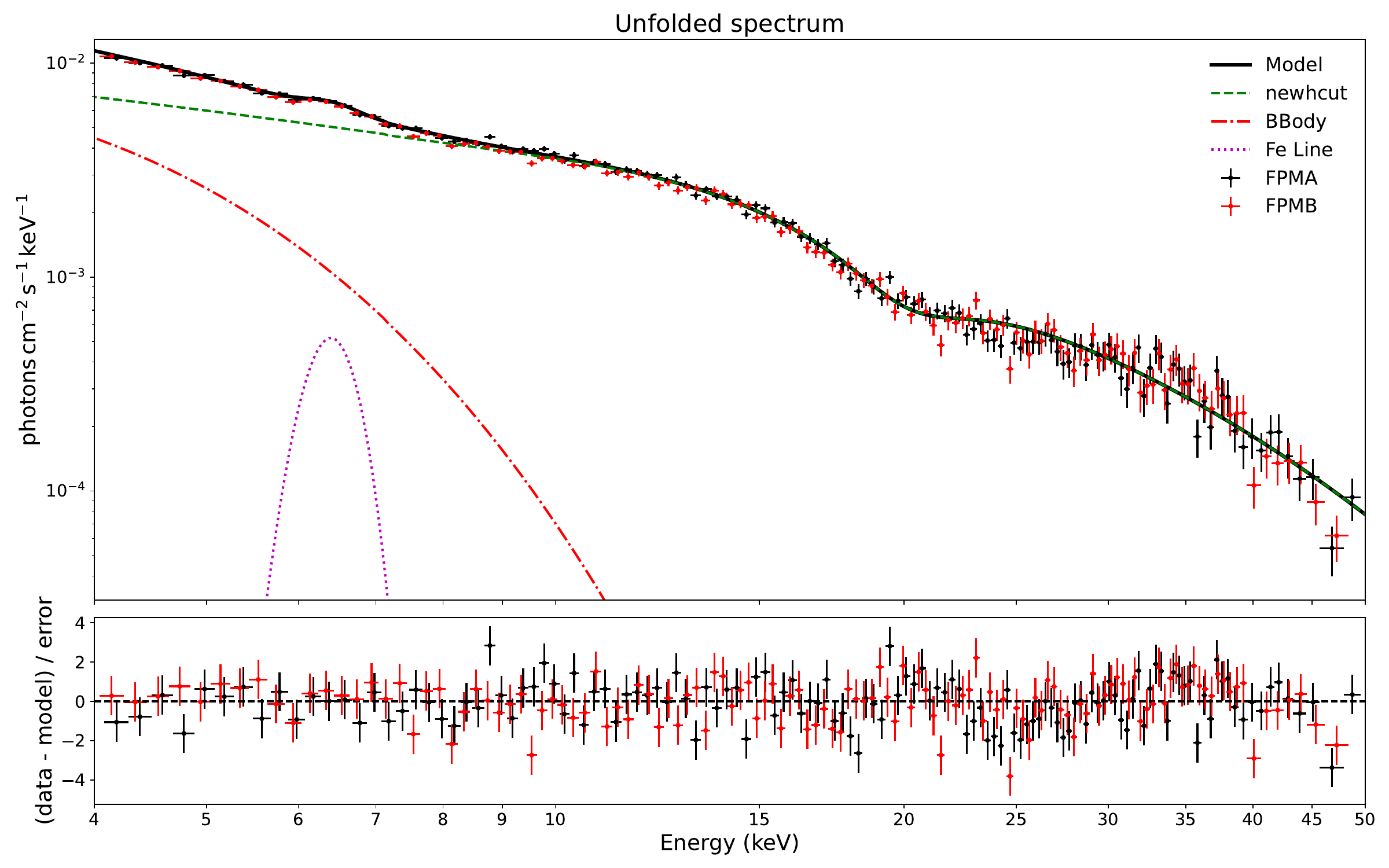}
        \end{subfigure}
    
        \vskip\baselineskip  
    
        \begin{subfigure}{0.45\textwidth}
            \centering
            \includegraphics[width=\linewidth]{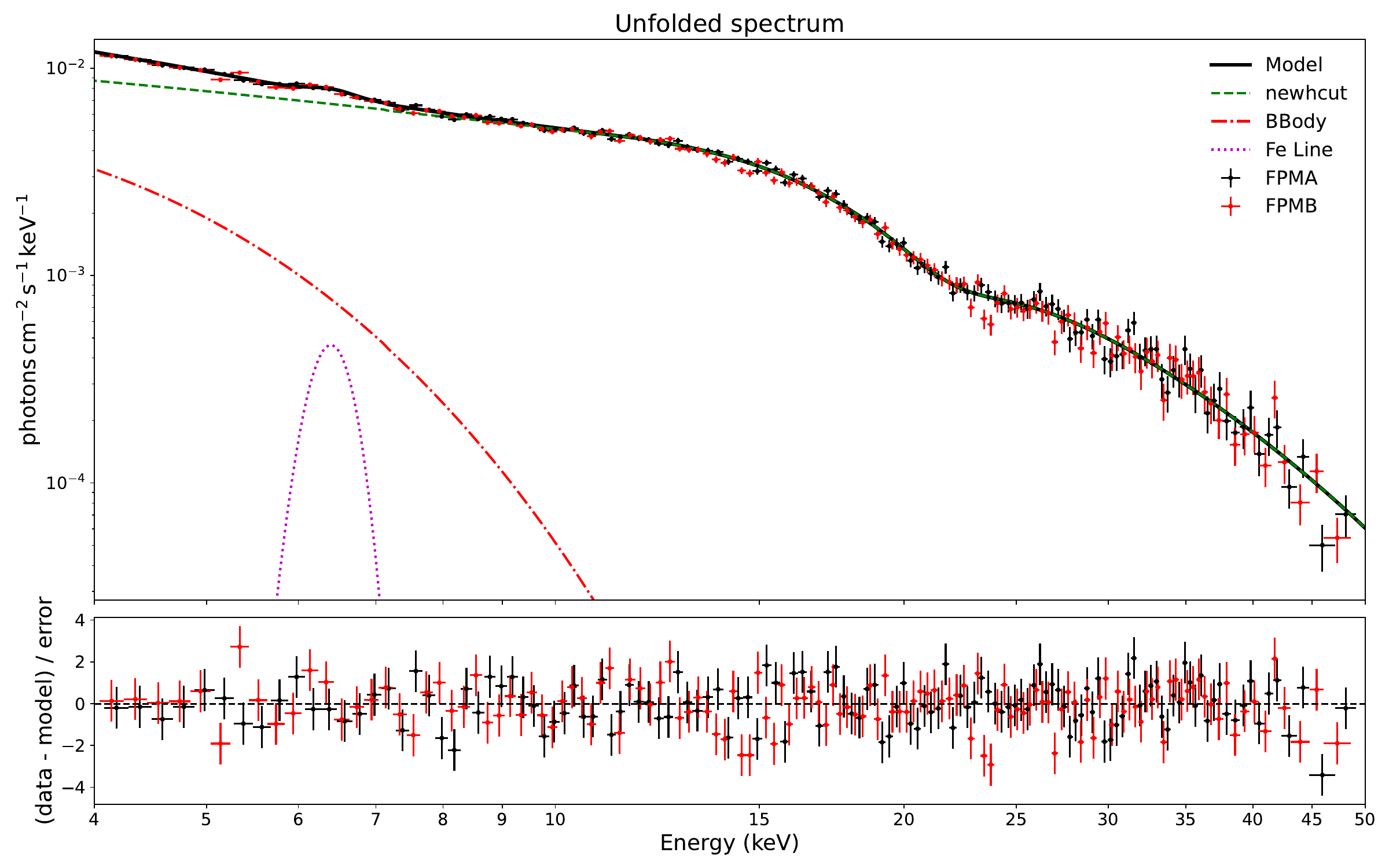}
        \end{subfigure}
        \hfill
        \begin{subfigure}{0.45\textwidth}
            \centering
            \includegraphics[width=\linewidth]{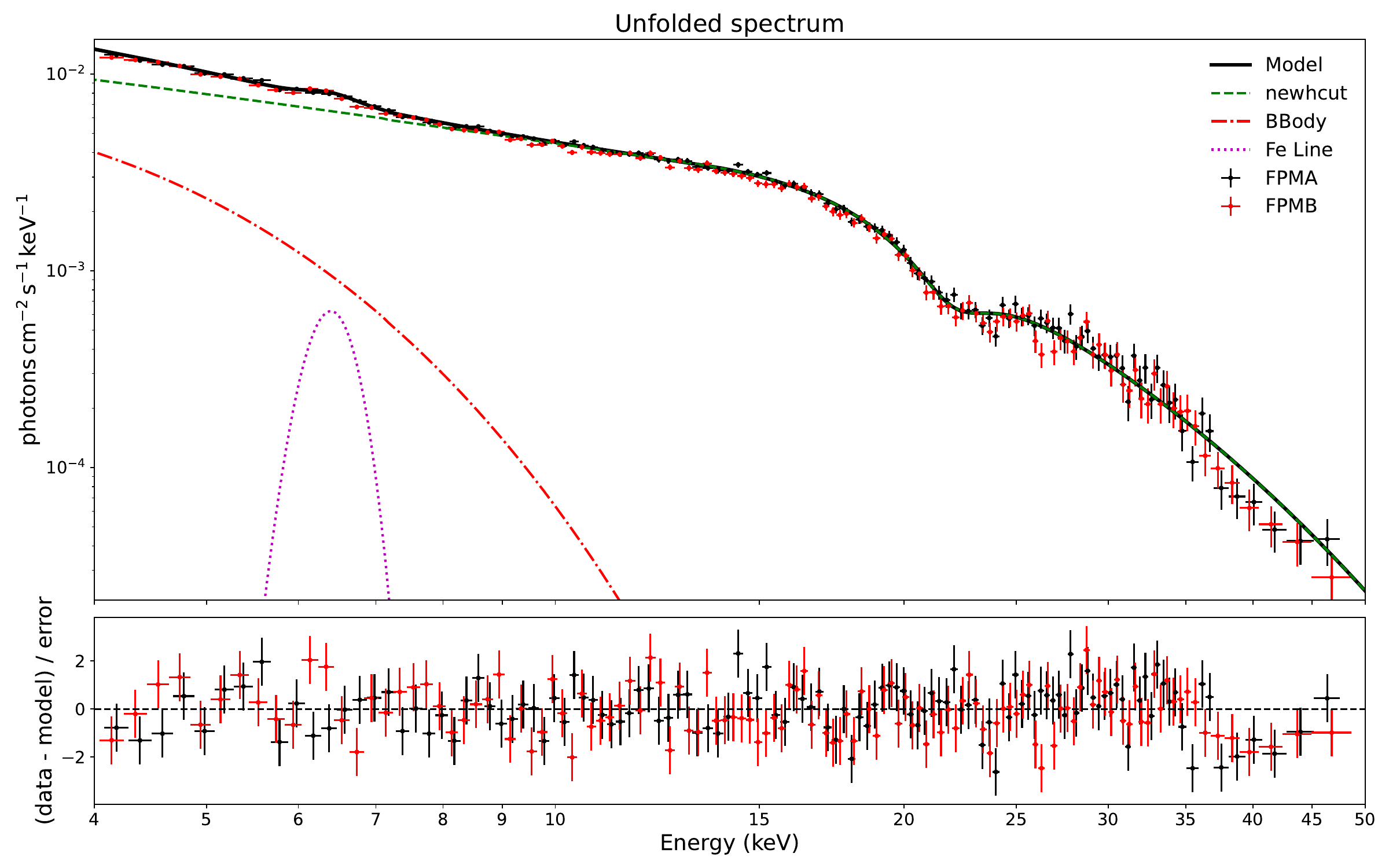}
        \end{subfigure}
    
        \vskip\baselineskip
    
        \begin{subfigure}{0.45\textwidth}
            \centering
            \includegraphics[width=\linewidth]{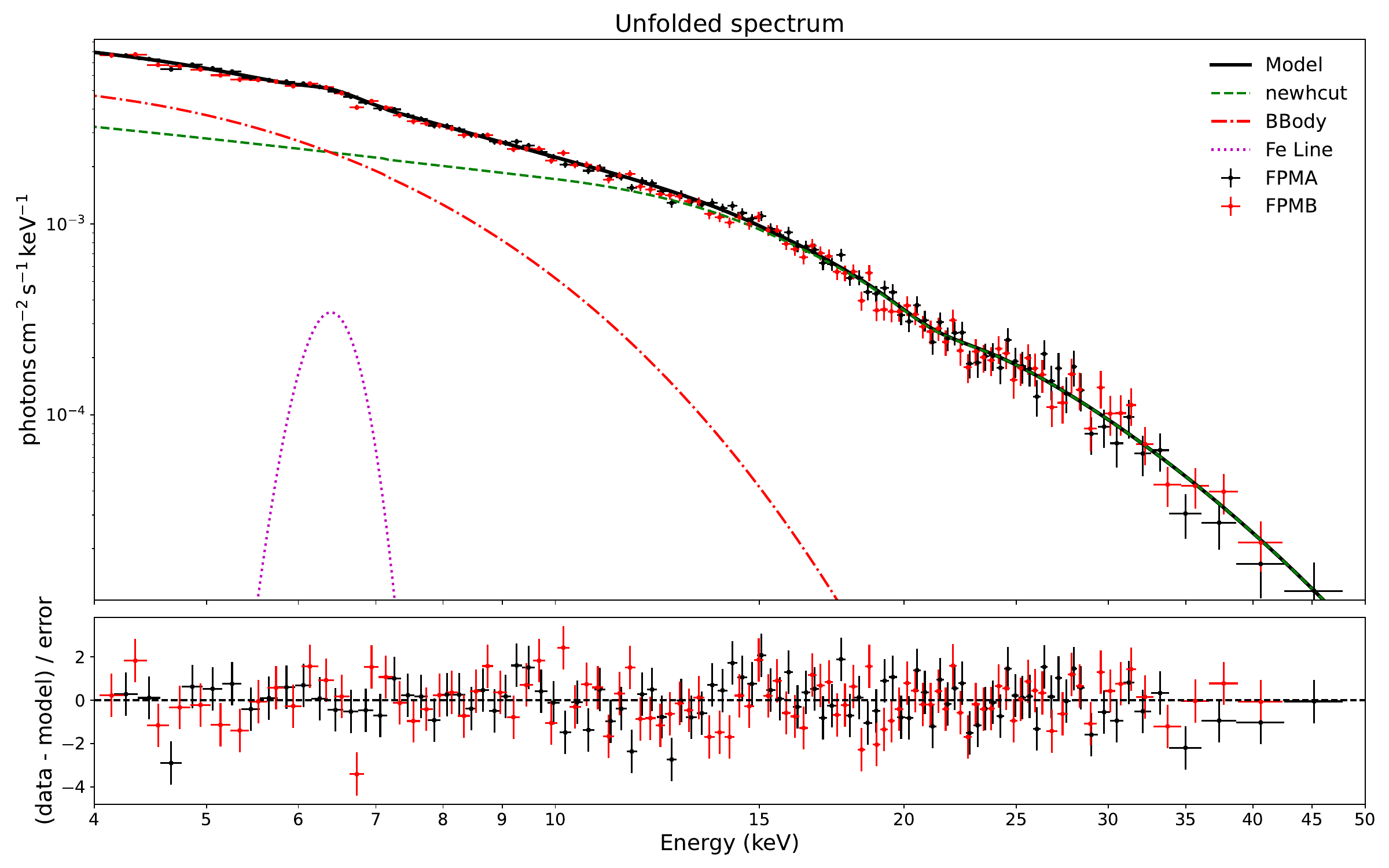}
        \end{subfigure}
        \hfill
        \begin{subfigure}{0.45\textwidth}
            \centering
            \includegraphics[width=\linewidth]{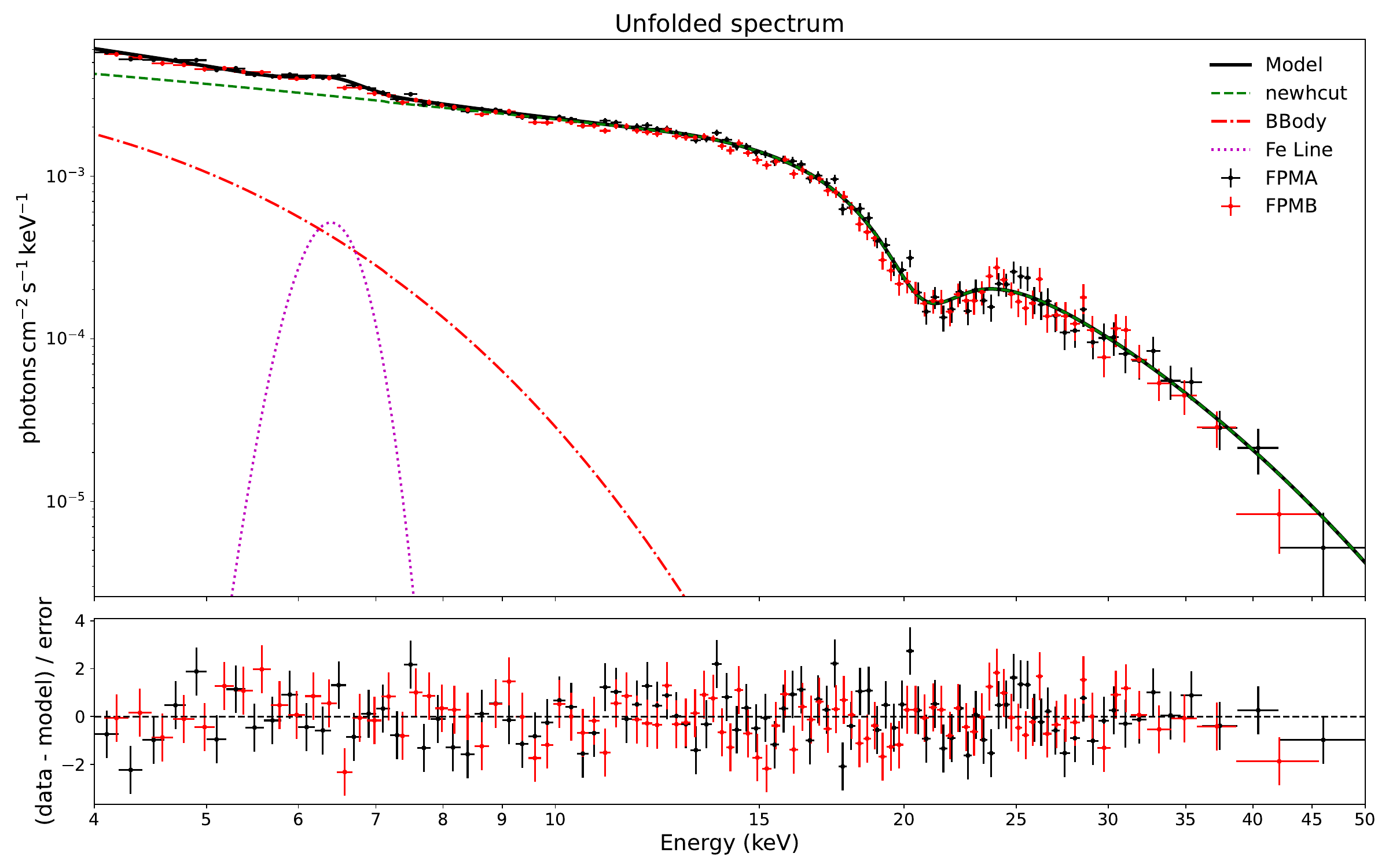}
        \end{subfigure}
    
        \vskip\baselineskip
    
        \begin{subfigure}{0.45\textwidth}
            \centering
            \includegraphics[width=\linewidth]{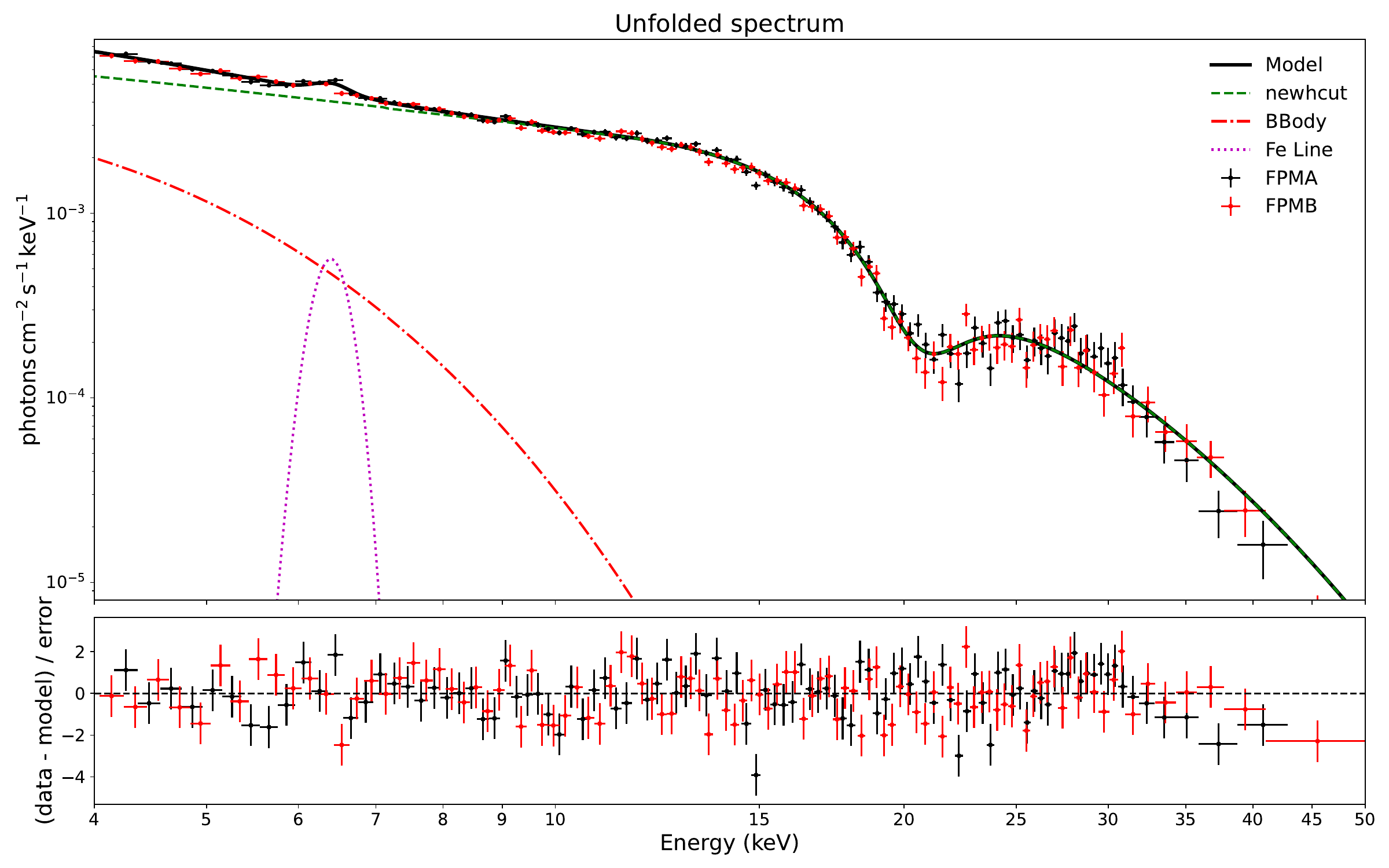}
        \end{subfigure}
        \hfill
        \begin{subfigure}{0.45\textwidth}
            \centering
            \includegraphics[width=\linewidth]{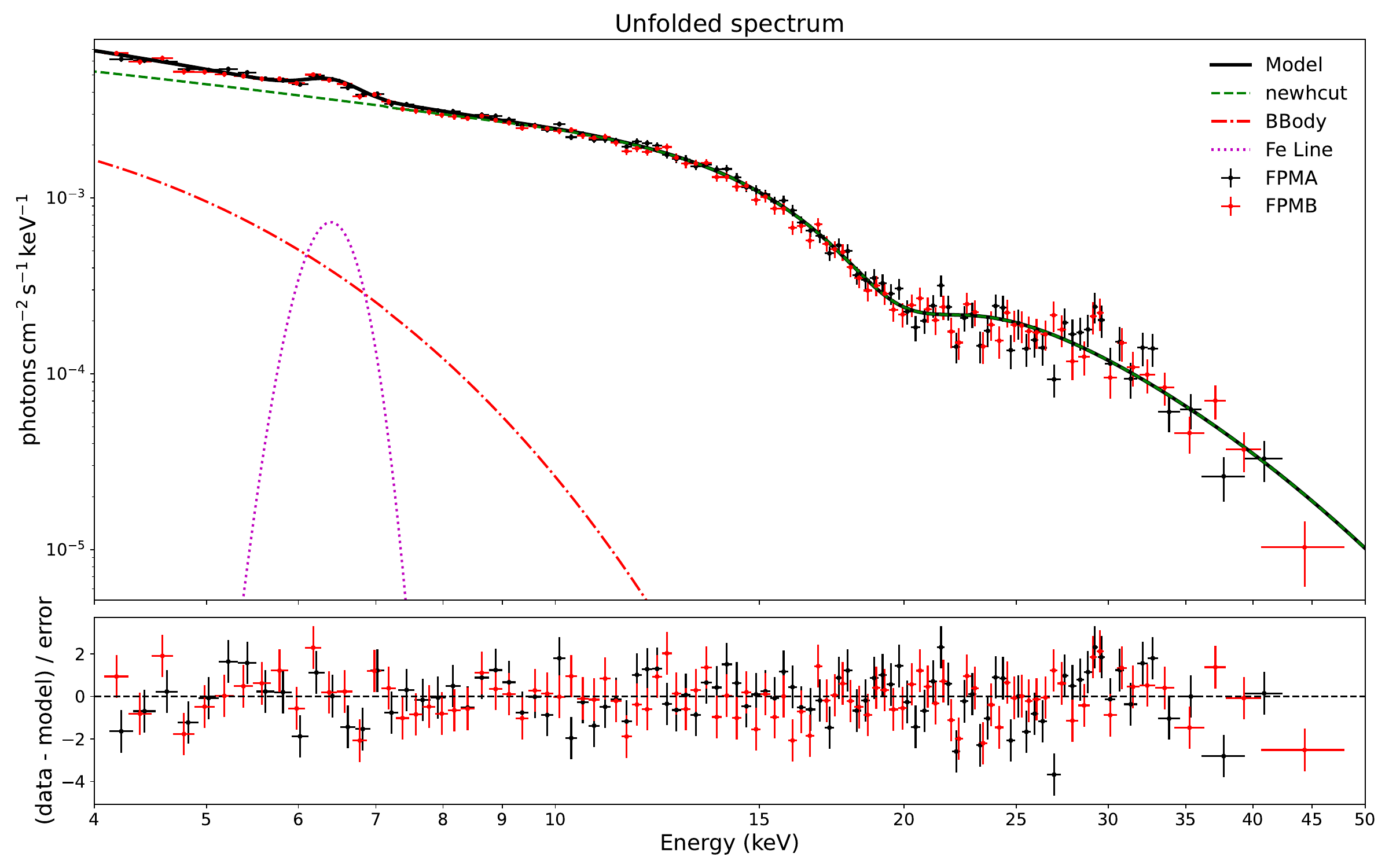}
        \end{subfigure}
    \caption{Observation 304: A grid of 8 figures in 4 rows and 2 columns, showing phase bins 0.000-0.125 to 0.875-1.000.}
    \label{fig:grid304}
\end{figure}

\subsection*{Phase-resolved spectral analysis of 306-2}
\begin{figure}[H]
    \centering
    \begin{subfigure}{0.45\textwidth}
        \centering
        \includegraphics[width=\linewidth]{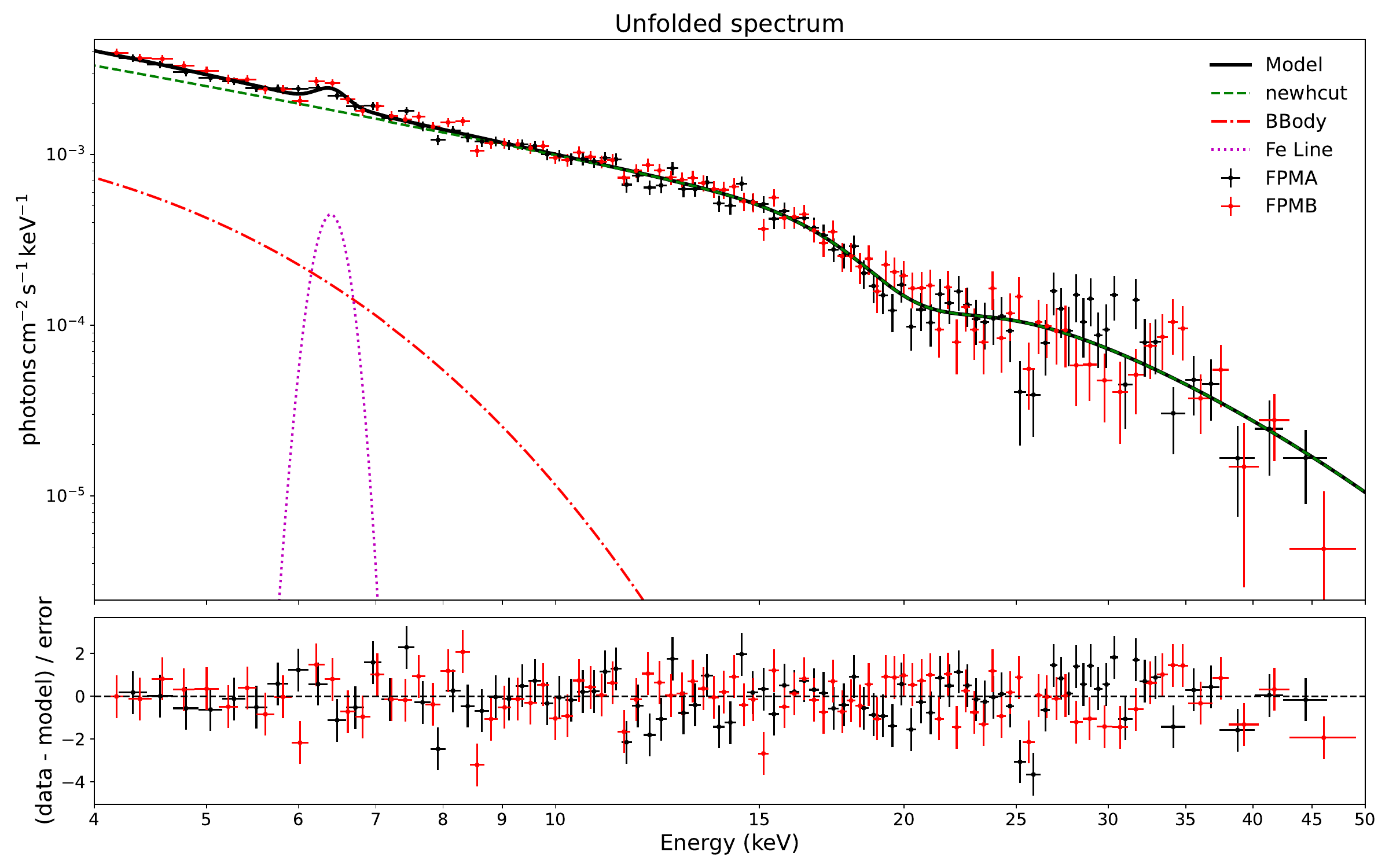}
    \end{subfigure}
    \hfill
    \begin{subfigure}{0.45\textwidth}
        \centering
        \includegraphics[width=\linewidth]{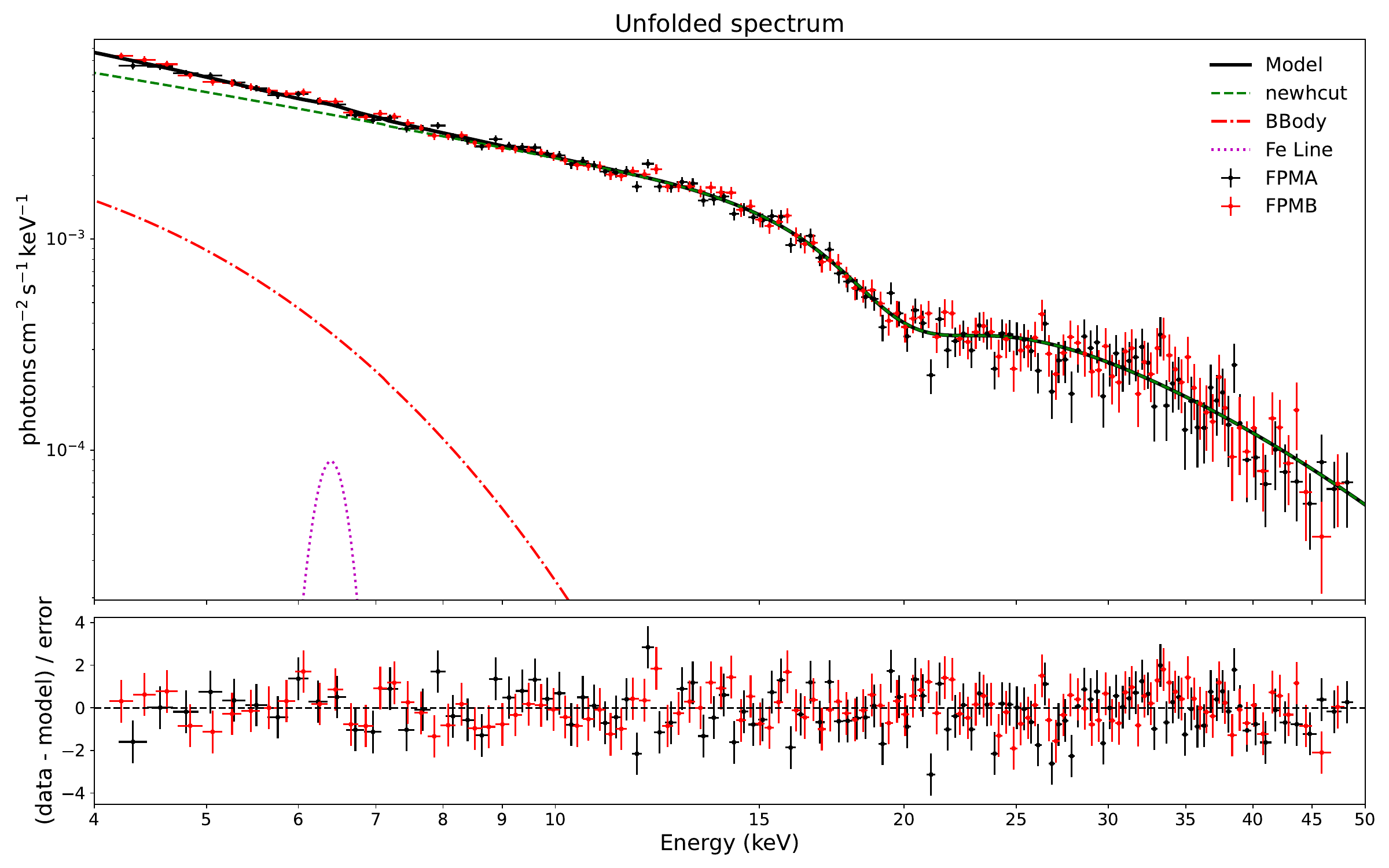}
    \end{subfigure}

    \vskip\baselineskip  

    \begin{subfigure}{0.45\textwidth}
        \centering
        \includegraphics[width=\linewidth]{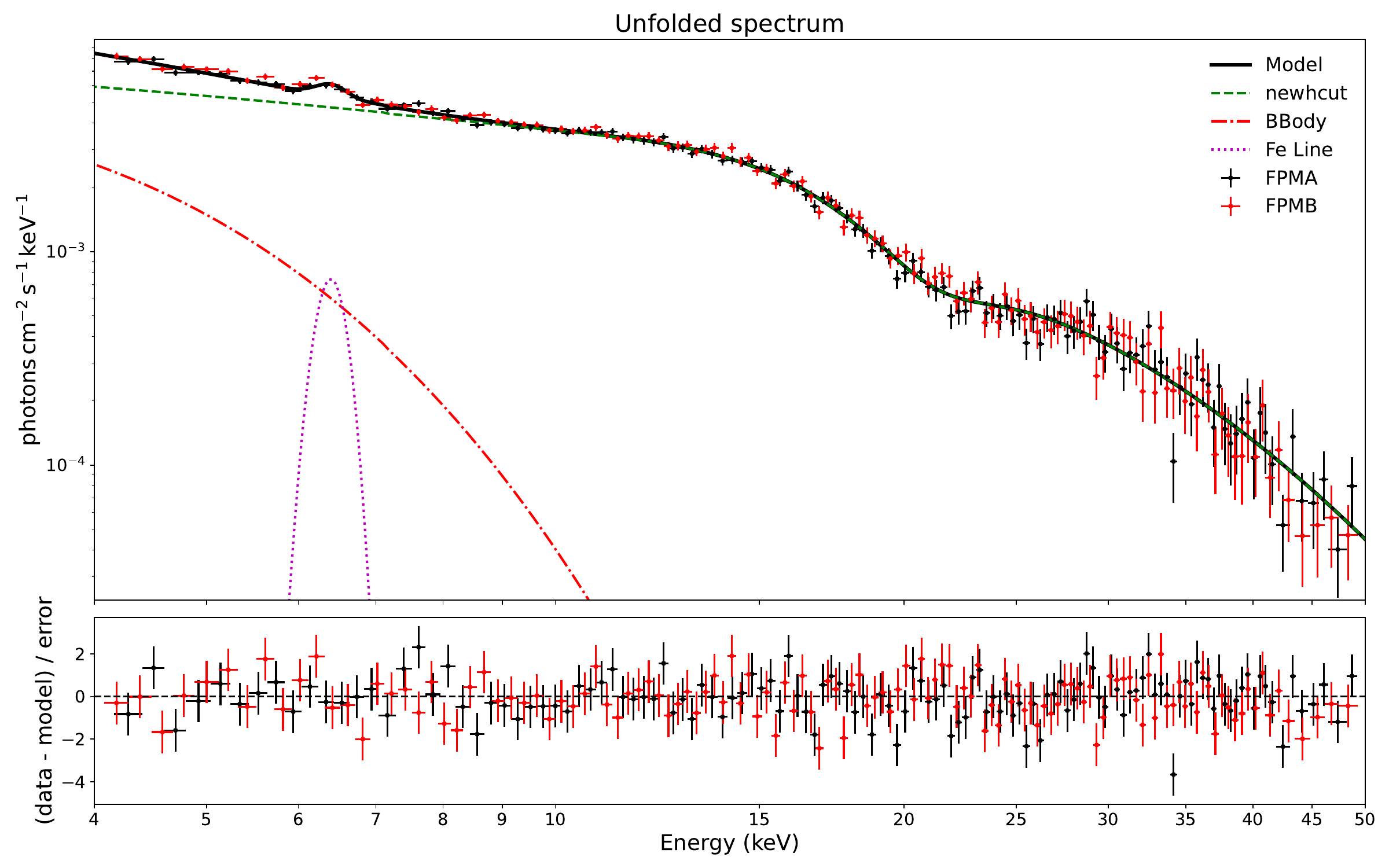}
    \end{subfigure}
    \hfill
    \begin{subfigure}{0.45\textwidth}
        \centering
        \includegraphics[width=\linewidth]{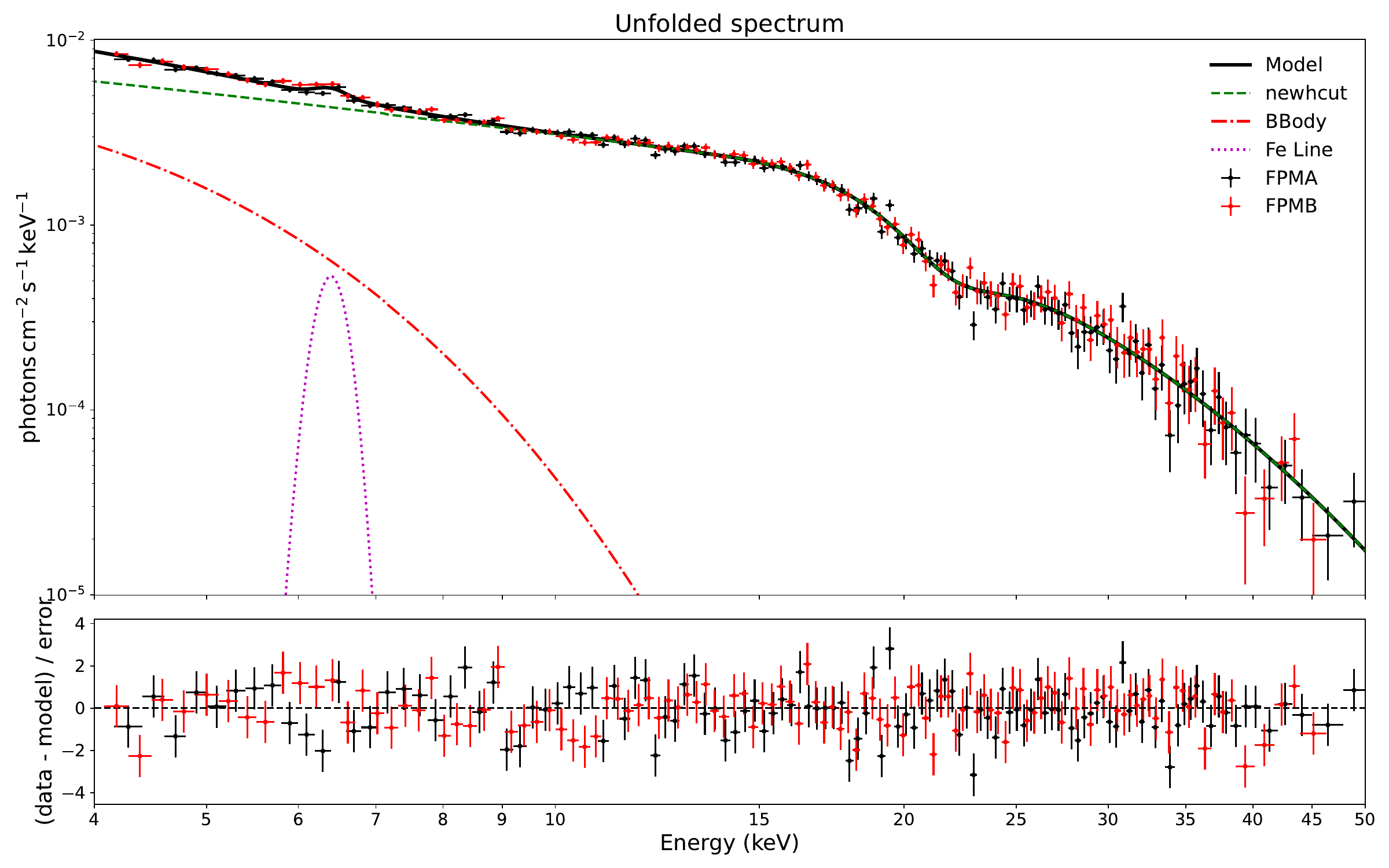}
    \end{subfigure}

    \vskip\baselineskip

    \begin{subfigure}{0.45\textwidth}
        \centering
        \includegraphics[width=\linewidth]{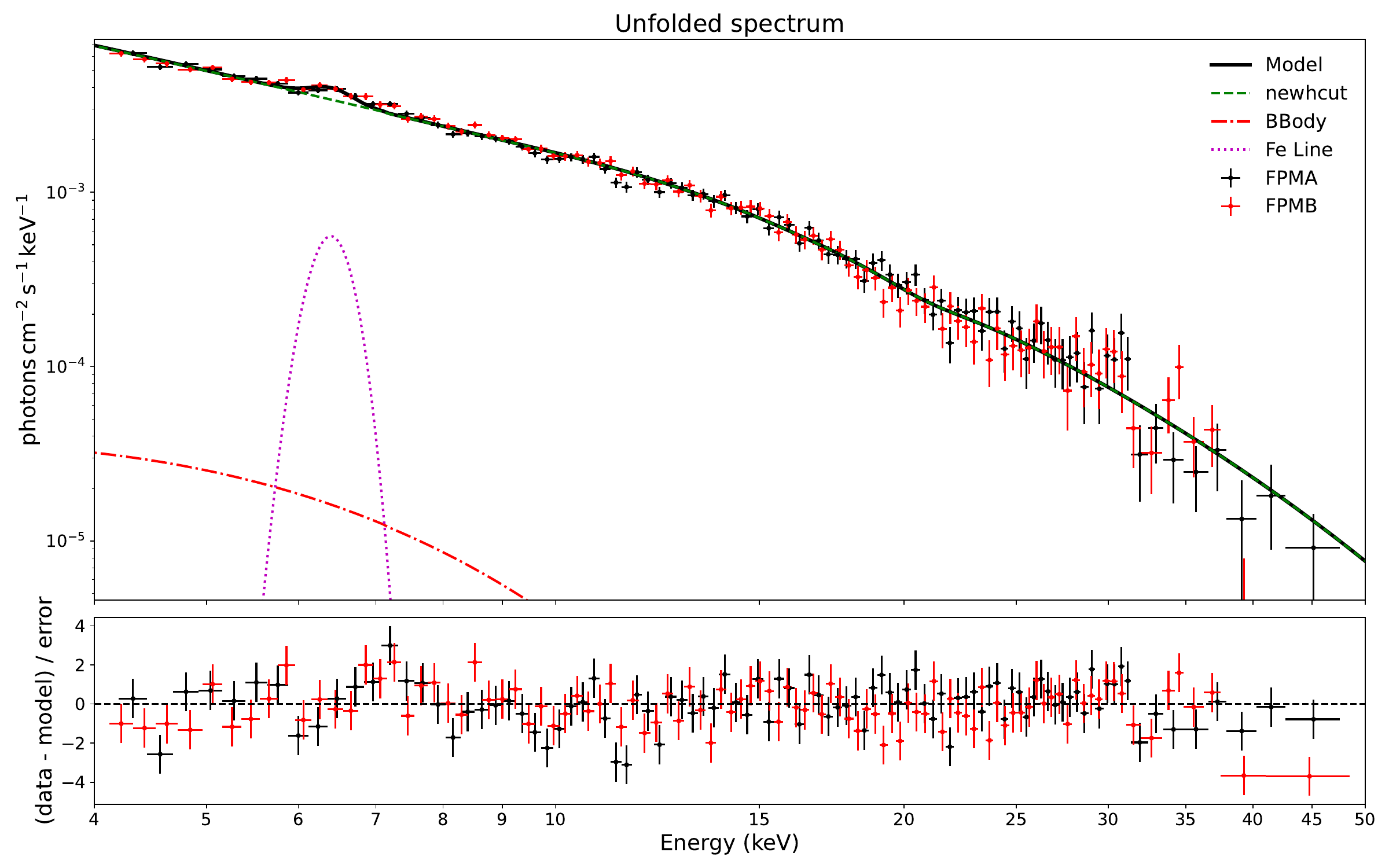}
    \end{subfigure}
    \hfill
    \begin{subfigure}{0.45\textwidth}
        \centering
        \includegraphics[width=\linewidth]{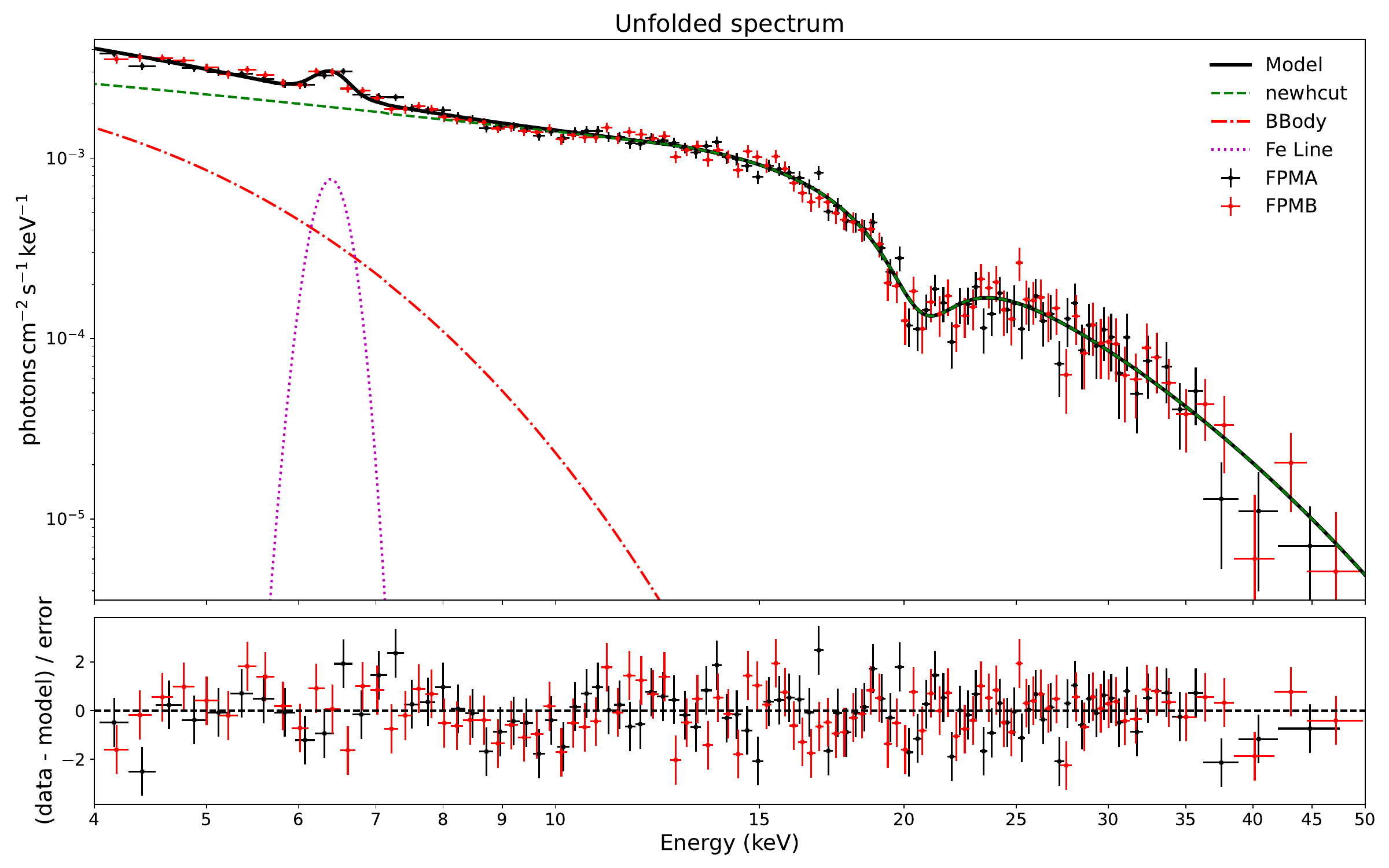}
    \end{subfigure}

    \vskip\baselineskip

    \begin{subfigure}{0.45\textwidth}
        \centering
        \includegraphics[width=\linewidth]{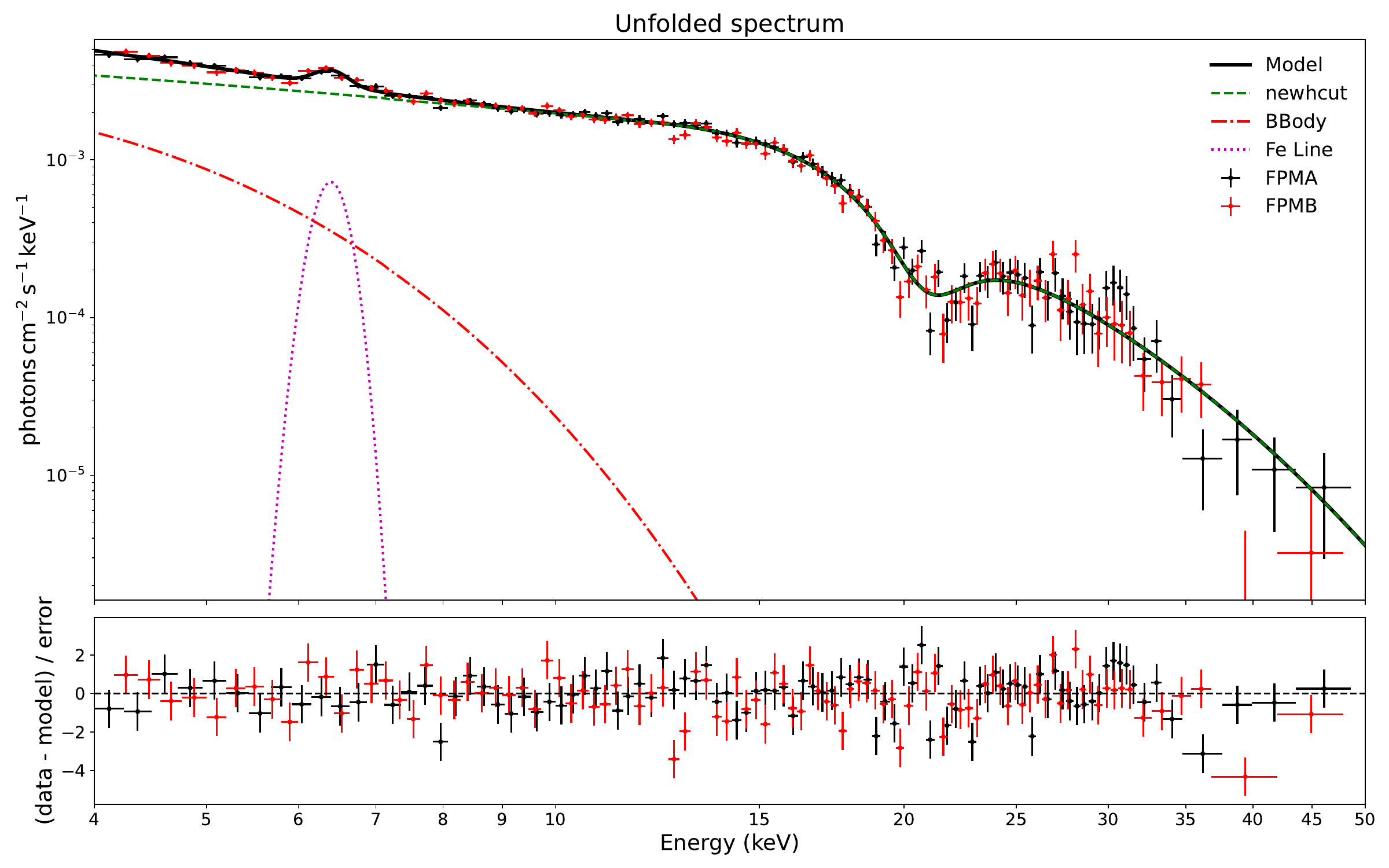}
    \end{subfigure}
    \hfill
    \begin{subfigure}{0.45\textwidth}
        \centering
        \includegraphics[width=\linewidth]{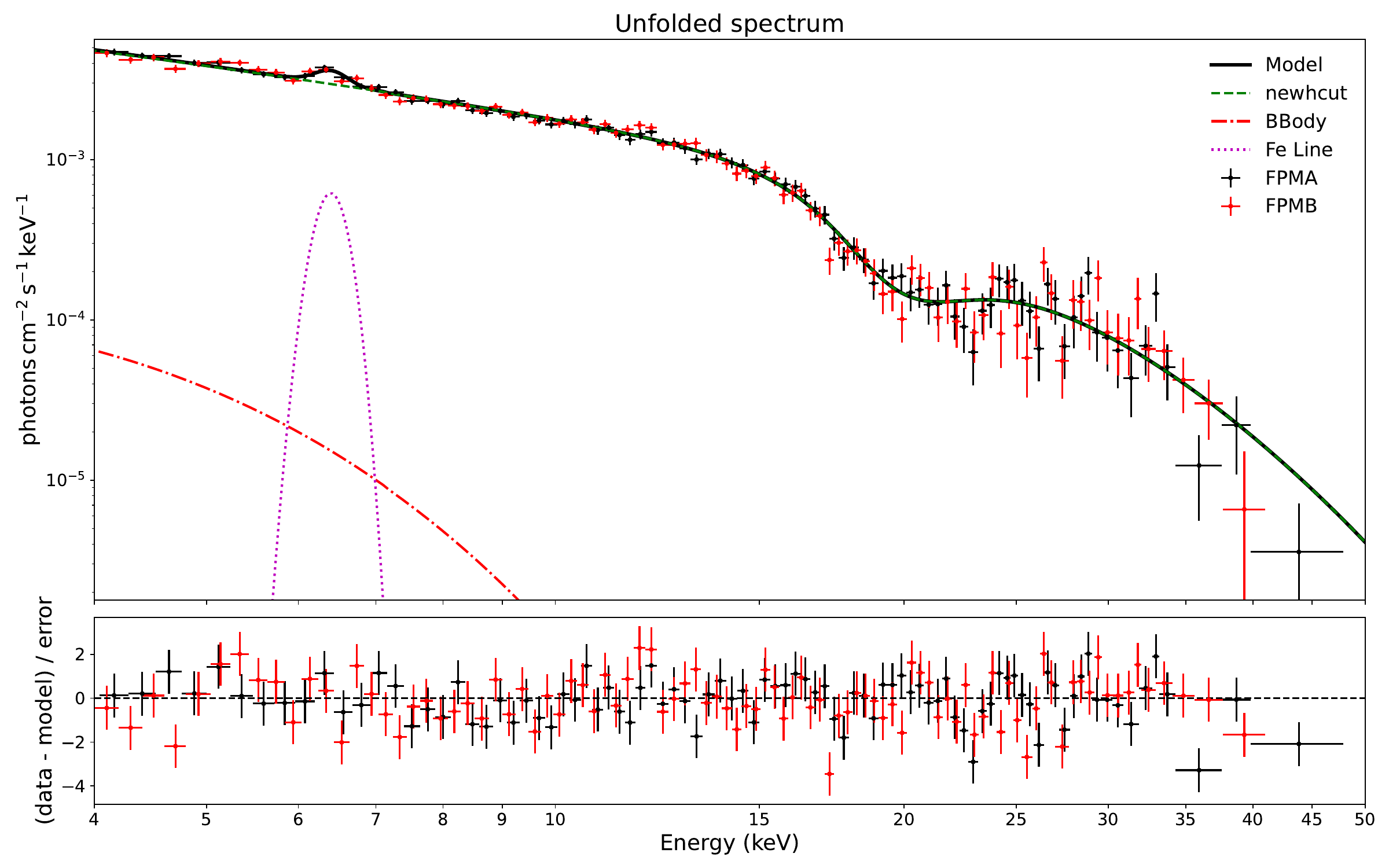}
    \end{subfigure}
    
    \caption{}
    \label{fig:grid3062}
\end{figure}

\onecolumn
\section{Tables of Spectral Analysis}
\label{spectral_tables}

\begin{table}[hbt!]
    \centering
    \caption{best fit parameters of the phase resolved analysis of the ObsID 304}
    \renewcommand{\arraystretch}{1.3} 
    \begin{tabular}{lcccc}
    \toprule
    \textbf{Parameter}  & \textbf{0.000-0.125} & \textbf{0.125-0.250} & \textbf{0.250-0.375} & \textbf{0.375-0.500} \\
    \toprule
    $\mathrm{nH}$ $(10^{22} cm^{-2})$ & (1) & (1)  & (1) & (1) \\
    $E_{\mathrm{cut}}$ (keV)    & $12.6_{-0.5}^{+0.7}$ & $12.3_{-0.3}^{+0.5}$ & $13.8_{-0.3}^{+0.4}$ & $16.0_{-0.2}^{+0.4}$ \\
    $E_{\mathrm{fold}}$ (keV)   & $16.5_{-0.8}^{+1.2}$ & $14.5_{-0.4}^{+0.6}$ & $10.6 \pm 0.2$       & $8.81_{-0.20}^{+0.11}$ \\
    $\Gamma$                    & $1.14_{-0.07}^{+0.09}$ & $0.75_{-0.05}^{+0.07}$ & $0.62 \pm 0.04$     & $0.85_{-0.03}^{+0.02}$ \\
    $E_{\mathrm{width}}$ & (5) & (5) & (5) & (5) \\
    $\mathrm{N}_{\mathrm{newhcut}}$ & $0.023_{-0.003}^{+0.005}$ & $0.021_{-0.002}^{+0.003}$ & $0.0214 \pm 0.0020$ & $0.0319_{-0.0025}^{+0.0019}$ \\
    $\mathrm{kT}$ (keV) & (1) & (1) & (1) & (1) \\
    $N_{\mathrm{bbody}}$        & $(9_{-2}^{+1}) \times 10^{-4}$ & $(1.94_{-0.21}^{+0.15}) \times 10^{-3}$ & $(1.42 \pm 0.17) \times 10^{-3}$ & $(1.75_{-0.13}^{+0.15}) \times 10^{-3}$ \\
    \bottomrule
    $D_\mathrm{cyc}$        & $0.42 \pm 0.04$        & $0.55 \pm 0.02$        & $0.47_{-0.02}^{+0.03}$ & $0.56_{-0.03}^{+0.04}$ \\
    $E_{cyc}$ (keV)               & $20.5 \pm 0.3$        & $19.76 \pm 0.16$       & $21.56_{-0.20}^{+0.17}$ & $21.90_{-0.17}^{+0.11}$ \\
    $\sigma_{\mathrm{cyc}}$   & (3.42)                     & (3.42)                     & $3.6_{-0.3}^{+0.4}$    & $2.18_{-0.16}^{+0.34}$ \\
    \bottomrule
    
    $E_{\mathrm{Fe}}$ (keV) & (6.4) & (6.4) & (6.4) & (6.4)\\ 

    $\sigma_{\mathrm{Fe}}$ (keV) & $0.34 \pm 0.03$     & $0.32 \pm 0.04$        & $0.27_{-0.08}^{+0.19}$ & $0.30_{-0.07}^{+0.24}$ \\
    $EW_\mathrm{Fe}$ & $0.210^{+0.037}_{-0.019}$ & $0.070^{+0.017}_{-0.016}$ & $0.043^{+0.016}_{-0.015}$ & $0.064^{+0.014}_{-0.013}$ \\
    $N_{\mathrm{Fe}}$        & $(6.8 \pm 0.8) \times 10^{-4}$ & $(4.3_{-1.0}^{+0.8}) \times 10^{-4}$ & $(3_{-1}^{+1}) \times 10^{-4}$ & $(4.8_{-0.9}^{+2.0}) \times 10^{-4}$ \\
    \bottomrule
    $\mathrm{C}_{\mathrm{FPMB}}$      & $0.987 \pm 0.009$      & $0.974 \pm 0.006$      & $0.986 \pm 0.005$      & $0.973 \pm 0.005$ \\
    W-stat/d.o.f             & $0.854/192$           & $1.049/249$            & $0.944/251$            & $0.883/234$ \\
    \toprule
    \textbf{Parameter} & \textbf{0.500-0.625} & \textbf{0.625-0.750} & \textbf{0.750-0.875} & \textbf{0.875-1.000} \\
    \bottomrule
    $\mathrm{nH}$ $(10^{22} cm^{-2})$ & (1) & (1)  & (1) & (1) \\
    $E_{\mathrm{cut}}$ (keV)    & $12.5_{-0.3}^{+0.3}$ & $14.9_{-0.3}^{+0.5}$ & $14.2_{-0.2}^{+0.3}$ & $12.1_{-0.3}^{+0.4}$ \\
    $E_{\mathrm{fold}}$ (keV)   & $8.5_{-0.2}^{+0.4}$ & $6.90_{-0.17}^{+0.14}$ & $7.17 \pm 0.15$       & $9.2_{-0.3}^{+0.4}$ \\
    $\Gamma$                    & $0.73_{-0.06}^{+0.10}$ & $0.73_{-0.04}^{+0.04}$ & $0.73 \pm 0.04$       & $0.84_{-0.06}^{+0.09}$ \\
    $E_{\mathrm{width}}$ & (5) & (5) & (5) & (5) \\
    $\mathrm{N}_{\mathrm{newhcut}}$ & $(9_{-1}^{+3}) \times 10^{-3}$ & $0.0123_{-0.0011}^{+0.0014}$ & $0.0158 \pm 0.0016$ & $0.018_{-0.002}^{+0.004}$ \\
    $\mathrm{kT}$ (keV)         & (1.5) & (1) & (1) & (1) \\
    $N_{\mathrm{bbody}}$        & $(2.59_{-0.18}^{+0.11}) \times 10^{-3}$ & $(7.9_{-1.1}^{+0.8}) \times 10^{-4}$ & $(9 \pm 1) \times 10^{-4}$ & $(7_{-2}^{+1}) \times 10^{-4}$ \\
    \bottomrule
    $D_\mathrm{cyc}$        & $0.24 \pm 0.04$        & $1.18_{-0.04}^{+0.06}$ & $1.34_{-0.04}^{+0.05}$ & $0.90 \pm 0.04$ \\
    $E_{\mathrm{cyc}}$ (keV)    & $20.9 \pm 0.5$        & $20.62_{-0.11}^{+0.09}$ & $20.44_{-0.10}^{+0.08}$ & $19.37 \pm 0.15$ \\
    $\sigma_{\mathrm{cyc}}$     & $2.7_{-0.5}^{+0.9}$    & $2.17_{-0.14}^{+0.23}$ & $2.52_{-0.13}^{+0.20}$ & (3.42)  \\
    \bottomrule
    
    $E_{\mathrm{Fe}}$ (keV)     & (6.4) & (6.4) & (6.4) & (6.4)\\ 
    $\sigma_{\mathrm{Fe}}$ (keV) & $0.33_{-0.03}^{+0.04}$ & $0.35_{-0.03}^{+0.04}$ & $0.22_{-0.08}^{+0.17}$ & $0.33 \pm 0.04$ \\
    $EW_\mathrm{Fe}$ & $0.060^{+0.014}_{-0.015}$ & $0.129^{+0.020}_{-0.018}$ & $0.070^{+0.018}_{-0.016}$ & $0.138^{+0.037}_{-0.023}$ \\
    $N_{\mathrm{Fe}}$        & $(2.8 \pm 0.7) \times 10^{-4}$ & $(4.6 \pm 0.7) \times 10^{-4}$ & $(3.2_{-0.7}^{+1.1}) \times 10^{-4}$ & $(6.0 \pm 0.8) \times 10^{-4}$ \\
    \bottomrule
    $\mathrm{C}_{\mathrm{FPMB}}$      & $0.982 \pm 0.007$      & $0.969 \pm 0.007$      & $0.986 \pm 0.006$      & $0.979 \pm 0.008$ \\
    W-stat/d.o.f               & $0.834/187$           & $0.736/186$            & $1.005/194$            & $1.008/185$ \\
    \bottomrule
    \end{tabular}
\end{table}

\begin{table}
    \centering
    \caption{Best fit parameters of the phase resolved analysis of ObsID 306-2}
    \renewcommand{\arraystretch}{1.3} 
    \begin{tabular}{lcccc}
    \toprule
    \textbf{Parameter} &\textbf{ 0.000-0.125} & \textbf{0.125-0.250} & \textbf{0.250-0.375} & \textbf{0.375-0.500} \\
    \toprule
    $\mathrm{nH}$ $(10^{22} cm^{-2})$            & (1) & (1) & (1) & (1) \\
    $E_{\mathrm{cut}}$ (keV) & $16.2_{-0.9}^{+1.7}$ & $14.5_{-0.6}^{+1.0}$ & $14.0_{-0.3}^{+0.4}$ & $16.5_{-0.4}^{+1.1}$ \\
    $E_{\mathrm{fold}}$ (keV)& $14.3_{-1.3}^{+1.0}$ & $17.1 \pm 0.8$       & $10.3 \pm 0.3$       & $8.5_{-0.5}^{+0.2}$ \\
    $\Gamma$                 & $1.34_{-0.03}^{+0.05}$ & $1.03 \pm 0.04$    & $0.54_{-0.04}^{+0.05}$ & $0.75_{-0.05}^{+0.03}$ \\
    $E_{\mathrm{width}}$ & (5) & (5) & (5) & (5) \\
    $\mathrm{N}_{\mathrm{newhcut}}$ & $0.0222_{-0.0016}^{+0.0030}$ & $0.027_{-0.002}^{+0.003}$ & $0.0131_{-0.0012}^{+0.0015}$ & $0.0178_{-0.0018}^{+0.0013}$ \\
    $\mathrm{kT}$ (keV)      & (1) & (1) & (1) & (1) \\
    $N_{\mathrm{bbody}}$     & $(3.2_{-0.9}^{+0.5}) \times 10^{-4}$ & $(7_{-2}^{+1}) \times 10^{-4}$ & $(1.11_{-0.16}^{+0.13}) \times 10^{-3}$ & $(1.18_{-0.11}^{+0.14}) \times 10^{-3}$ \\
    \bottomrule
    $\tau$                  & $0.74_{-0.06}^{+0.12}$ & $0.78_{-0.04}^{+0.07}$ & $0.59_{-0.03}^{+0.04}$ & $0.57_{-0.04}^{+0.11}$ \\
    $E_{\mathrm{cyc}}$ (keV) & $19.99_{-0.17}^{+0.42}$ & $19.71 \pm 0.19$  & $20.8 \pm 0.2$       & $21.8_{-0.4}^{+0.2}$ \\
    $\sigma_{\mathrm{cyc}}$ & (3.82) & (3.82) & (3.82) & $3.1_{-0.4}^{+0.7}$ \\
    \bottomrule

    $E_{\mathrm{Fe}}$ (keV)  & (6.4) & (6.4) & (6.4) & (6.4)\\ 
    $\sigma_{\mathrm{Fe}}$ (keV) & $0.19_{-0.02}^{+0.03}$ & $0.20 \pm 0.04$   & $0.19_{-0.03}^{+0.04}$ & $0.19 \pm 0.04$ \\
    $EW_\mathrm{Fe}$ & $0.107^{+0.037}_{-0.023}$ & < 0.007 & $0.067^{+0.015}_{-0.015}$ & $0.052^{+0.016}_{-0.015}$ \\
    $N_{\mathrm{Fe}}$     & $(2.2_{-0.3}^{+0.4}) \times 10^{-4}$ & $< 22 \times 10^{-5}$ & $(3.6 \pm 0.8) \times 10^{-4}$ & $(2.6 \pm 0.8) \times 10^{-4}$ \\
    \bottomrule
    $\mathrm{C}_{\mathrm{FPMB}}$ & $1.034 \pm 0.012$ & $1.010 \pm 0.010$ & $1.020 \pm 0.008$ & $1.011 \pm 0.008$ \\
    W-stat/d.o.f     & $0.785/175$ & $0.686/231$ & $0.747/246$ & $0.826/224$ \\
    \hline
    \textbf{Parameter} & \textbf{0.500-0.625} & \textbf{0.625-0.750} & \textbf{0.750-0.875} & \textbf{0.875-1.000} \\
    \hline
    $\mathrm{nH}$ $(10^{22} cm^{-2})$ & (1) & (1) & (1) & (1) \\
    $E_{\mathrm{cut}}$ (keV) & $12.2 \pm 0.5$ & $14.8_{-0.3}^{+0.9}$ & $14.9_{-0.2}^{+1.6}$ & $18.2_{-1.0}^{+1.5}$ \\
    $E_{\mathrm{fold}}$ (keV)& $13.3 \pm 0.6$ & $7.8 \pm 0.3$       & $6.67_{-0.42}^{+0.16}$ & $7.5_{-0.8}^{+0.5}$ \\
    $\Gamma$                 & $1.583_{-0.024}^{+0.018}$ & $0.70_{-0.05}^{+0.07}$ & $0.63_{-0.05}^{+0.06}$ & $1.07_{-0.07}^{+0.02}$ \\
    $E_{\mathrm{width}}$ & (5) & (5) & (5) & (5) \\
    $\mathrm{N}_{\mathrm{newhcut}}$ & $0.065_{-0.003}^{+0.002}$ & $(7.1_{-0.8}^{+1.3}) \times 10^{-3}$ & $(8.6_{-0.8}^{+1.3}) \times 10^{-3}$ & $0.0220_{-0.0032}^{+0.0011}$ \\
    $\mathrm{kT}$ (keV)      & (1.5) & (1) & (1) & (1) \\
    $N_{\mathrm{bbody}}$     & $(1.8_{-1.7}^{+0.7}) \times 10^{-5}$ & $(6.4_{-1.0}^{+0.7}) \times 10^{-4}$ & $(6.5_{-1.3}^{+0.8}) \times 10^{-4}$ & $< 3.9 \times 10^{-4}$ \\
    \bottomrule
    $\tau$   & $0.14_{-0.02}^{+0.03}$ & $1.05_{-0.07}^{+0.10}$ & $1.24_{-0.05}^{+0.16}$ & $1.52_{-0.11}^{+0.19}$ \\
    $E_{\mathrm{cyc}}$ (keV) & $20.63_{-0.83}^{+0.16}$ & $20.61_{-0.19}^{+0.12}$ & $20.74_{-0.28}^{+0.10}$ & $19.27_{-0.15}^{+0.26}$ \\
    $\sigma_{\mathrm{cyc}}$ & (3.82) & $1.93_{-0.18}^{+0.48}$ & $2.24_{-0.15}^{+0.60}$ & (3.82) \\
    
    \bottomrule
    $E_{\mathrm{Fe}}$ (keV)  & (6.4) & (6.4) & (6.4) & (6.4)\\ 
    $\sigma_{\mathrm{Fe}}$ (keV) & $0.260_{-0.038}^{+0.020}$ & $0.22 \pm 0.04$   & $0.21 \pm 0.03$       & $0.20_{-0.03}^{+0.04}$ \\
    $EW_\mathrm{Fe}$ & $0.107^{+0.020}_{-0.021}$ & $0.188^{+0.028}_{-0.026}$ & $0.129^{+0.024}_{-0.021}$ & $0.111^{+0.021}_{-0.021}$ \\
    $N_{\mathrm{Fe}}$     & $(3.7_{-0.9}^{+0.7}) \times 10^{-4}$ & $(4.3 \pm 0.6) \times 10^{-4}$ & $(3.9 \pm 0.7) \times 10^{-4}$ & $(3.2 \pm 0.7) \times 10^{-4}$ \\
    \bottomrule
    $\mathrm{C}_{\mathrm{FPMB}}$ & $1.014 \pm 0.007$ & $1.000 \pm 0.012$ & $0.995 \pm 0.010$ & $1.002 \pm 0.012$ \\
    W-stat/d.o.f   & $0.976/185$ & $0.732/187$ & $0.875/184$ & $0.847/174$ \\
    \bottomrule
    \end{tabular}
\end{table}

\clearpage

\onecolumn
\section{Good Time Intervals}
\label{gtis_app}

\begin{figure}[!hbt]
    \centering
    \begin{tabular}{cc}
    \includegraphics[width=0.48\textwidth]{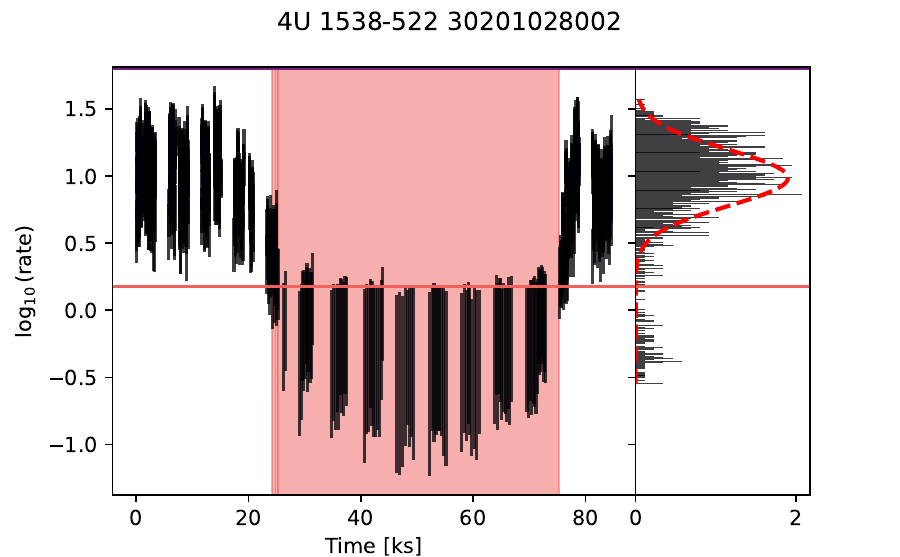}  & 
    \includegraphics[width=0.48\textwidth]{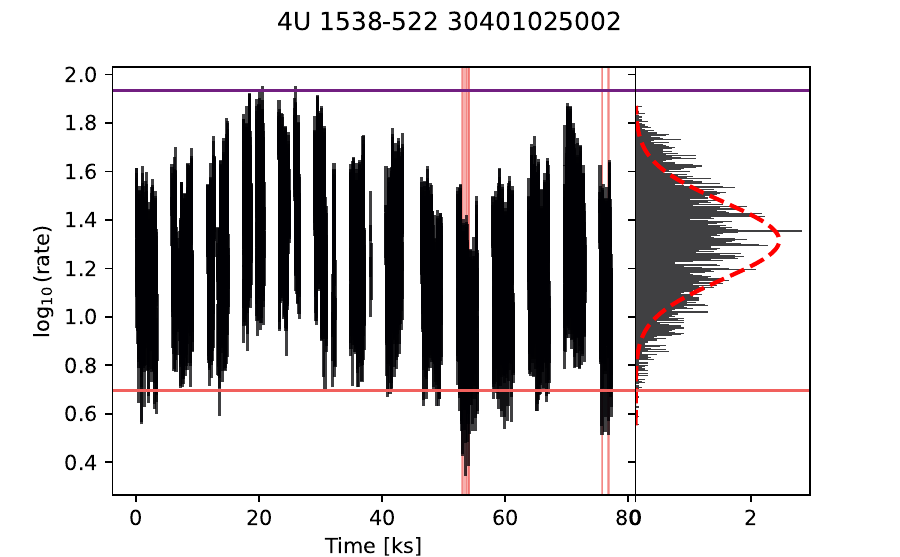} \\
    \includegraphics[width=0.48\textwidth]{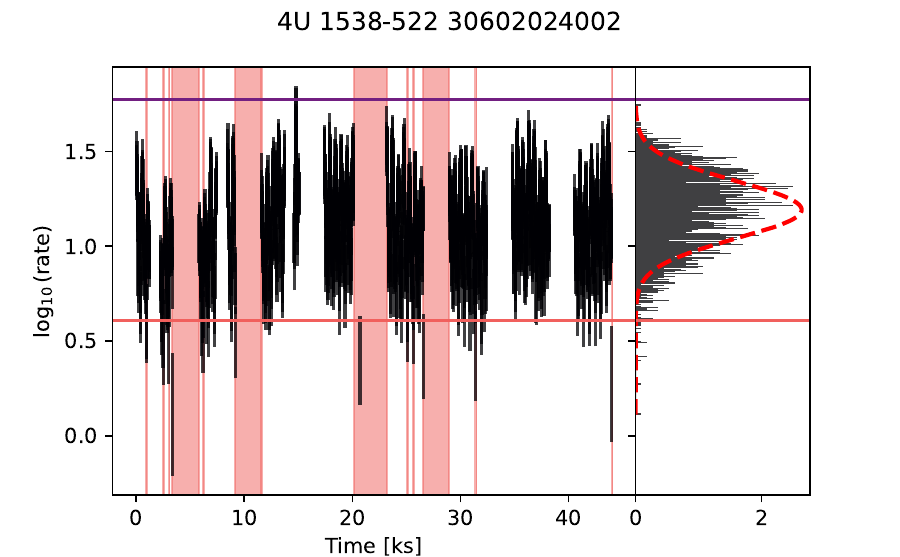} & 
    \includegraphics[width=0.48\textwidth]{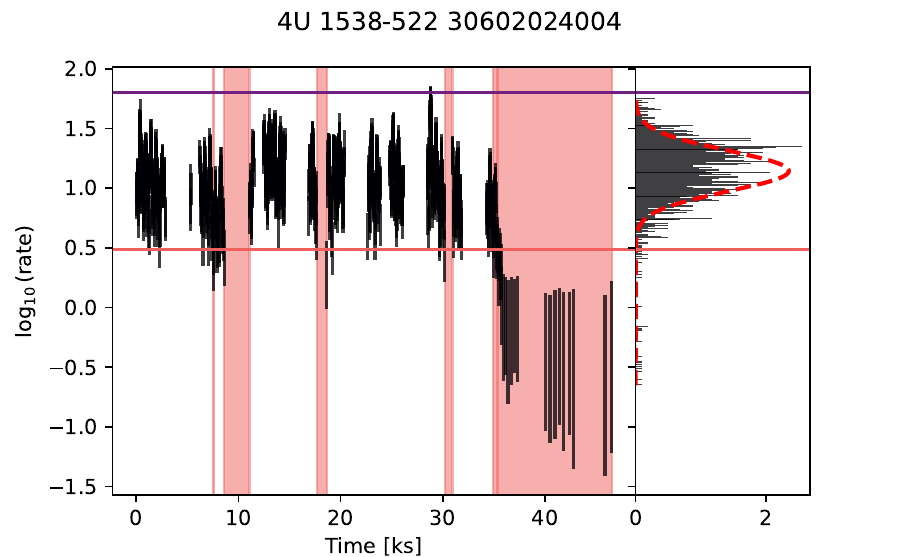} \\
    \end{tabular}
    \caption{Filtering of light curves using a log-normal distribution of counts in the 3–70 keV band. Left panels: Total light curve of each observation. The horizontal lines mark the adopted count rate limits, while the pink vertical bands indicate time intervals that were excluded from further analysis. Right panels: Logarithmic histograms of count rates, with horizontal lines denoting the same adopted limits, where good time intervals were defined as those corresponding to count rates within the \(\pm5\sigma\) of the central peak of the log-normal distribution.}
    \label{fig:enselectedpp}
\end{figure}

\end{document}